\documentclass{aa}  

\usepackage{graphicx}
\usepackage{txfonts}
   \newcommand{\unit}[1]{\ensuremath{\, \mathrm{#1}}}
   \newcommand{\mesa}{{\sc mesa}}          % code names
   \newcommand{\ramses}{{\sc ramses}}          % code names
   \newcommand{\vapor}{{\sc vapor}}          % code names
   \newcommand{\athena}{{\sc athena}}          % code names
   \newcommand{\cloudy}{{\sc cloudy}}          % code names
   \newcommand{\polaris}{{\sc polaris}}          % code names
   \newcommand{\Fig}[1]{Fig.~\ref{fig:#1}}    % Fig. reference
   \newcommand{\Figure}[1]{Figure~\ref{fig:#1}}    % Figure reference

\begin{document}

   \title{Linear dust polarization during the embedded phase of protostar formation}

    \titlerunning{Toward a complete understanding of dust polarization}

   \subtitle{Synthetic observations of bridge structures}

   \author{M. Kuffmeier
          \inst{1}\fnmsep\thanks{\email{ru151@uni-heidelberg.de}, International Postdoctoral Fellow of Independent Research Fund Denmark (IRFD)},
          S. Reissl \inst{1}, S. Wolf \inst{2}, I. Stephens \inst{3} \and H. Calcutt \inst{4}
          }

\institute{\label{inst1} Zentrum für Astronomie der Universität Heidelberg, Institut für Theoretische Astrophysik, Albert-Ueberle-Str. 2, 69120 Heidelberg\and
\label{inst2} Institut für Theoretische Physik und Astrophysik, Christian-Albrechts-Universität zu Kiel, Leibnizstraße 15, 24118 Kiel, Germany \and
 \label{inst3} Harvard-Smithsonian Center for Astrophysics, 60 Garden Street, Cambridge, MA 02138, USA \and  \label{inst4} Department of Space, Earth and Environment, Chalmers University of Technology, H\"{o}rsalsv\"{a}gen 11, 412 96, Gothenburg, Sweden
}

   \date{Received \today}

% \abstract{}{}{}{}{} 
% 5 {} token are mandatory
 
  \abstract
  % context heading (optional)
  % {} leave it empty if necessary  
   {Measuring polarization from thermal dust emission can provide important constraints on the magnetic field structure around embedded protostars. However, interpreting the observations is challenging without models that consistently account for both the complexity of the turbulent protostellar birth environment and polarization mechanisms.}
  % aims heading (mandatory)
   {We aim to provide a better understanding of dust polarization maps of embedded protostars with a focus on bridge-like structures such as the structure observed toward the protostellar multiple system IRAS 16293--2422 by comparing synthetic polarization maps of thermal reemission with recent observations.}
  % methods heading (mandatory)
   {We analyzed the magnetic field morphology and properties associated with the formation of a 
   protostellar multiple
   based on ideal magnetohydrodynamic 3D zoom-in simulations carried out with the \ramses\ code. 
   To compare the models with observations, we postprocessed a snapshot of a bridge-like structure that is associated with a forming triple star system with the radiative transfer code \polaris\ and produced multiwavelength dust polarization maps.
   }
  % results heading (mandatory)
   {The typical density in the most prominent bridge of our sample is about $10^{-16} \unit{g}\unit{cm}^{-3}$, and the magnetic field strength in the bridge is about 1 to 2 mG. Inside the bridge, the magnetic field structure has an elongated toroidal morphology, and the dust polarization maps trace the complex morphology. In contrast, the magnetic field strength associated with the launching of asymmetric bipolar outflows is significantly more magnetized ($\sim$100 $\unit{mG}$).
   At $\lambda=1.3$ mm, and the orientation of the grains in the bridge is very similar for the case accounting for radiative alignment torques (RATs) compared to perfect alignment with magnetic field lines. However, the polarization fraction in the bridge is three times smaller for the RAT scenario than when perfect alignment is assumed.    
   At shorter wavelength ($\lambda \lesssim 200 \mu$m), however, dust polarization does not trace the magnetic field because other effects such as self-scattering and dichroic extinction dominate the orientation of the polarization.
   }
  % conclusions heading (optional), leave it empty if necessary 
   {Compared to the launching region of protostellar outflows, the magnetic field in bridge-like structures is weak. 
   Synthetic dust polarization maps of ALMA Bands 6 and 7 (1.3 mm and 870 $\mu$m, respectively) can be used as a tracer of the complex morphology of elongated toroidal magnetic fields associated with bridges. 
}


   \maketitle
%
%________________________________________________________________

\section{Introduction}
Theory suggests that magnetic fields play an important role in star formation \citep[see the review by][and references therein]{PudritzRay2019}. For example, magnetic fields can transport angular momentum from the forming disk to the larger scales through a process called magnetic braking during the protostellar collapse phase \citep{LuestSchlueter1955,Mestel1956}. Moreover, magnetic fields can lead to the launching of outflows such as bipolar jets and disk winds \citep{BlandfordPayne1982,PudritzNorman1983}. With (sub-)millimeter long-baseline interferometers, it became possible to resolve low- and high-velocity outflows together with Keplerian disks of protostellar objects \citep{SargentBeckwith1987,Agra-Amboage2011,Bjerkeli2016,Chen2016,Hirota2017,LeeJet2017}. 
%These results suggest that magnetic fields are important during the early phase of star and disk formation. 
In the classical model of low-mass star formation, a single protostar forms from the collapse of a spherical prestellar core \citep{Shu1977}. As the magnetic field is coupled to the gas, models predict a characteristic hour-glass shape during the collapse, which is followed by the the launching of symmetrical bipolar outflows with a corresponding symmetrical magnetic field structure \citep[e.g.,][]{Allen2003}. 
However, high-resolution observations of protostars show a more complex pattern during the star formation process \citep[e.g.,][]{LeGouellec2019}.
The reason is that stars neither form as isolated entities nor in a static medium, as assumed in classical models. Stars form from collapsing prestellar cores that are the densest parts of a filamentary giant molecular cloud (GMC) \citep{Andre2010}, and turbulence causes significant deviations from symmetry \citep{Padoan_turbfrag,MacLowKlessen2004}.
Moreover, observations show that a significant portion of stars are parts of binary or higher order systems \citep{Duquennoy-Mayor1991,Connelley2008,Raghavan2010}, indicating that stars often form together with companions, as seen in surveys of Class 0 young stellar objects \citep[YSOs;][]{Chen2013,Tobin2016_multisurvey,Tobin2020,Maury2019}.

However, constraining the role of magnetic fields observationally is challenging as the magnetic fields themselves are invisible. Therefore we are forced to trace magnetic fields indirectly in observations. A powerful method for constraining the magnetic field structure is through polarization observations of thermal dust emission, at least at lower densities beyond $\sim$100 AU from the star \citep{Girart2006,Rao2009,Stephens2013,Qiu2014,Hull2014,Hull2017,Sadavoy2019}. At smaller radial distances from the star, dust grows in the disk to $\gtrsim$10 $\mu$m, and hence the polarization of dust grains is likely a result of self-scattering \citep{Kataoka2015,Yang2016}.

\citet{KuffmeierBridge} presented the first zoom-in simulations of the formation of a protostellar triple system, where the companions form with a wide separation of $\sim$1000 AU in distance. In agreement with observations of objects such as the protostellar multiple IRAS 16293--2422 \citep{Pineda2012,Jacobsen2018-IRAS16293,vanderWiel2019}, two of the protostellar companions are connected with transient bridge structures.
In this paper we follow up on that work and analyze the magnetic properties of the forming triple system. Furthermore, we produce synthetic polarization maps using the radiative transfer code \polaris\footnote{http://www1.astrophysik.uni-kiel.de/$\sim$polaris/}\ \citep{Reissl2016} to allow appropriate comparisons of the bridge in our model with observations.

Section 2 describes the magnetohydrodynamical (MHD) zoom-in simulations using the adaptive mesh refinement (AMR) code \ramses\ and the postprocessing using the radiative transfer code \polaris.  
The magnetic field structure associated with the prominent bridge-structure and the corresponding synthetic maps of polarized dust for different wavelengths are presented in section 3. In section 4 we discuss the limitations of our model and elaborate on the implications by comparing our results with observations.
Section 5 presents a summary of the key results and the conclusions of this study.

\begin{figure}
    \includegraphics[width=\columnwidth]{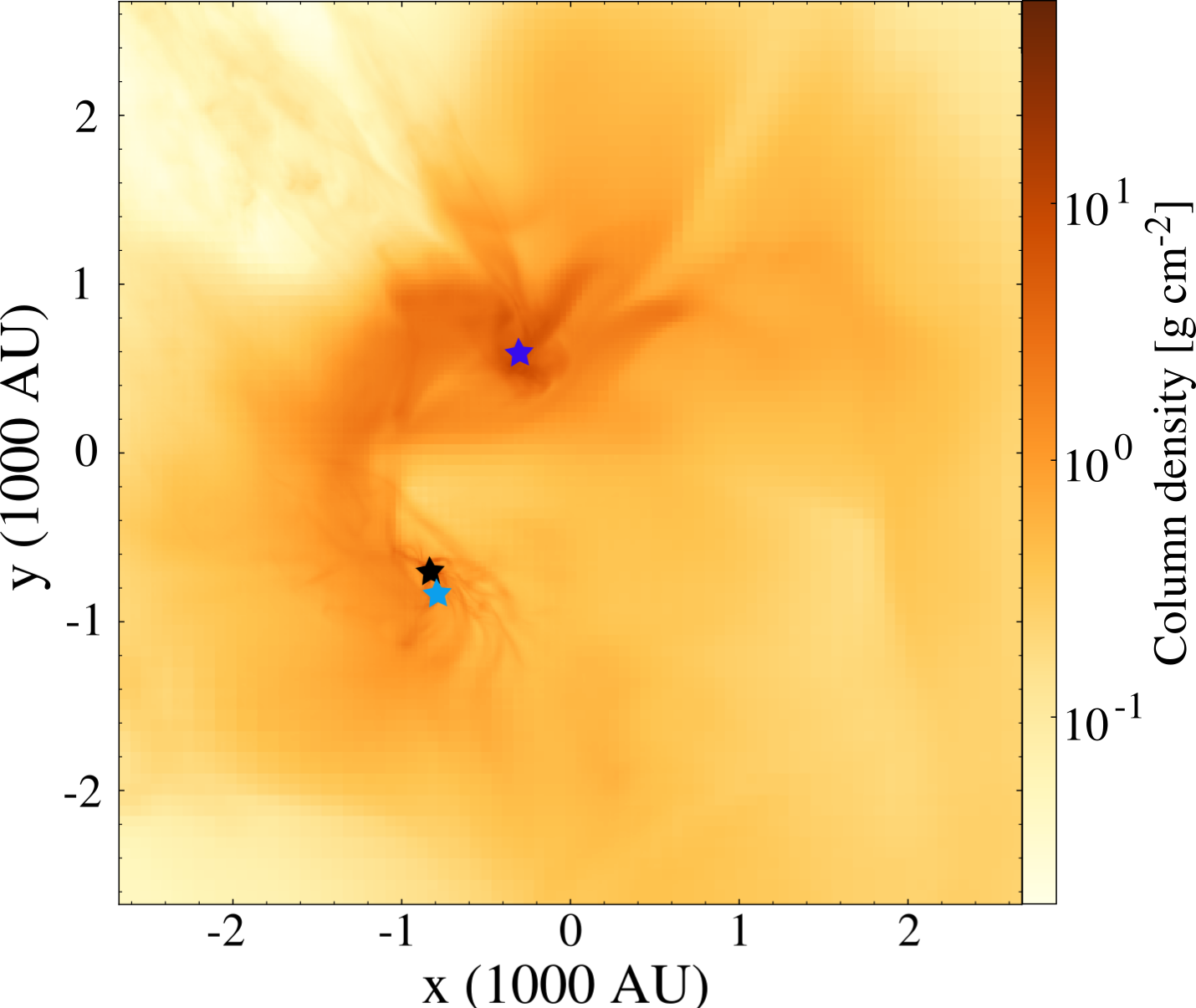} 
        \caption{Column density seen along the z-axis of the coordinate system $\approx$ 4 kyr prior to the formation of the third protostar (blue star), when the primary S1 (black star) is $\approx$70 kyr old and the secondary S2 (cyan star) is $\approx$27 kyr old. }
        \label{fig:bridge_Sigma_repro}
\end{figure}

\begin{figure*}
\begin{center}
    \includegraphics[width=0.49\textwidth]{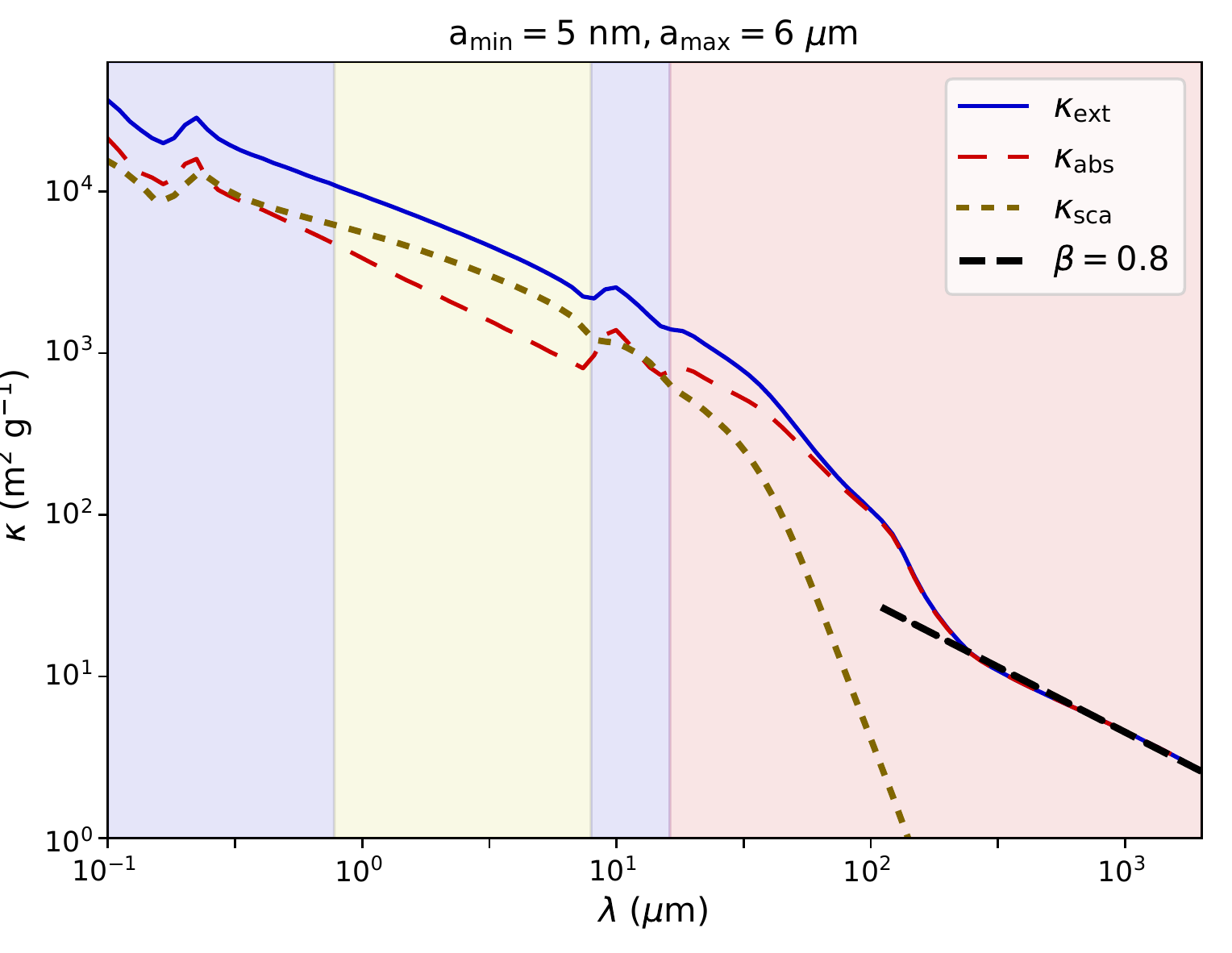}
    \includegraphics[width=0.49\textwidth]{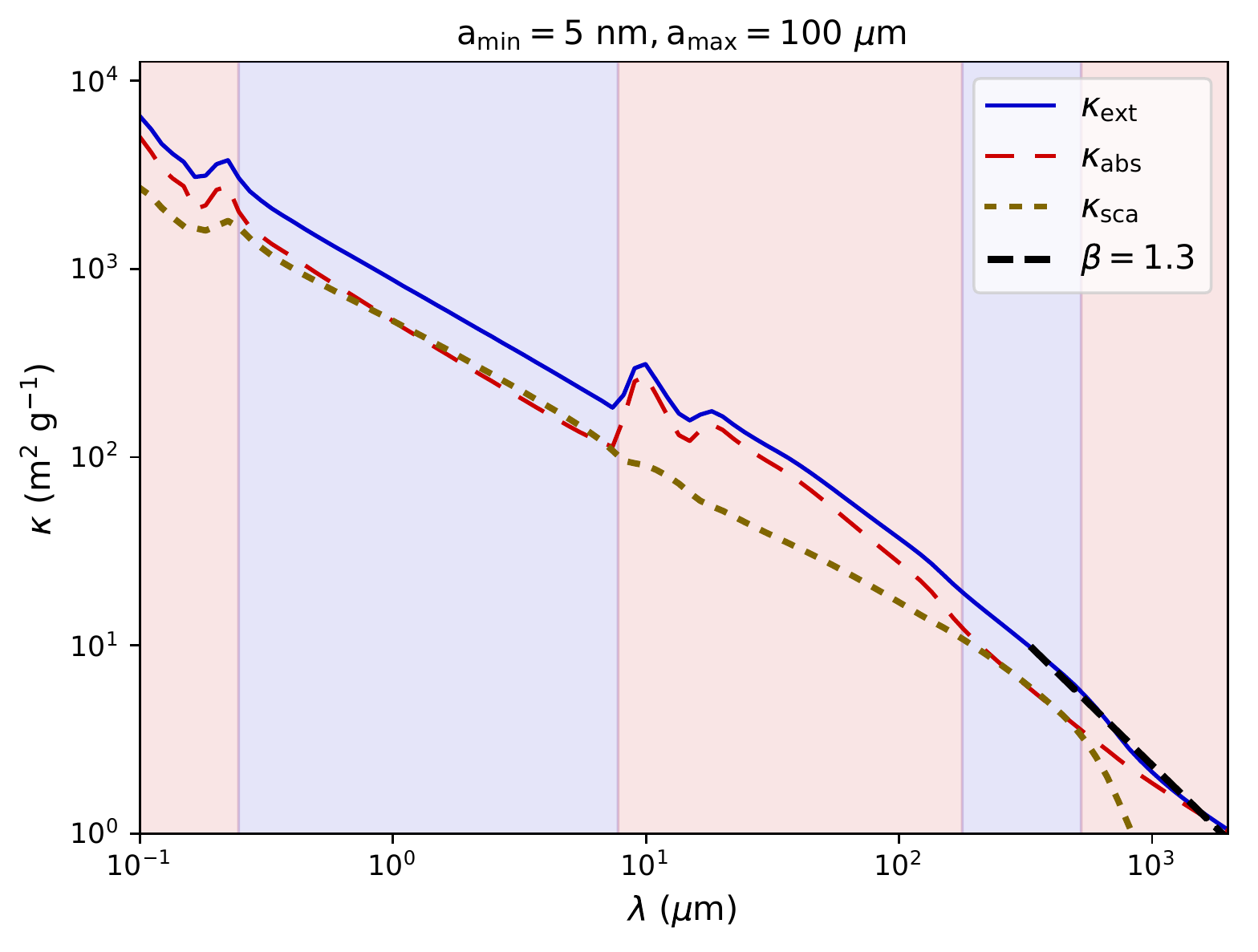}
\end{center}
\caption{Left: Mass opacities for scattering $\kappa_{\mathrm{sca}}$ (short-dashed ocher), absorption $\kappa_{\mathrm{abs}}$ (long-dashed red), and extinction $\kappa_{\mathrm{ext}} = \kappa_{\rm abs} + \kappa_{\rm sca}$ (solid blue) for a dust model with grain sizes between $a_{\mathrm{min}}=5\ \mathrm{nm}$ and $a_{\mathrm{max}}=6\ \mu$m, leading to an opacity index of $\beta = 0.8$.  The shaded areas indicate the wavelength regimes where dust polarization may be dominated by dichroic extinction (light blue), scattering (light yellow), and emission (light red). The shaded areas are estimated under the assumption that the product of dust number density $\rho_{\mathrm{dust}}$ and path length $\ell$ gives
$\rho_{\mathrm{dust}} \times \ell \approx 1\ \mathrm{g\ cm}^{-2}$ (see Sect. \ref{sect:ObsConditions}  and \citet{Reissl2014} for further details). Right: Same as the left panel, but with an upper grain size of   $a_{\mathrm{max}}=100\ \mu\mathrm{m,}$ corresponding to an opacity index of $\beta = 1.3$.}
\label{fig:DustModel}
\end{figure*}

\section{Methods}
\subsection{MHD zoom-in simulations}
We carried out the MHD zoom-in simulations with a modified version of the AMR code \ramses\ \citep{Teyssier2002,Fromang2006}. 
A detailed description of the simulations can be found in \citet{Kuffmeier2016} and \citet{Kuffmeier2017}; we briefly summarize the main parameters of the zoom-in model here.
The cubical box size of the GMC is (40 pc)$^{3}$, and the cloud evolves for about 5 Myr with a highest resolution of 16 levels of refinement with respect to the box length, which corresponds to a minimum cell size of $\Delta x = 2^{-16}\times40$ pc $\approx 126$ AU. 
The stars are modeled as sink particles, and we use an accretion recipe similar to that of \citet{Federrath2011}, as explained in detail in \cite{Haugboelle2018}. 
The turbulence in our box is driven by supernova explosions of type II from massive stars \citep{Kuffmeier2016}. At the end of their mass-dependent lifetime, energy is injected into the box at the location of the type II supernova explosion.  
To account for the thermodynamics in the GMC, we use heating and cooling tables that are based on models of \cite{Gnedin_Hollon_2012}, who used \cloudy\ models \citep{Ferland1998} (optically thin cooling), as explained in \citet{Kuffmeier2017,KuffmeierBridge}.
To account for UV shielding of photoelectric heating at higher densities, the temperature is tapered down exponentially to $T=10 \unit{K}$ for number densities $n>200 \unit{cm}^{-3}$ \citep[see also][]{Padoan2016}. 
As protostellar heating is ignored in our model, most of the gas in the densest regions is quasi-isothermal and cold.

For the zoom-in, we select individual stars that have accreted less than 2 M$_{\odot}$ by the end of the simulation. 
For each zoom-in simulation, we run the simulation again from the snapshot prior to the formation of the selected sink and apply higher resolution in the region where this star forms, while the rest of the box is evolved at coarser resolution. We do not excise part of the box, but keep the full domain of (40 pc)$^3$.
In the case of the zoom-in simulation of the protostellar multiple in this paper, we used a minimum cell size of $\Delta x \approx 2$ AU until about $t = 43$ kyr, and then continued with a coarser resolution of $\Delta x \approx 4$ AU afterward. Generally, using the zoom-in method circumvents the simplified assumption of modeling a dense core as a collapsing sphere that is detached from the GMC environment. The zoom-in procedure provides adequate initial conditions of the dynamically evolving prestellar cores, and prevents assuming possibly ill-defined boundary conditions.

In this paper, we focus on the most prominent bridge-structure introduced in \citet{KuffmeierBridge} (see \Fig{bridge_Sigma_repro}). 
At this snapshot, the primary protostar is $\approx$70 kyr old and the secondary protostar is $\approx$27 kyr old. At the time of its formation, the secondary protostar is located at a distance of $\approx$1500 AU from the primary. Afterward, it migrates toward the primary, and the two protostars orbit each other with high eccentricities and a separation of $\sim$100 AU at this time. The mass of the primary is $\approx$0.49 M$_{\odot}$ and the secondary has a mass of $\approx$0.25 M$_{\odot}$.
The blue star illustrates the location at which the third protostar forms 4 kyr after this snapshot.

\subsection{Dust model}
The dust in the interstellar medium (ISM) is canonically modeled by a dust mixture of $62.5\ \%$ silicate and $37.5\ \%$ graphite following a power-law size distribution of ${ N(a)\propto a^{-q} }$ \citep[][]{Mathis1977,LiDraine2001}. Within the ISM, the range of grain sizes is usually assumed to be between a minimum and maximum size cutoff of $a_{\mathrm{min}}=5\ \mathrm{nm}$ and $a_{\mathrm{max}}=250\ \mathrm{nm,}$ where the quantity $a$ is the effective radius of a dust grain corresponding to a sphere of equivalent volume. 

However, dust scattering models of circumstellar disks and protostellar sources  suggest maximum grain sizes of about $50\ \mu\mathrm{m} - 300\ \mu\mathrm{m}$ \citep[][]{Kataoka2015,Kataoka2017,Hull2017,Ueda2020}. While studies based on the dust opacity index ${ \beta =   \left( \log_{\mathrm{10}}\kappa(\lambda_1)-\log_{\mathrm{10}}\kappa(\lambda_1) \right)/ \left( \log_{\mathrm{10}}\lambda_1 - \log_{\mathrm{10}}\lambda_2 \right) }$, originally suggested millimeter- to centimeter-sized grains, these grain sizes are expected to be erroneous because the studies ignored the scattering opacity \citep[][]{Lin2019,Liu2019,Zhu2019}. However, millimeter-sized grains are believed to play a role in shielding from line emission in disks. 

A further complication is that the growth of grains is a complex process, and the redistribution of the size distribution remains a field of ongoing research that has many unanswered questions \citep{Ossenkopf1994,Hoang2019,McKinnon2019,Kannan2020,Vogelsberger2020}. We therefore applied a two-component dust model to distinguish between dense and ambient regions. Dense regions are defined by the threshold of the number density $n=3\times10^7\ \mathrm{cm}^{-3}$ ($\rho_{\mathrm{gas}} \approx 1.2 \times 10^{-16} \mathrm{g\ cm}^{-3}$  for a mean molecular mass of $\mu=2.37$. The maximum cutoff size is $a_{\mathrm{max}}=100\ \mu\mathrm{m}$ in dense regions with a mixture of graphite and silicate of $1:1$ and a power-law index of $q=3.9$.
For the surrounding, we assumed ISM conditions, but with an upper cutoff of $a_{\mathrm{max}}=6\ \mu\mathrm{m}$ \citep[see the review by][]{Draine2003}. In detail, we assumed that grain growth takes place predominantly within the densest regions. We therefore chose the exact value of the density such that most of the large dust grains are located within these dense regions and not in the surrounding material. Consequently, the disks are more saturated with micrometer-sized grains and nanometer-sized grains are sparse. However, micrometer-sized grains are also assumed to be sparsely present within the bridge itself. For the dust-to-gas mass ratio we applied the canonical value of ${ \rho_{\mathrm{dust}}/ \rho_{\mathrm{gas}}\ \hat{=}\ 1\ \% }$ for both components \citep[][]{Mathis1977,Bohlin1978}. The field of measuring grain sizes is rapidly developing, therefore we also produced synthetic maps assuming an upper grain size of %\ref{app:B}
$a_{\mathrm{max}}=3 \mathrm{mm}$ for the dense regime.
In appendix \ref{app:A} we show results for synthetic maps of emitted radiation, scattered radiation, polarization including radiative alignment torques (RATs), and perfect alignment for 53 $\mu$m, 214 $\mu$m, and $1.3$ mm wavelength as based on an upper grain size for the dense regime of $a_{\mathrm{max}}=100 \mu$m (\Fig{AppSyntheticDust100}) and $a_{\mathrm{max}}=3 \mathrm{mm}$ (\Fig{AppSyntheticDust3mm}).

In \Fig{DustModel} we show the resulting opacities $\kappa$ of extinction, absorption, and scattering for the dust components with an upper dust radius of ${ a_{\mathrm{max}}=6\ \mu\mathrm{m} }$ and ${ a_{\mathrm{max}}=100\ \mu\mathrm{m} }$. The dust components have an opacity index $\beta$ of $0.8$ and $1.3$, respectively, which is typical for the natal environment of young stars \citep[][]{Lommen2007,Lommen2009}.

\subsection{Synthetic simulations with the radiative transfer code \polaris}
\label{sect:RTPostProcessing}
To compare the MHD simulation with observations of dust polarization, we postprocessed the simulation data with the publicly available radiative transfer code \polaris\ \citep[][]{Reissl2016}. 
The code performs photon propagation by means of the Monte Carlo (MC) method and includes dust scattering and absorption considering various photon-emitting sources. 
To calculate the dust temperature and the grain alignment efficiency,  \polaris\ keeps track of the magnitude, direction, and isotropy of the radiation field per cell. 
By default, we used a range of wavelength of $0.9\ \mu\mathrm{m} - 3 \mathrm{mm}$ logarithmically distributed over 100 wavelengths bins. 
\polaris\ can run the radiative transfer simulations on an octree grid. 
This enabled us to keep and adopt the native grid structure of the \ramses\ simulations. 
 MC runs that consider the full radiation field have a high cost in memory.
To avoid extensive memory use, we excised a subregion in which the protostars form with a side length of about $3.232\times 10^4\ \mathrm{AU, which}$ corresponds to an octree refinement level of 13. 

We considered as protostellar heating sources the properties of the protostars, which we call S1 and S2 hereafter. We computed their temperature and radii based on the accretion rates with the stellar evolution code \mesa\ \citep{Paxton2011}. \citep[For more details on using \mesa\ for \ramses\ simulations, see][]{Kuffmeier2018,Jensen2018}. The resulting effective temperatures and subsequent luminosities are  ${ T_* = 4776\ \mathrm{K} }$ and ${ L_* = 4.27\  L_\odot }$ for source S1 and ${ T_* = 4336\ \mathrm{K} }$ and ${ L_* = 2.57\ L_\odot }$ for S2. In addition to the protostars, we assumed a diffuse interstellar radiation field (ISRF) in order to obtain realistic dust temperatures at larger distances from the protostars. 
The ISRF uses a parameterization of the spectral energy distribution (SED) as presented in \cite{Mathis1983}, which is typical for the ISM. 
To guarantee an optimal signal-to-noise ratio, we performed the MC simulations with $3\times 10^8$ photons per wavelength and source.

\begin{figure*}
    \includegraphics[width=\textwidth]{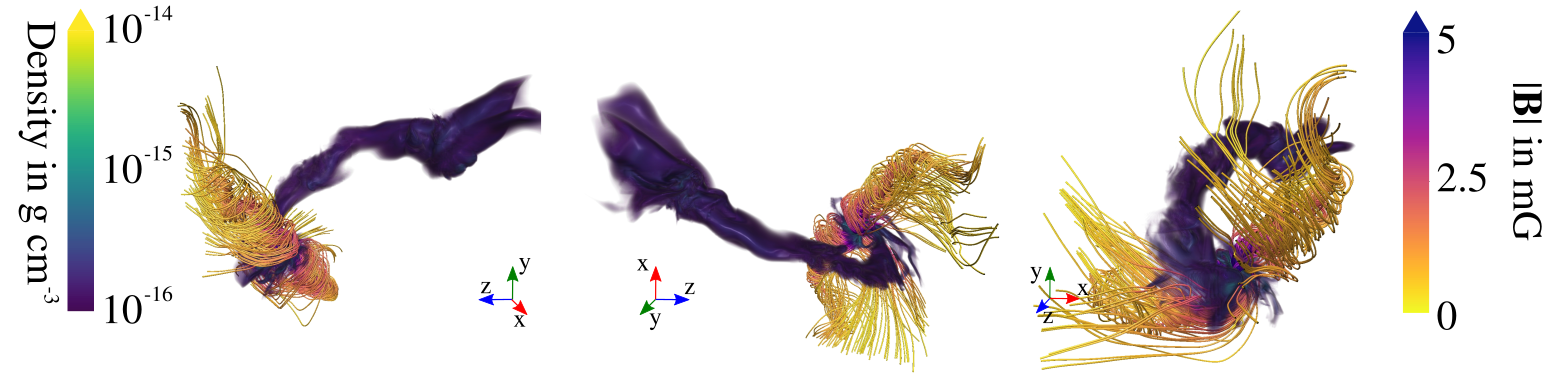} 
        \caption{Density distribution in the bridge and around forming protostars. The panels illustrate the density distribution and the pattern of the strong magnetic fields associated with the primary in the negative direction of the axes. The bridge is shown in the left panels along the x-axis, in the middle panels along the y-axis, and in the right panels along the z-axis. The orientation is also illustrated with the red, green, and blue axes (for the  x, y, and z direction). 
        To show the bridge more clearly, we only plot densities above $10^{-16} \unit{g}\unit{cm}^{-3}$ and gradually increase the opacity of the gas from 0 to 100 \% for densities $\rho_{\mathrm{gas}}> 10^{-14} \unit{g}\unit{cm}^{-3}$. The illustration is based on a cubical box of 2048 AU in length around the center, computed as $(\mathbf{r}_{\rm A} + \mathbf{r}_{\rm C})$ / 2 at the time of formation of the third companion $\approx 4$ kyr later.}
        \label{fig:bridge_vapor}
\end{figure*}

Using the dust as introduced in the previous section, we ran the radiative transfer postprocessing with \polaris\ for the snapshots shown in Figs. \ref{fig:bridge_Sigma_repro}, \ref{fig:bridge_vapor}, and \ref{fig:bridge_iso_vapor} as input to determine the radiation field. \polaris\ assumes an energy equilibrium between absorbed radiation and grain emission in order to calculate the dust temperature \citep[][]{Lucy1999,BjorkmanWood2001,Reissl2016}. Furthermore, knowing the radiation field allows us to determine
the efficiency of grain alignment according to the RAT theory\footnote{A detailed description of the latest implementation of the RAT alignment theory in \polaris\ is outlined in \cite{Reissl2020}.} \citep[][]{LazarianHoang2007,Hoang2014}. According to the RAT, irregular grains experience a net torque when exposed to an anisotropic radiation field, and the grains start to spin up with an angular velocity of $\omega_{\mathrm{rad}}$. In order to determine whether a grain may align with its minor principal axis with the magnetic field orientation, we must account for random collisions with the gas. The latter process results in a grain rotation with an angular velocity $\omega_{\mathrm{gas}}$ , whereas the direction of rotation remains randomized for each grain. A common parameterization for a stable grain alignment is ${ \omega_{\mathrm{rad}}^2/\omega_{\mathrm{gas}}^2 > 3}$ \citep[see][and references therein]{Hoang2014}. This parametrization enables us to calculate a characteristic grain-size threshold $a_{\mathrm{alg}}$  at which all paramagnetic grains may have a stable alignment \citep[][]{Hoang2014,Reissl2020}. 

A second criterion for alignment with the magnetic field direction is related to the Larmor procession timescale. This criterion may prevent grains at the larger end of the size distribution to randomize. However, within the MHD zoom-in simulation, we have a field strength of up to $100 \unit{mG}$ and grains can always align as long as ${ a>a_{\mathrm{alg}} }$ \citep[see][for details]{Reissl2020}.  We also note that the paramagnetic properties of silicate and graphite materials are different by about six orders of magnitude \citep[see, e.g.,][]{Draine1996,Hoang2014A}. We therefore considered the graphite to be randomized in our radiative transfer postprocessing. In principle, the direction of grain alignment may also be dominated by the radiation field, as reported by \cite{LazarianHoang2007}. However, such a change in direction typically occurs only in close proximity of a star on scales of the innermost $\sim$10 AU \citep{LazarianHoang2007,Tazaki2017}. We did not take this effect into account for our synthetic observations, where we observe an object spanning thousands of AU. 

Aligned dust grains may contribute to polarization in two different ways. For  dichroic extinction, any background radiation is most efficiently blocked by the grain in the direction of its major principal axis. If the grain were aligned with its minor principal axis with the magnetic field orientation, the radiation would become polarized along the field lines \citep[see, e.g.,][]{Martin1974}. For a wavelength where dust emission becomes relevant, the dust grain preferentially emits thermal  radiation along its major axis. This means that the emitted radiation traces the magnetic field orientation rotated by $90^{\circ}$ \citep[for details, we refer to][and references therein]{Brauer2016}. \polaris\ solves the radiative transfer problem by simultaneously taking  grain alignment, dichroic extinction, and thermal emission into account \citep{Reissl2016,Reissl2020}.

Finally, we created dust intensity and polarization maps with \polaris\ using two different modes. For pure dust extinction and emission, we considered the dust grains to be oblate spheroids with an aspect ratio of $1/2$. For dust self-scattering processes, we assumed spherical grains and applied the Mie scattering theory, where we also considered multiple scattering events\footnote{A code with a consistent treatment of scattering on nonspherical dust grains that are partially aligned with the magnetic field direction is not yet available. However, a paper providing information on this is in preparation.}. In this case, the polarization signal carries no information about the magnetic field morphology. We assumed an object-observer distance of $120\ \mathrm{pc}$ \citep[see][and references therein]{Jacobsen2018-IRAS16293} for the synthetic maps and smoothed them with a Gaussian beam with a full width at half-maximum of $8^\mathrm{''}\times 8^\mathrm{'',}$ corresponding to $9.6\ \mathrm{AU}\times 9.6\ \mathrm{AU}$.

\polaris\ delivers its results as a four-component Stokes vector ${ S=\left(I,Q,U,V\right)^T }$ , where $I$ is the total intensity, $Q$ and $U$ are the components of the linear polarization, and $V$ is the circular polarization. Consequently, linear dust polarization is completely described by the degree of polarization  ${ P_{\rm frac}=(Q^2+U^2)^{1/2}/I }$ and its orientation angle ${ \phi_{\mathrm{pol}} = 0.5\arctan \left(-U/Q\right) }$. We emphasize that this angle needs to be rotated by $90^{\circ}$ in the far-IR, submillimeter, and millimeter wavelength regime in order to infer the magnetic field orientation from the polarization signal. However, in the densest regions, scattering might dominate the polarization angle.

\section{Results}
In this section, we present the properties associated with the magnetic field in the bridge structure. 
Subsection 3.1 provides the magnetic field properties as directly obtained from the MHD simulation,
and in subsection 3.2 we present the results of the synthetic observations.

\subsection{Magnetic field structure}

\subsubsection{Magnetic tower around the primary and bipolar outflow}
In \Fig{bridge_vapor}, we show 3D visualizations of the density and the strongest magnetic fields that are associated with the bridge shown in \Fig{bridge_Sigma_repro}  that was first presented in \citet{KuffmeierBridge}, seen along each of the three coordinate axes as lines of sight.
We used the visualization tool \vapor\ \citep{clyne2005prototype,clyne2007interactive} for the illustrations. 
We excised a cubical region of $\approx 4100$ AU in length around the bridge. 
%To visualize the data, we converted the AMR grid to a data cube with $512\times512\times512$ resolution by extrapolating cells of size $\Delta x < 8$ AU and interpolating cells larger than $\Delta x > 8$.  
To show the structure of the magnetic field, we used the bias function of flow lines in \vapor.
The software computes $N\times 2^{|b|}$ field lines of which the strongest or weakest $N$ are visualized for positive or negative bias $b$. 
For the visualizations shown in \Fig{bridge_vapor}, we chose $N=100$ and set the bias $b$ to the strongest possible value of $b=15$. 

The plot shows that the strongest field lines are associated with characteristic bipolar magnetic towers that are due to the winding-up of the magnetic field lines during the formation and evolution of the primary protostar. 
The highest outflow speeds are $\gtrsim 10 \unit{km}\unit{s}^{-1}$, 
which is consistent with the highest resolution of $4 \unit{AU}$ and the corresponding Kepler speed at the launching point for the protostellar mass of $M_{\rm A} = 0.47 \unit{M}_{\odot}$ at $t=70 \unit{kyr}$ after the formation of the primary protostar. 
The analysis of previous higher resolution runs suggests that the outflow speeds would be higher if we applied higher resolution because the launching radius of the outflow speed would be smaller at the footpoint of the outflow \citep{Kuffmeier2017}. 
The visualizations also show that the orientation of the outflow is perturbed from pure symmetry as a result of the turbulent protostellar birth environment. We discuss the asymmetry of outflows in more detail in section 4.1.

\begin{figure*}
    \includegraphics[width=\columnwidth]{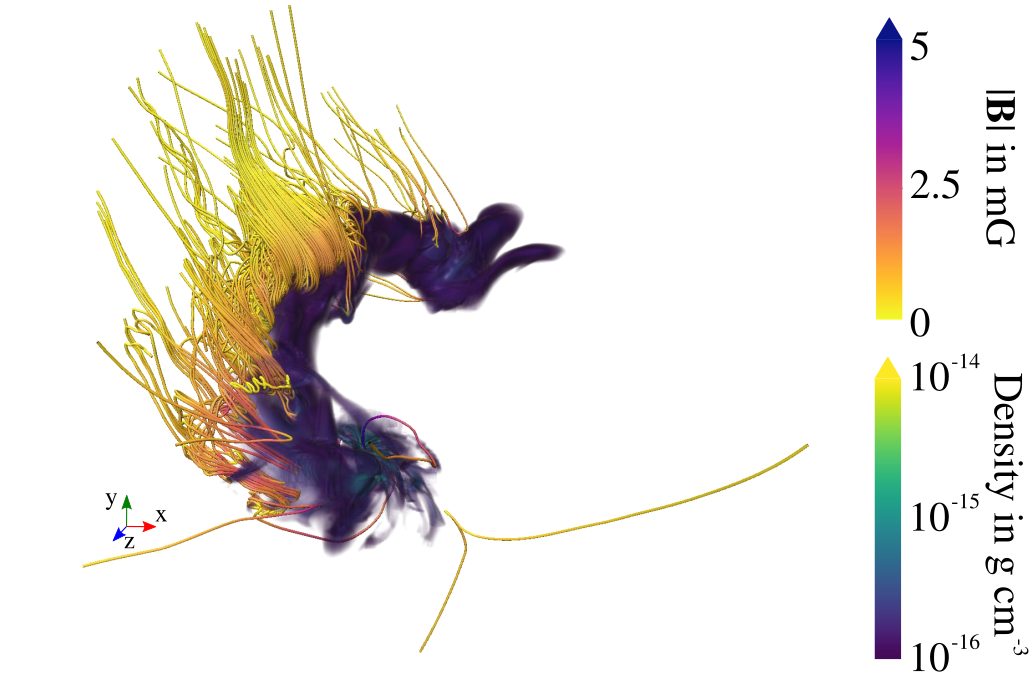} 
    \includegraphics[width=\columnwidth]{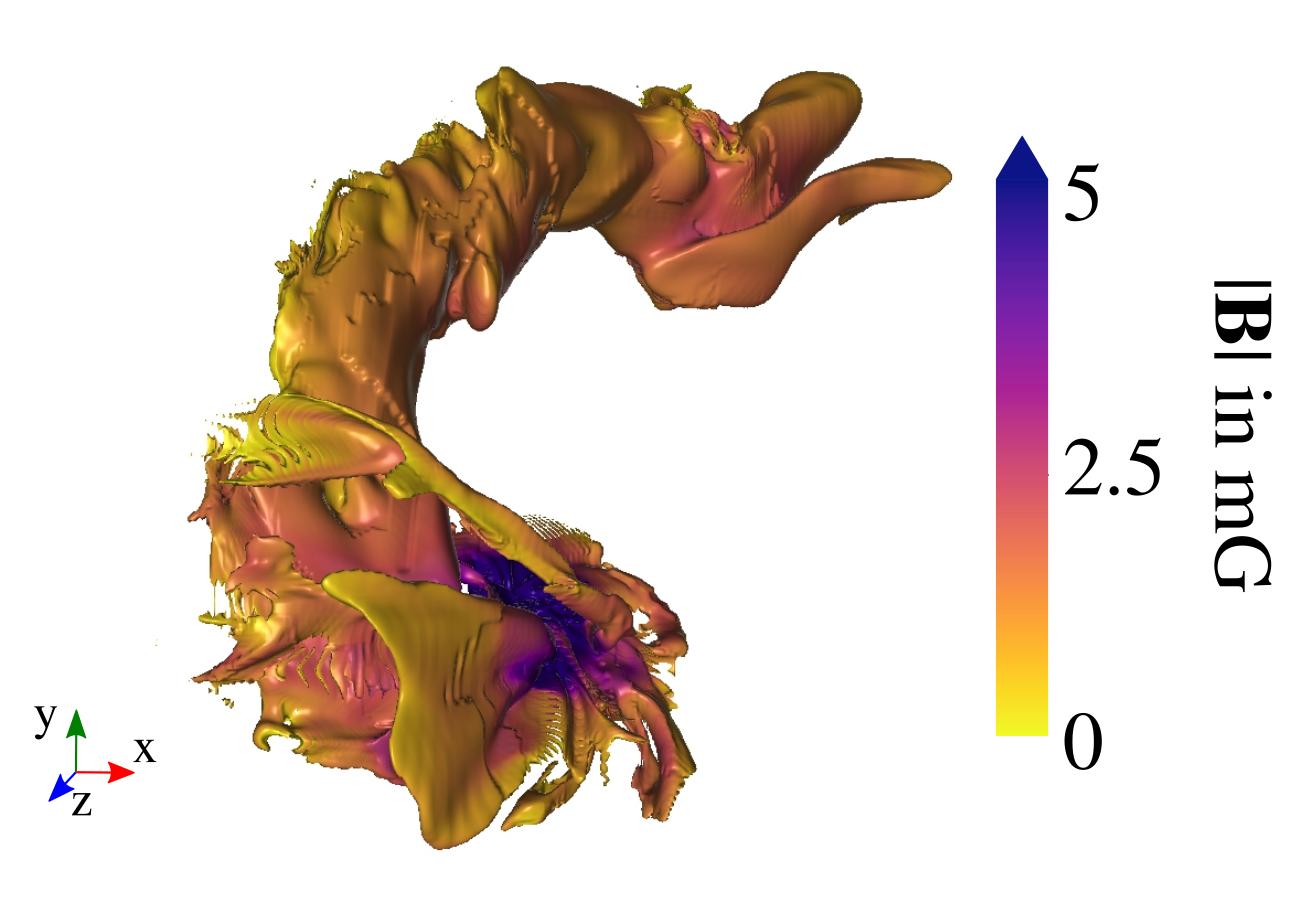} 
        \caption{Left panel: Contour shows the density above a threshold of $10^{-16} \unit{g}\unit{cm}^{-3}$ in the bridge, and the flow lines correspond to the 100 weakest flow lines of $100\times2^{10}$ computed flow lines in the region of the bridge. 
        Right panel: Contour shows $|\mathbf{B}|$ on an isosurface of density $10^{-16} \unit{g}\unit{cm}^{-3}$. }
        \label{fig:bridge_iso_vapor}
\end{figure*}

\begin{figure}
    \includegraphics[width=\columnwidth]{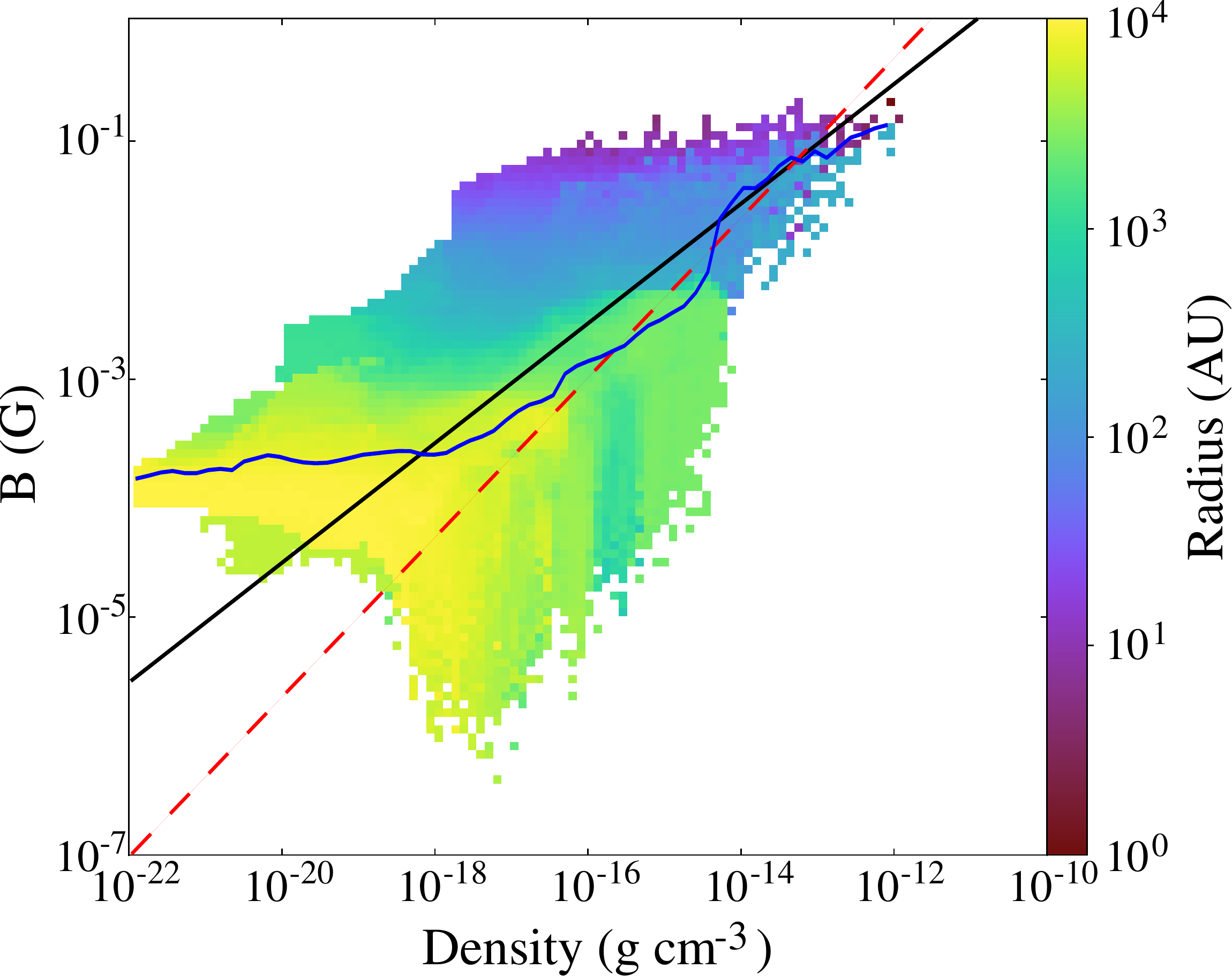} 
        \caption{Distribution of the magnetic field strength over density around the primary at $t=70$ kyr. 
        The color indicates the radial distance from the primary at this time.
        The solid blue line shows the volume-averaged magnetic field strength.
        For comparison, we show the scaling of $B\propto \rho_{\mathrm{gas}}^{\frac{2}{3}}$ (dashed red line) and $B\propto \rho_{\mathrm{gas}}^{\frac{1}{2}}$ (solid black line).}
        \label{fig:rho_B_r}
\end{figure}

\begin{figure*}
\begin{center}
     \includegraphics[width=0.49\textwidth]{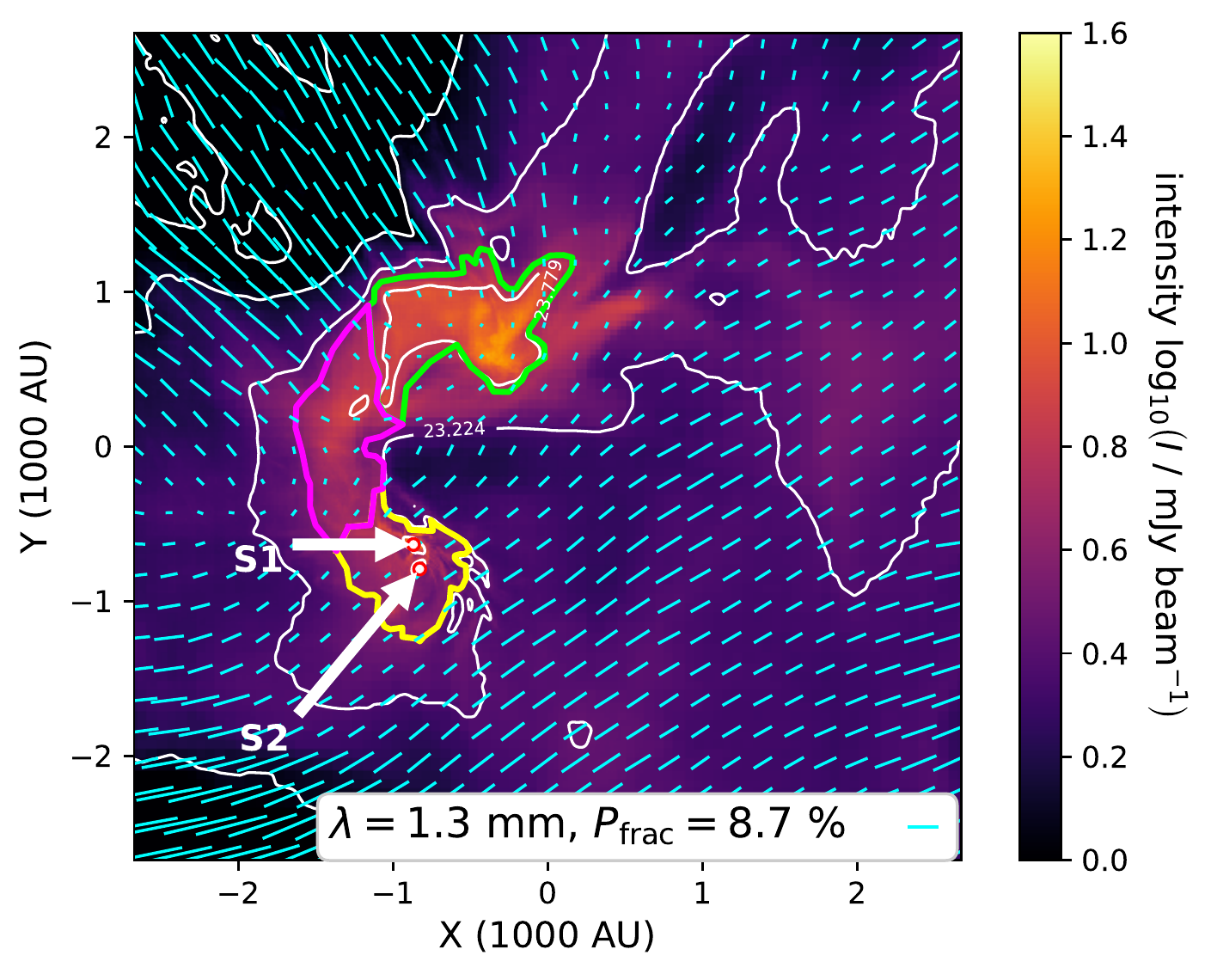}
     \includegraphics[width=0.49\textwidth]{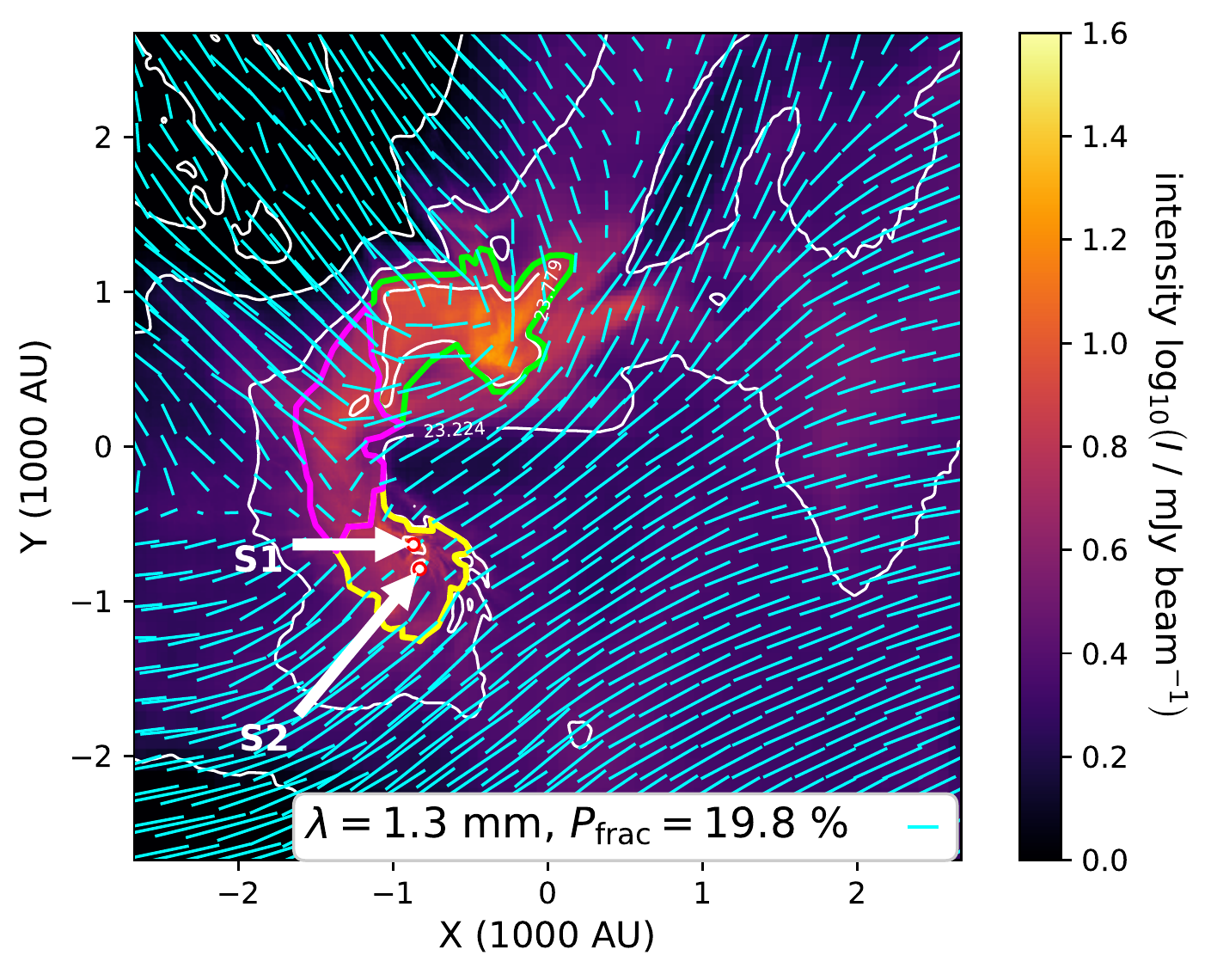}
\end{center}

\begin{center}
     \includegraphics[width=0.49\textwidth]{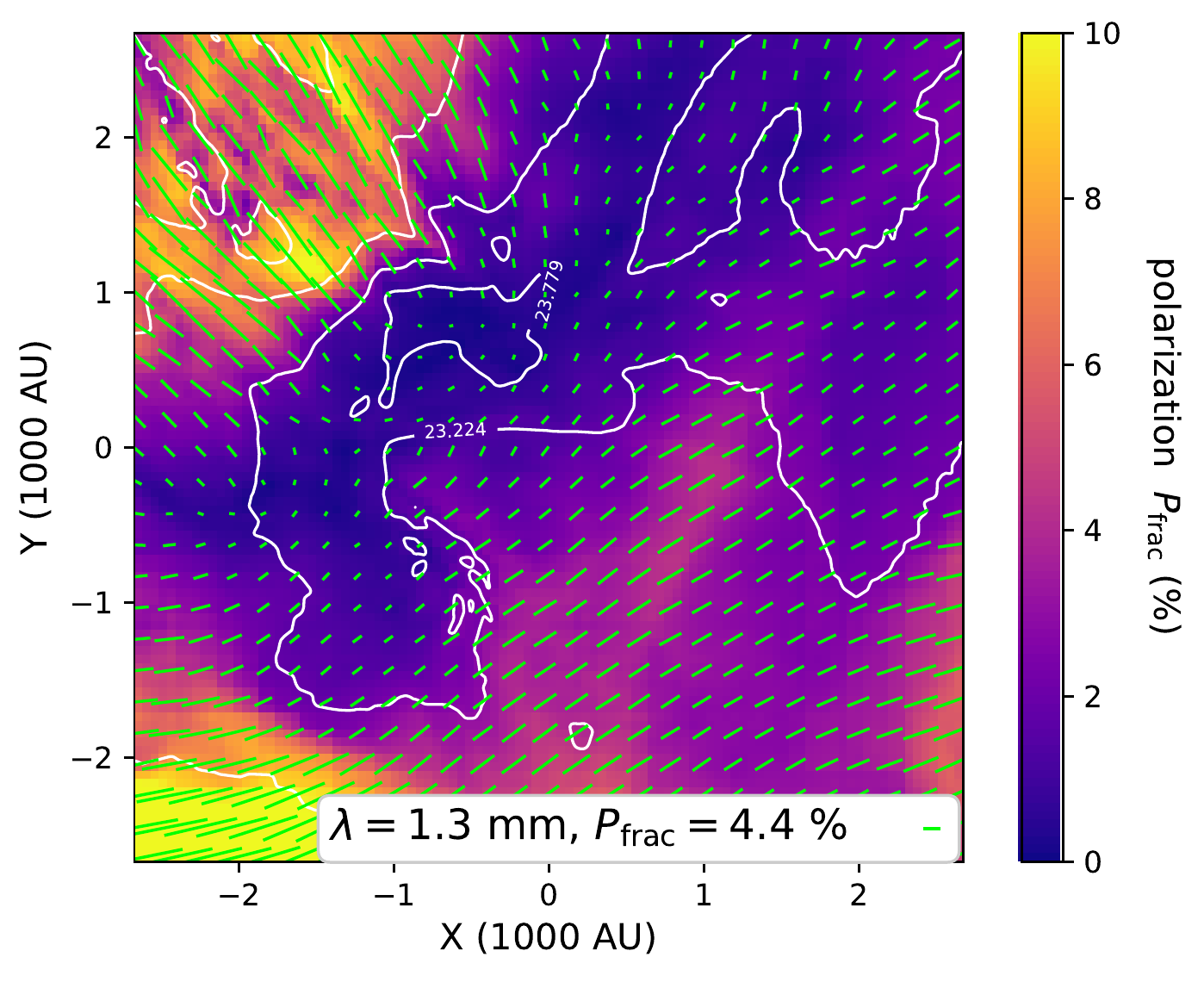}
     \includegraphics[width=0.49\textwidth]{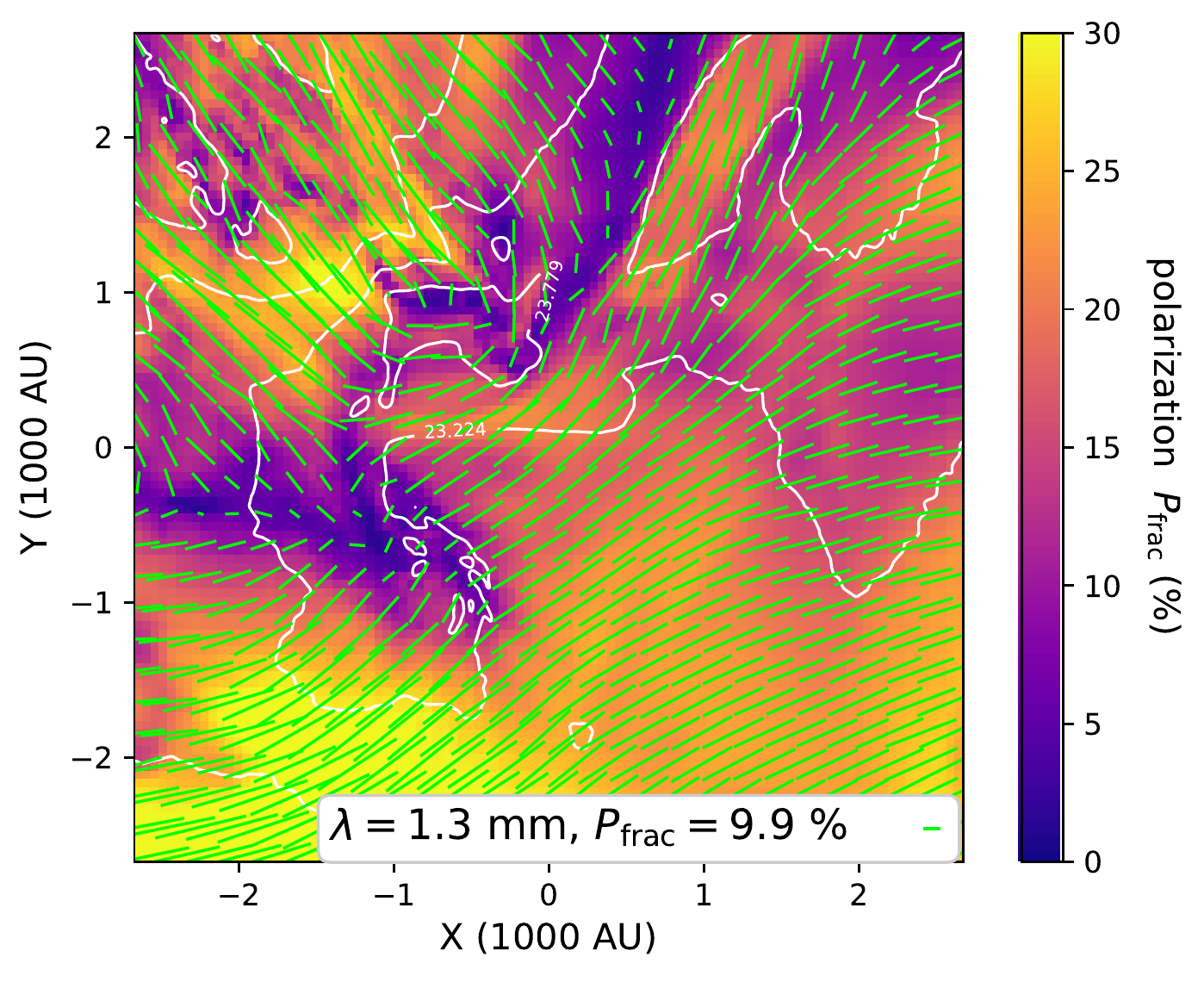}
\end{center}

\caption{Top row: Stokes $I$ component of the $1.3\ \mathrm{mm}$ dust emission considering RAT grain alignment (left) in comparison with perfect alignment (right). White contours show the logarithmic column density, while the length and orientation of the pseudovectors represent the degree and angle of the linear dust polarization. The polarization vectors are rotated by $90^\circ$ to match the actual magnetic field direction.
The area of the bridge structure is enclosed by blue lines, while the natal structures that harbor the primary and secondary (D1) are shown in yellow and the natal structure of the tertiary (D2) is plotted in green. The borders follow the contour line of $I=4.24\ \mathrm{mJy\ beam}^{-1}$. 
%The panels are to be compared with the right panel of \Fig{rho_B_r} for interpretation. 
Bottom row: 
Same as the top row, but the color-scale now shows the polarized fraction, $P_{\rm frac}$.}
\label{fig:SyntheticDustEM}
\end{figure*}

\begin{figure*}
\begin{center}
     \includegraphics[width=0.48\textwidth]{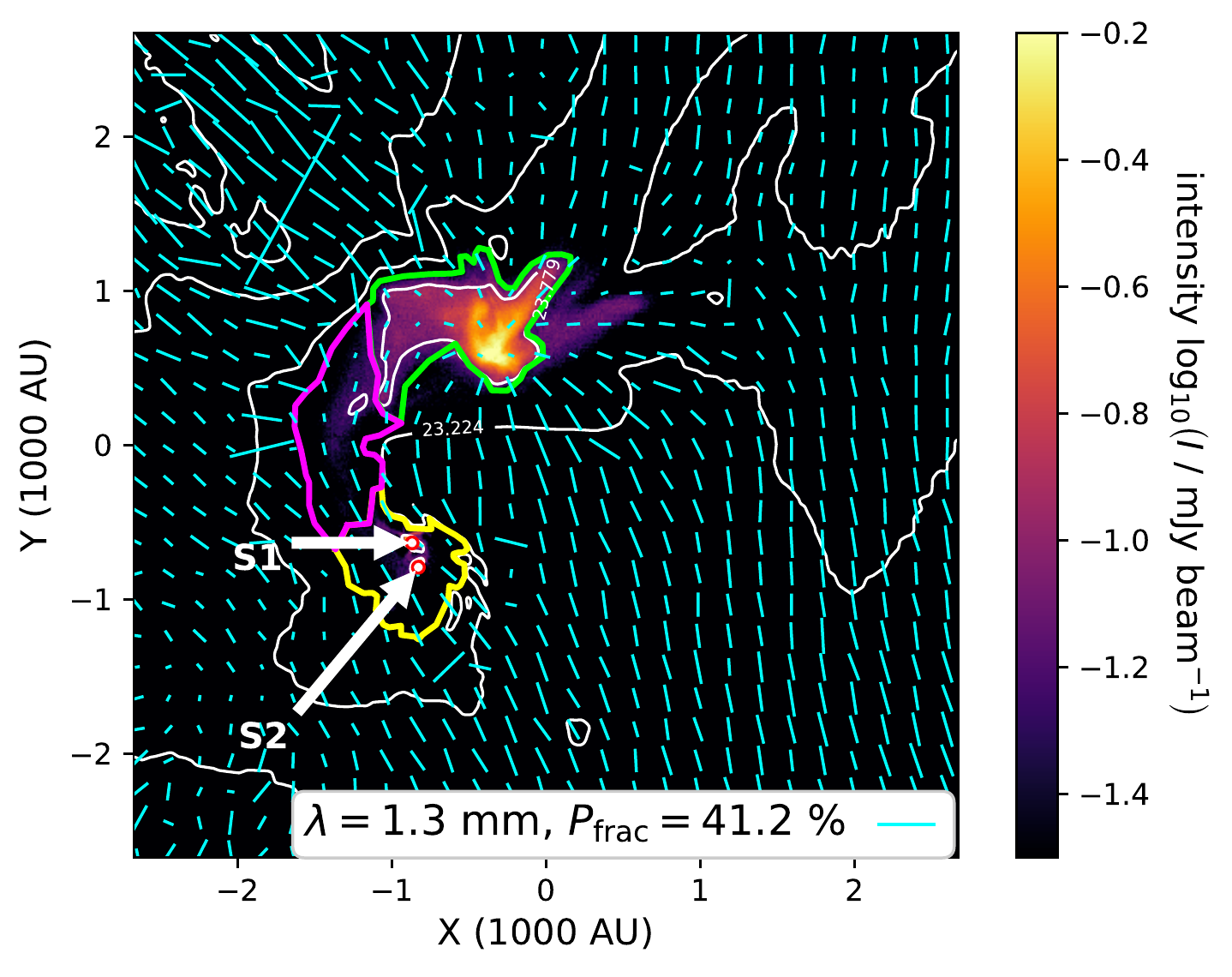}
     \includegraphics[width=0.49\textwidth]{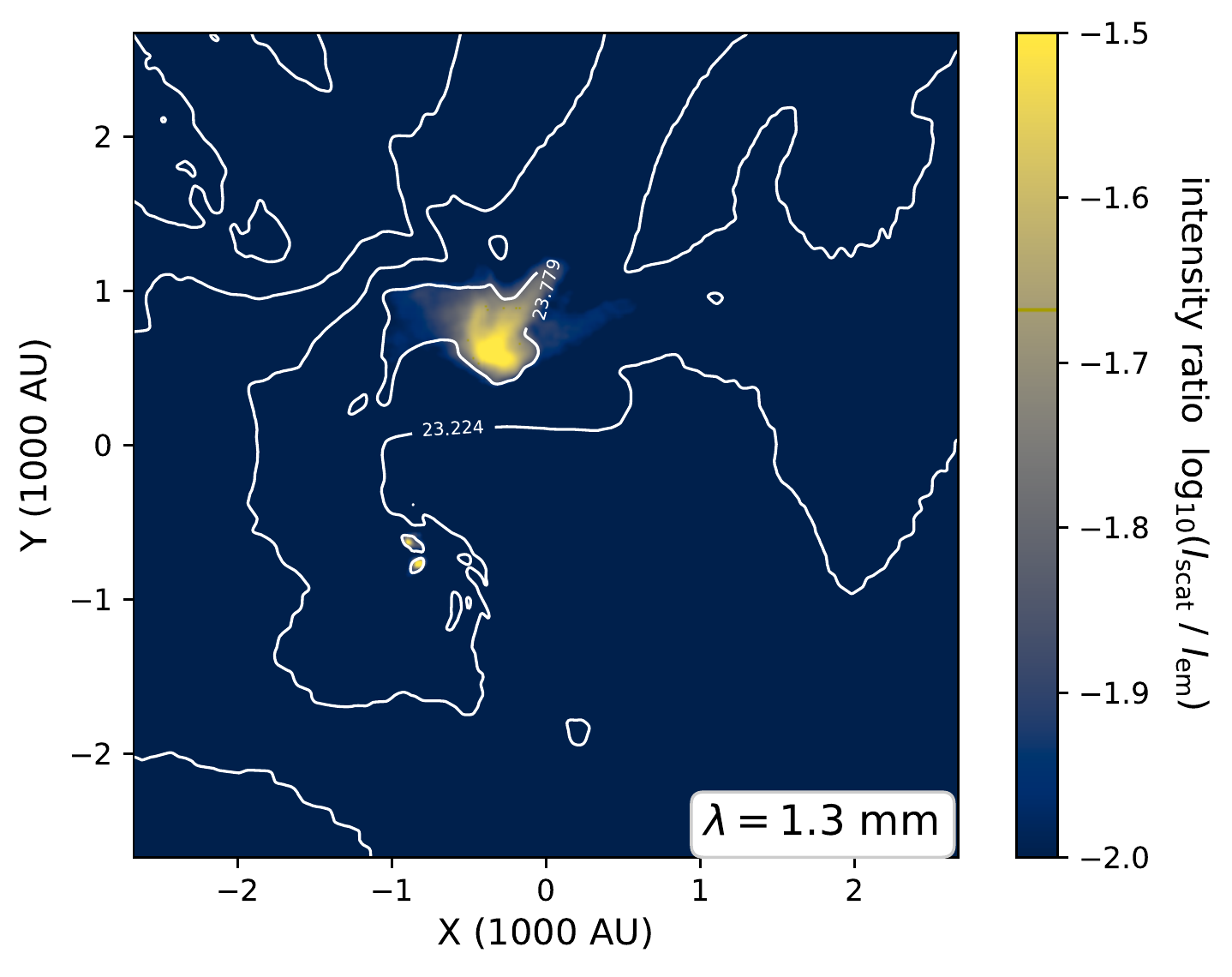}
\end{center}
\caption{Left panel: Stokes I and the polarization result considering dust self-scattering at $1.3\ \mathrm{mm}$ only. 
Right panel: Logarithmic ratio of scattered radiation $I_\mathrm{scat}$ to the emitted radiation $I_\mathrm{em}$ corresponding to the intensity maps shown in Fig. \ref{fig:SyntheticDustEM} and on the left side.}
\label{fig:SyntheticDustSCA}
\end{figure*}

\begin{figure*}
\begin{center}
     \includegraphics[width=0.49\textwidth]{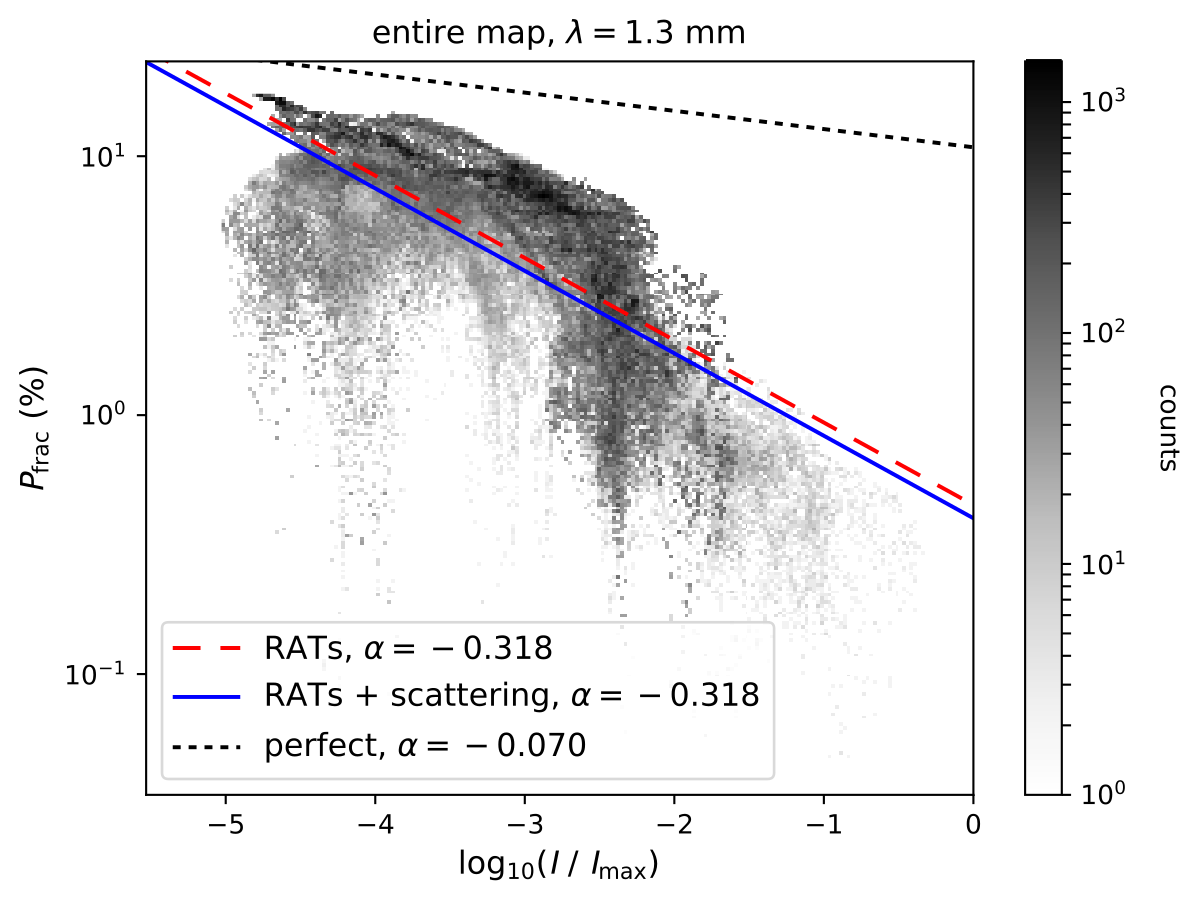}
     \includegraphics[width=0.49\textwidth]{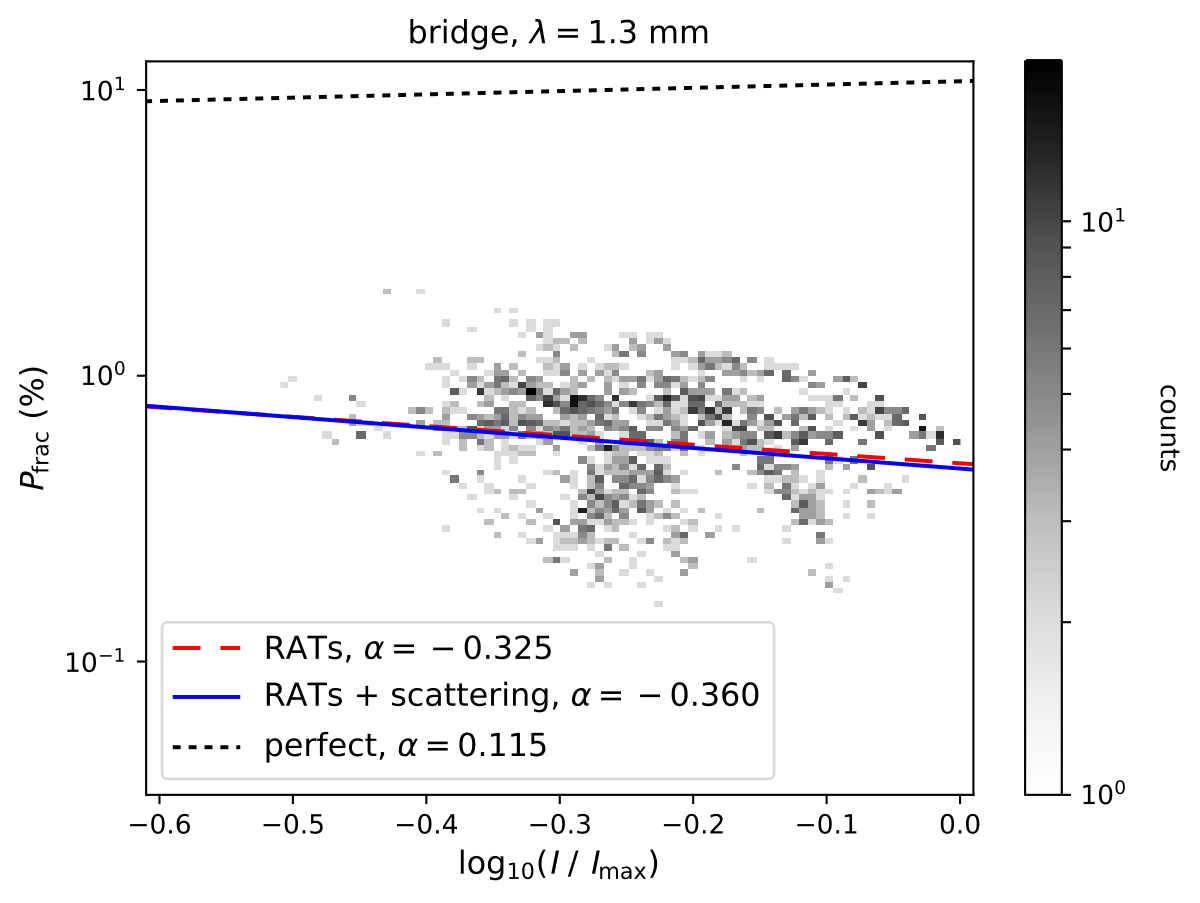}
\end{center}
\begin{center}
     \includegraphics[width=0.49\textwidth]{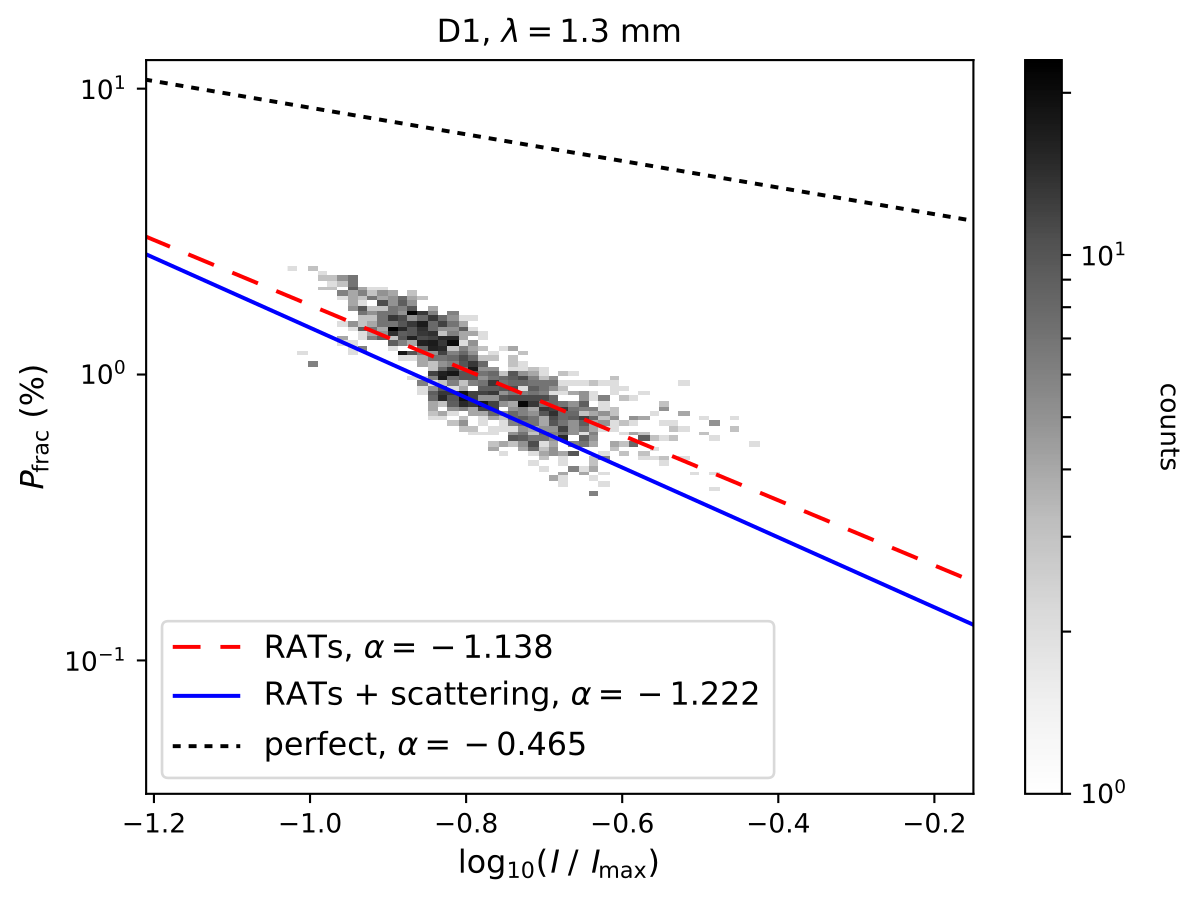}
     \includegraphics[width=0.49\textwidth]{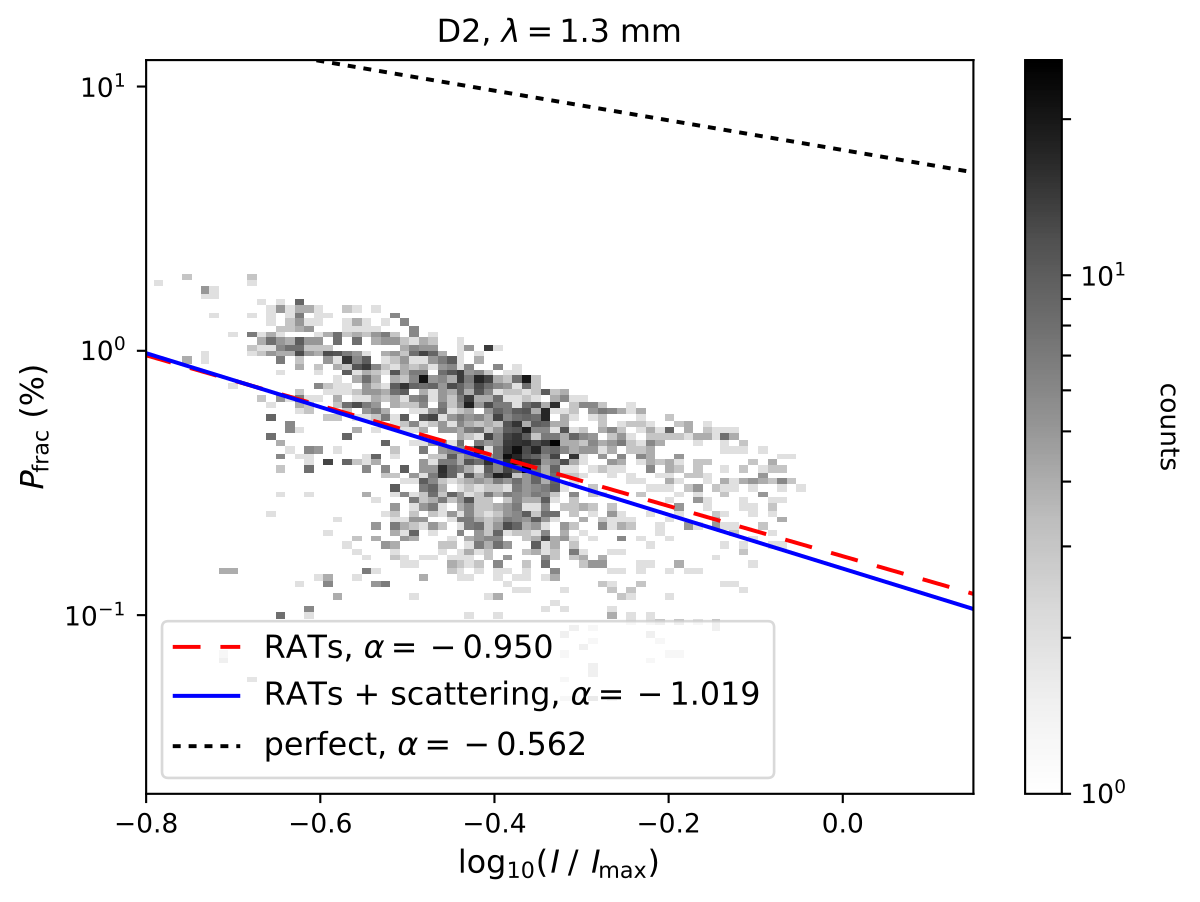}
\end{center}
\caption{PI relation at a wavelength of $1.3\ \mathrm{mm}$ for the entire dust emission map (top left) shown in \Fig{SyntheticDustEM} and the regions of the bridge (top right), D1 (bottom left), and D2 (bottom right). We compare the fitted trends of the PI relation $P_{\rm frac}\propto I^{\alpha}$ for the cases of RAT alignment (long-dashed red), RAT plus scattering (solid blue), and perfect grain alignment (short-dashed black). The gray-scaled counts represent the pixels of the maps in \Fig{SyntheticDustEM} considering only RAT alignment.}
\label{fig:PIRelation}
\end{figure*}

\begin{figure*}
\begin{center}
     \includegraphics[width=0.48\textwidth]{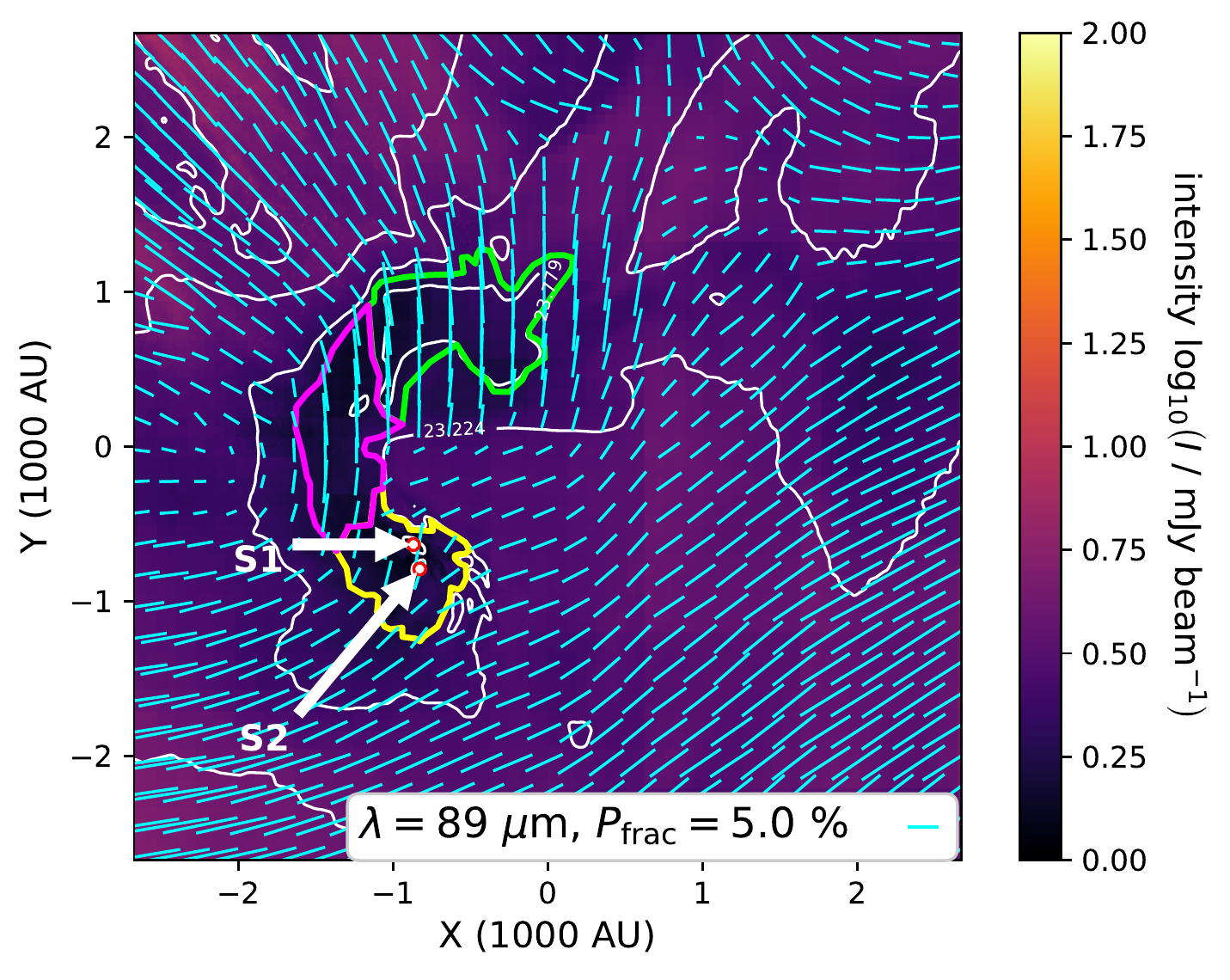}
     \includegraphics[width=0.5\textwidth]{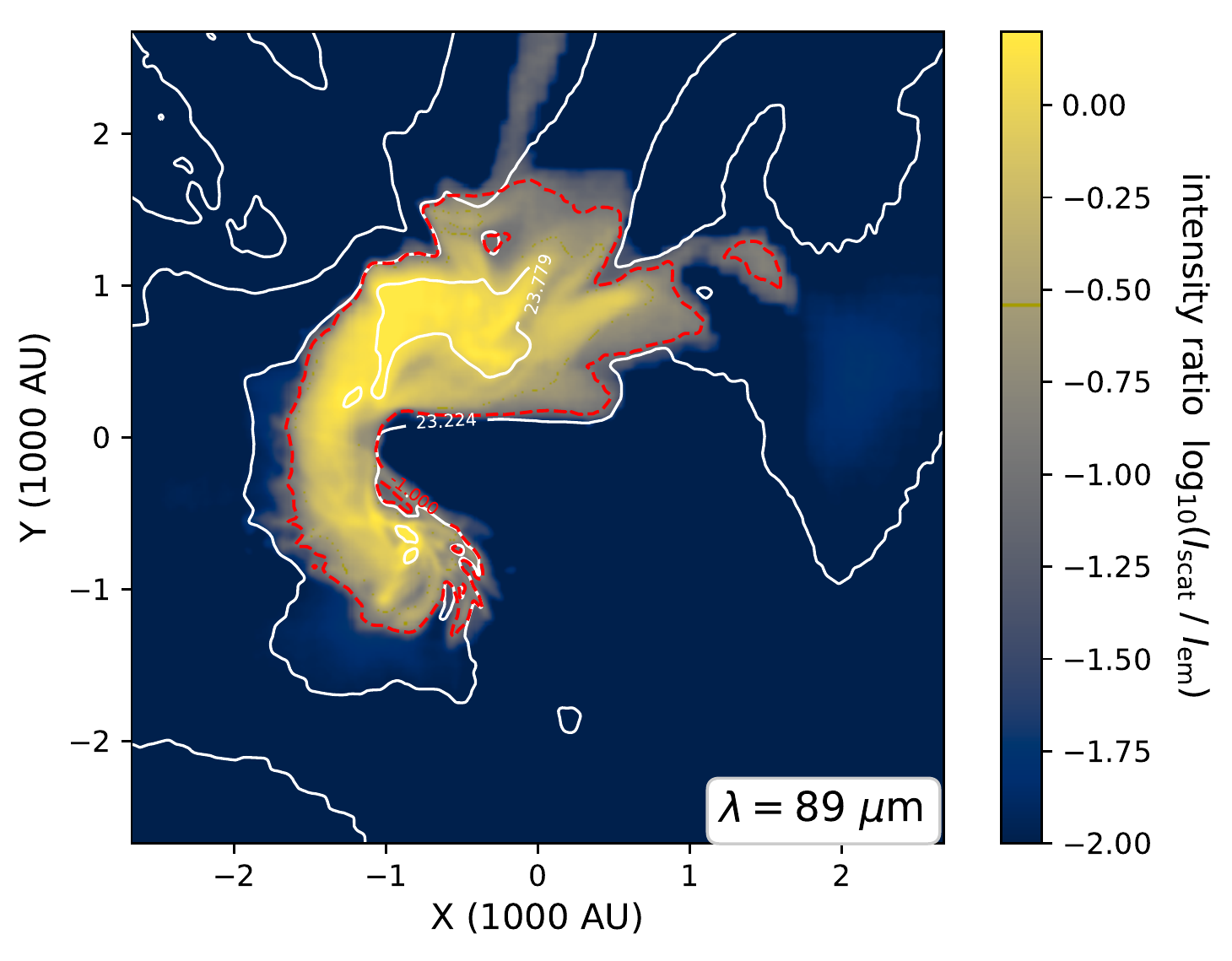}
\end{center}

\caption{Left: Same as the dust emission map in \Fig{SyntheticDustEM} considering only RAT alignment, but for a wavelength of $89\ \mu\mathrm{m}$. Right: Corresponding  ratio of scattered $I_\mathrm{scat}$ to emitted radiation $I_\mathrm{em}$. Red contours indicate the ratio $I_\mathrm{scat}/I_\mathrm{em}=0.1$. The polarization vectors are rotated by a considerable amount in D1, D2, and the bridge compared to the $1.3\ \mathrm{mm}$ observations in \Fig{SyntheticDustEM}.}
\label{fig:SyntheticDustEMWave}
\end{figure*}

\begin{figure*}
\begin{center}
     \includegraphics[width=0.49\textwidth]{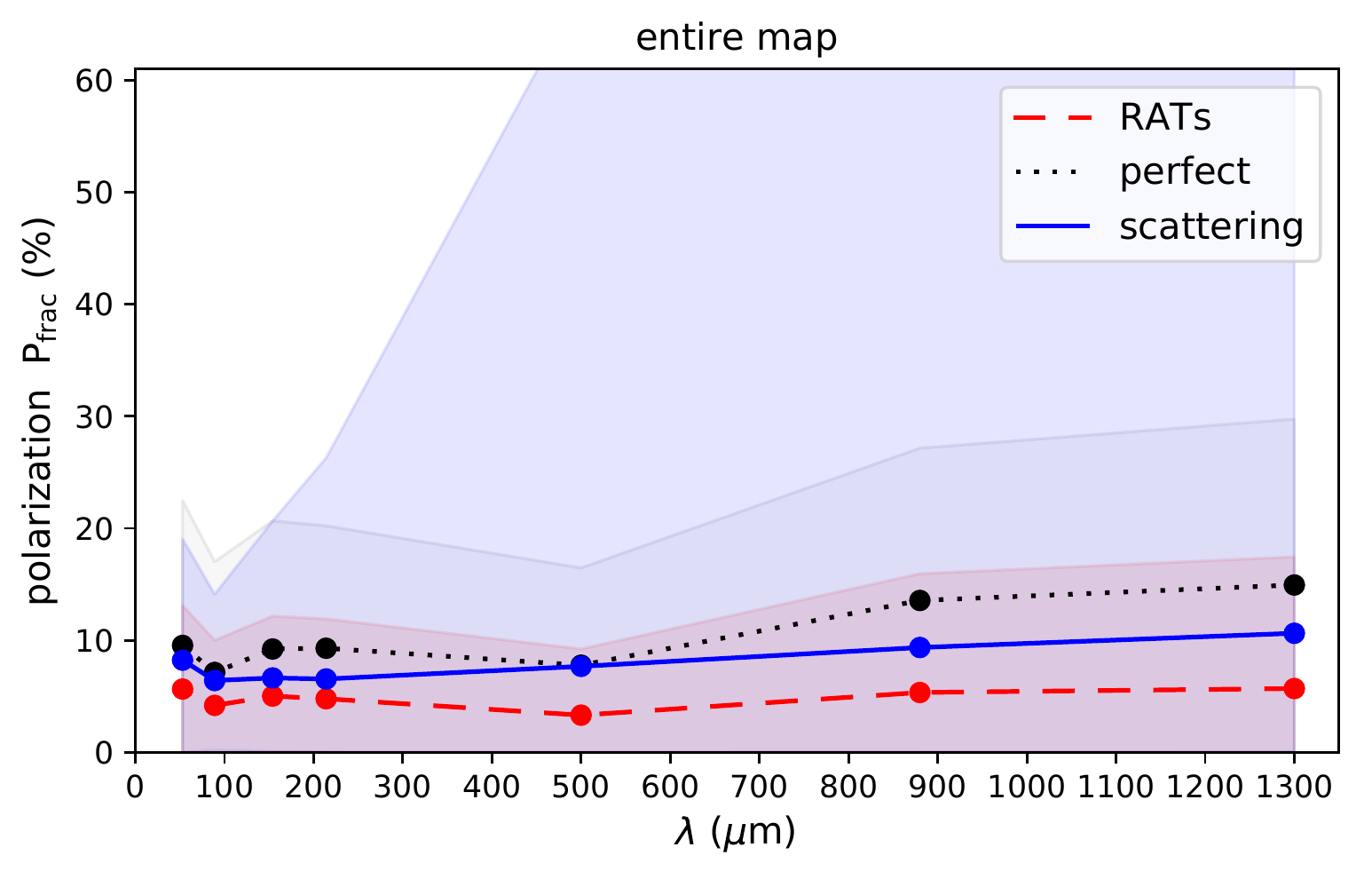}
     \includegraphics[width=0.48\textwidth]{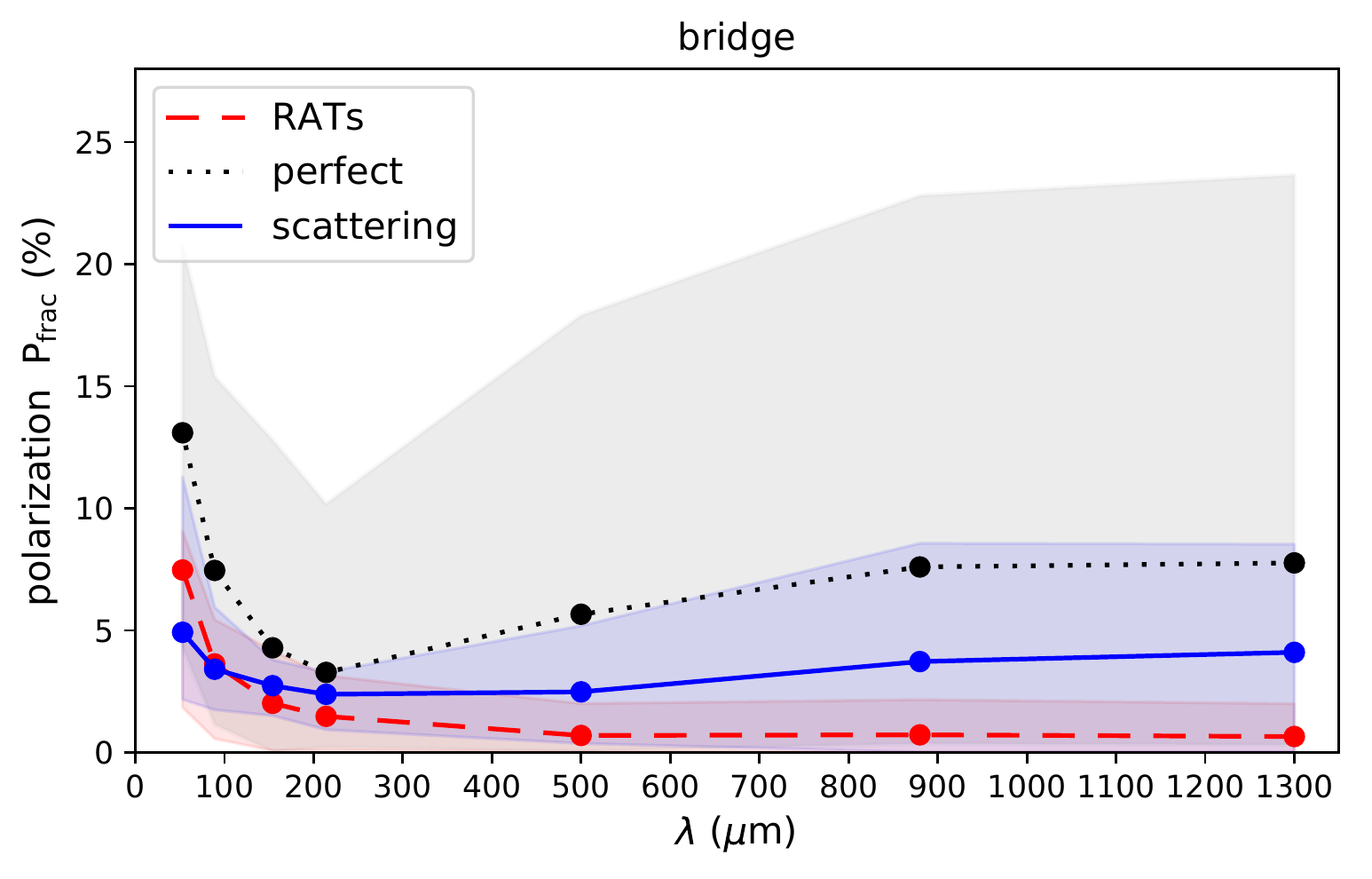}
\end{center}

\begin{center}
     \includegraphics[width=0.49\textwidth]{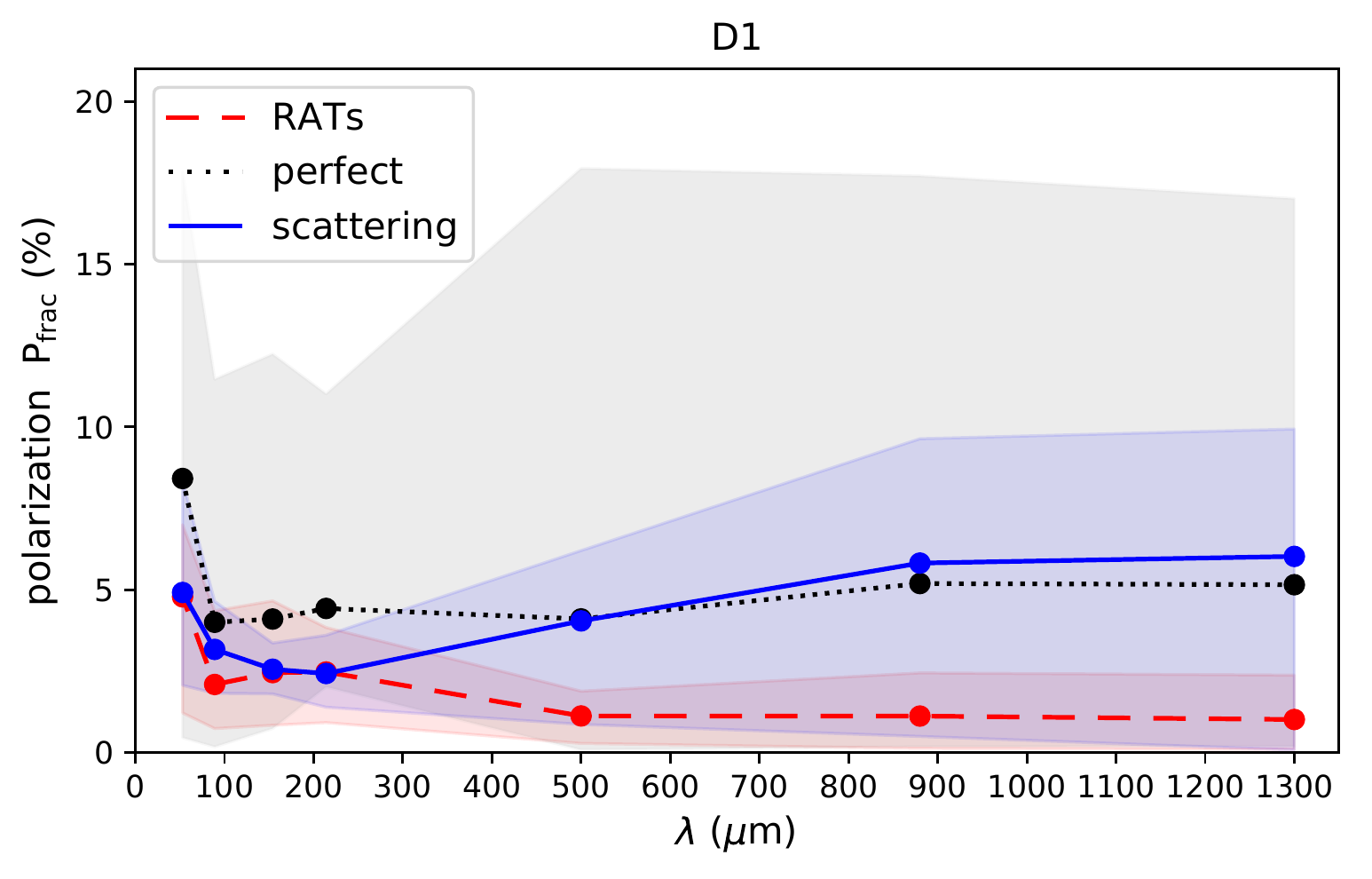}
     \includegraphics[width=0.48\textwidth]{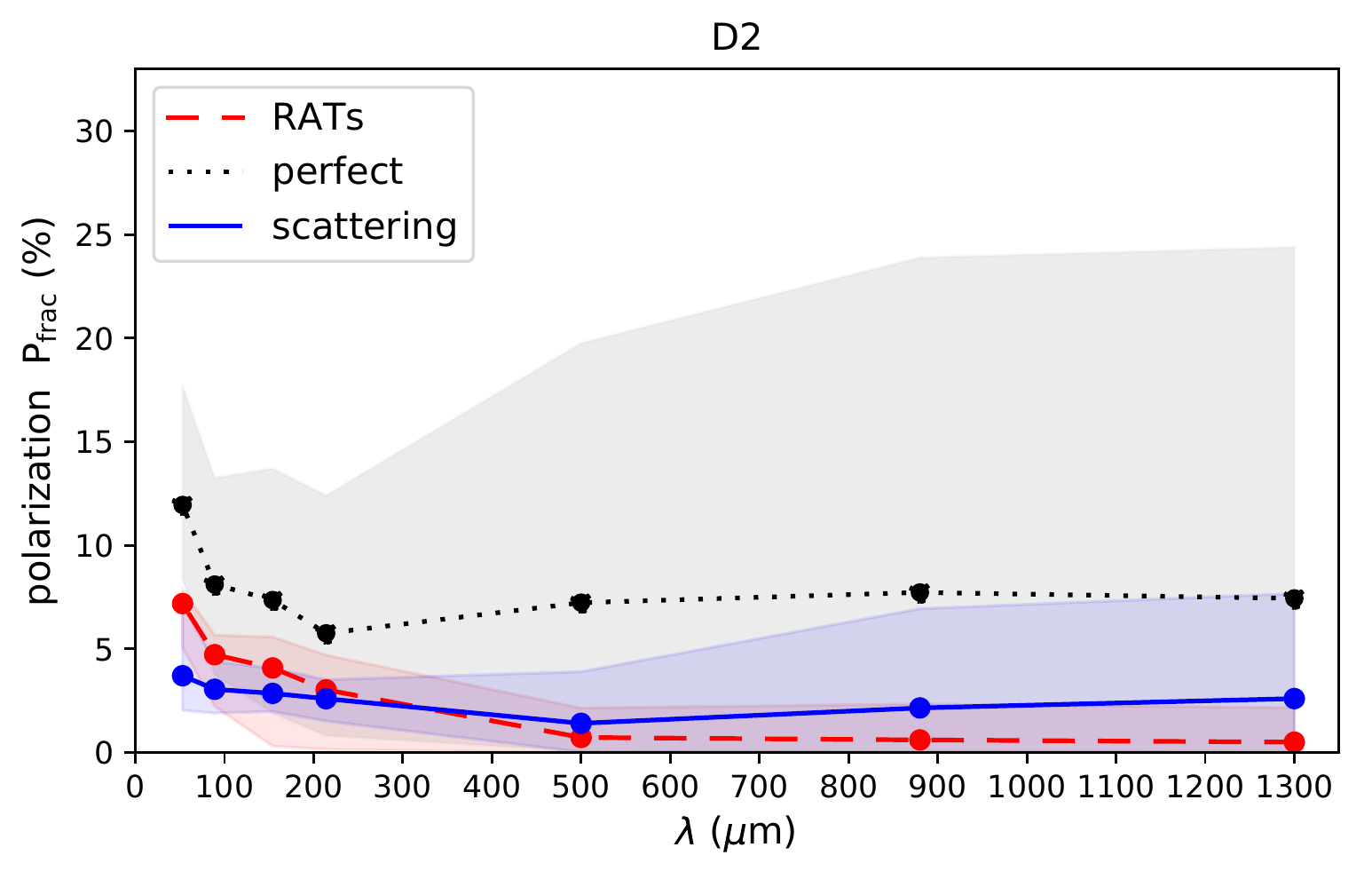}
\end{center}

\caption{Degree of linear polarization $P_{\mathrm{frac}}$ dependent on wavelength assuming RAT-aligned grains (long-dashed red), perfect alignment (short-dashed black), and scattering (solid blue) for the entire maps of intensity $I$ (top left) shown in \Fig{SyntheticDustEM} and \ref{fig:SyntheticDustSCA}, together with  the bridge (top right) and the regions D1 (bottom left) and D2 (bottom right). The lines represent the average values of the entire region, while the corresponding shaded areas indicate the range between maximum and minimum values.}
\label{fig:PolWave}
\end{figure*}

\begin{figure*}
\begin{center}
     \includegraphics[width=0.49\textwidth]{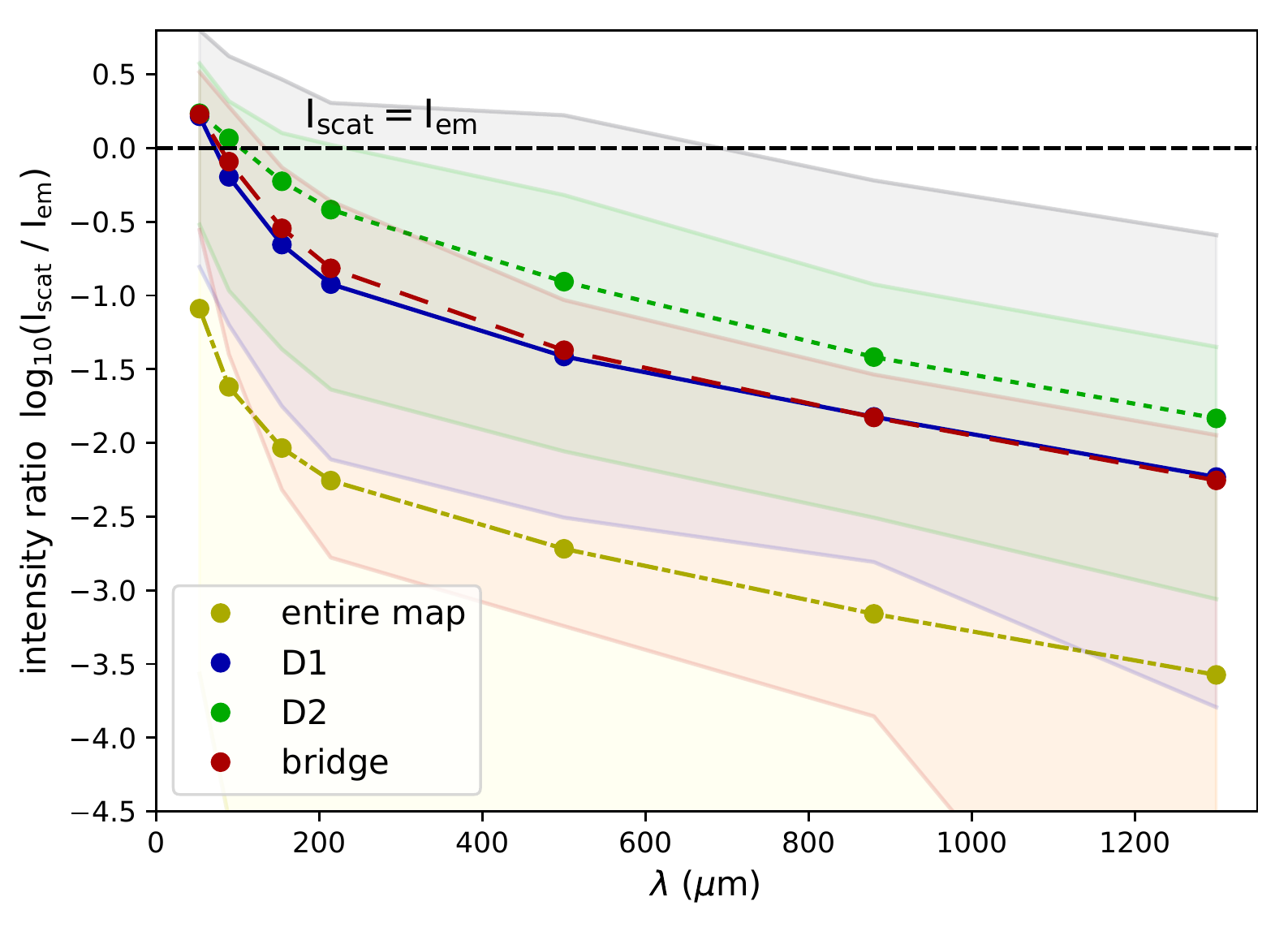}
     \includegraphics[width=0.49\textwidth]{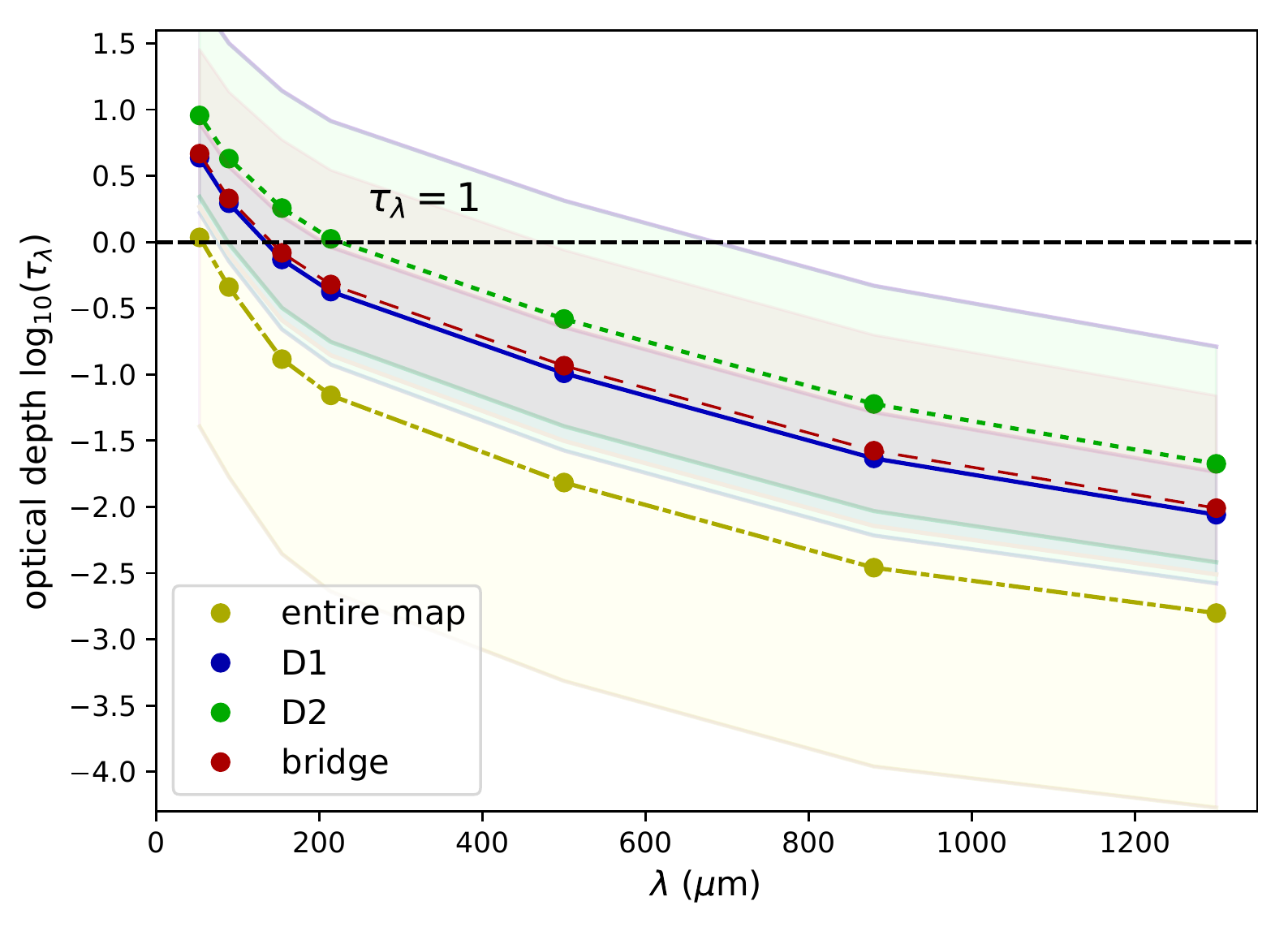}
\end{center}

\caption{Left panel: Wavelength-dependent ratio of scattered intensity $I_{\mathrm{scat}}$ to emitted intensity $I_{\mathrm{em}}$ of the entire map (dash-dotted yellow) shown in \Fig{SyntheticDustEM} assuming RAT alignment in comparison with the regions D1 (solid blue), D2 (short-dashed green), and the bridge structure (long-dashed red). Right panel:  Same as the right panel, but for the optical depth $\tau_\lambda$.}
\label{fig:RatTauWave}
\end{figure*}

\subsubsection{Weakly magnetized bridge}
\Fig{bridge_iso_vapor} illustrates the magnetic properties of the bridge in more detail.
In the left panel, we show the density together with the magnetic field structure associated with the bridge. 
For this purpose, we visualized 100 flow lines by again using the bias function in \vapor. Here, the bias was set to $b=-10$, such that the weakest $N=100$ of $N\times2^{|b|}$ computed flow lines are displayed.   
To emphasize the magnetization of the bridge, we map the magnetic field strength $|\mathbf{B}|$ on an isosurface of the typical density in the bridge of $10^{-16}\unit{g}\unit{cm}^{-3}$ in the right panel of \Fig{bridge_iso_vapor}.  
The visualizations show that the bridge is weakly magnetized, with a typical field strength of 1 to 2 mG, while the gas located closer to the protostars ($r<100$ AU) is much more magnetized, with a field strength of $\gg$5 mG (see \Fig{bridge_vapor}). 

\subsubsection{Distribution of the $B$-field strength} 
\Figure{rho_B_r} shows the magnetic field strength $|\mathbf{B}_{\rm rms}|$ of cells located within 4000 AU from the primary over cell density.
The color indicates the radial distance from the individual cells to the primary, and the plot shows a wide distribution of the field strength. 
As multiple cells can fall in the same bin of density $\rho$ and magnetic field strength $B$, 
a companion diagram in Appendix \ref{app:B} (\Fig{rho_B_count}) shows how many cells are located in each bin.    
We show the volume-averaged field strength per density bin. 
The overall trend of increasing field strength with increasing density in the vicinity of the primary protostar is consistent with flux-freezing.
This means that the highest field-strengths occur at the highest densities close to the protostar.
At lower densities $\rho < \sim 10^{-18}$, the profile is noticeably more shallow. 
Such a shallow profile at lower densities and a steeper $B$-$\rho$ relation at higher densities is in good agreement with results from colliding-flow simulations presented in Fig.\, 2 of \cite{ChenKingLi2016}. 
The profile at lower densities indicates that gravity is subdominant at lower densities until the magnetic field has funneled enough material into dense filaments. 
The prestellar cores form in the filaments \citep{Andre2010}, and self-gravity of the gas becomes dominant, leading to a steeper $B$-$\rho$ relation as a result of flux-freezing.
This profile is in good agreement with observations \citep{Crutcher2010} and also with recent results by \cite{Kuznetsova2020}.

For comparison, we also show the scaling of $B\propto \rho_{\mathrm{gas}}^{\frac{1}{\xi}}$, where $\xi$ is a characteristic parameter typically in the range of $\frac{3}{2}$ and  $2$ that is expected for the collapse of an isolated sphere \citep{Hennebelle&Fromang2008,Masson2016}. 
As the protostar is embedded in a turbulent filamentary environment and does not form from a collapsing isolated symmetrical core, there is significant scatter in the magnetic field strength at given density and no distinct narrow distribution along neither of the lines at this stage \citep[for a discussion of the observed large spreads in magnetic field strength, see][]{Crutcher2012}. 
The maximum field strength in the model occurs close to the primary protostar, and it is $\approx 0.1$ G, which is similar to the upper limit determined for protostellar collapse with ambipolar diffusion \citep{Hennebelle2016}. 
\citet{Masson2016} showed that in core-collapse simulations without turbulence and when the equations of ideal MHD are solved, the magnetic field strength rises to values of $>1 $ G at densities $\rho_{\mathrm{gas}} > 10^{-15} \unit{g}\unit{cm}^{-3}$,
while the field strength reaches a plateau value of $\approx 0.1$ G at most when ambipolar diffusion is accounted for.

The field strength is not as high in our simulation because the field is quenched as a result of the limited resolution for a minimum cell size of $4$ AU, and turbulence hampers the pile-up of the magnetic field in our model, similar to the recent results by \citet{Guszejnov2020}. 
If we resolved the vicinity of the young star in more detail, we would reach higher densities and higher magnetic field strengths as a consequence of flux-freezing. 
However, considering the low levels of ionization on these scales, field strength of $|\mathbf{B}| > 0.1$ G are unrealistic on radial distances in the range of $10 \unit{AU} \lesssim r \lesssim 0.1$ AU from the individual protostars as shown in non-ideal MHD simulations \citep[e.g.,][]{Tsukamoto2015a,Masson2016,Vaytet2018}.    
\subsection{Synthetic dust-polarization maps}
\subsubsection{Dust emission and polarization}
\Figure{SyntheticDustEM} illustrates the dust emission map of the natal structure of the bridge that connects the forming tertiary protostar (location: $x \approx -500 \unit{AU}$, $y\approx 1000 \unit{AU}$) with the secondary (location: $x\approx -1250 \unit{AU}$, $y\approx -1100 \unit{AU}$) along the $z$-axis. 
The upper panels show the Stokes $I$ component of the $1.3\ \mathrm{mm}$ dust emission, and the lower panels show the linear dust polarization $P_{\rm frac}$. 
The panels on the left-hand side correspond to the case of RAT grain alignment, and the plots on the right show the scenario where all grain sizes are perfectly aligned. 
The length and orientation of the cyan pseudovectors show the alignment and polarization degree of the dust emission. 
To allow a more direct comparison with the $B$ field, the displayed pseudovectors are rotated by 90$^{\circ}$ with respect to the orientation of the grains.

The density enhancement in the bridge also correlates with a higher intensity measured by the Stokes $I$ component. 
Considering the orientation of the dust grains, we find a pattern that is consistent with the projected morphology of the toroidal magnetic field lines associated with the the bridge (see \Fig{bridge_iso_vapor}). 

When we compare the scenarios of RAT alignment with perfect alignment (all grain sizes are aligned), the orientation angle $\phi_{\mathrm{pol}}$ of the dust grains is virtually unaffected.
However, the relative degree of dust polarization is typically lower for high column densities when RAT is accounted for. 
The relative difference in polarization between the two scenarios is even more pronounced inside the bridge. 
The lower panels of \Fig{SyntheticDustEM} show that we would expect a maximum polarization fraction of $P_{\rm frac}$ of up to 30 $\%$ in the bridge for the case of perfect alignment, but only up to $10 \%$ when the RAT effect is accounted for, which is expected from theory at these spatial scales. 
In detail, the RAT alignment and the subsequent polarization increase with a larger radiation field and decreases with higher densities. As shown in \Fig{SyntheticDustEM} the dust polarization with RATs shows no particular response to the presence of S1 and S2 in the disk. The RATs seem to be driven by the ISRF and the density alone, but not by the local radiation field of the protostars. Because the polarization fraction from RATs is substantially smaller than perfect alignment, this would indicate that grains are significantly more randomized toward the denser regions. This would explain the lower degree of polarization for RAT alignment within the vicinity of the protostars and the bridge compared to perfect alignment. 

\subsubsection{Dust self-scattering}
The polarization of dust grains may not be exclusively caused by the magnetic field, but also by self-scattering of large dust grains. Especially, the dust polarization on scales $<100$ AU, that is, in disks, is typically caused by self-scattering rather than by dust grains that are aligned with the magnetic field in the disk \citep[][]{Kataoka2015}. These are also the regions where our dust component with a $a_{\mathrm{max}}=100 \mu$m is situated (see Sect.\ref{sect:RTPostProcessing}).  
As expected from theory, radiation due to self-scattering is strongest in the vicinity of the forming stars, especially in the region where the third companion is about to form.
In \Fig{SyntheticDustSCA} we show the intensity purely caused by self-scattering and the intensity ratio of self-scattering to dust emission at $1.3$ mm wavelength.

The right panel in \Fig{SyntheticDustSCA} also demonstrates that the relative radiation due to self-scattering is highest in the vicinity of the forming protostars. Compared to the radiation that is induced by dust emission, radiation from self-scattering is only a minor contributor at $1.3$ mm wavelength. 
Self-scattering is only responsible for $<$1 $\%$ of the polarization in the bridge, and even in the vicinity of the protostars, it only contributes $\sim$ 10 $\%$ at most to the polarized emission at $1.3$ mm. 

\subsubsection{Polarization-intensity dependences}
Numerous observations of cores in the submillimeter regime show a significant depolarization in the most luminous regions \citep[][]{Henning2001,Wolf2003,Goncalves2005,Brauer2016}. This (anti)correlation is commonly fitted by the polarization-intensity (PI) relation ${ P_{\rm frac} \propto I^{\alpha} }$ \citep{Henning2001}. In this section we investigate how the polarization in distinct regions behaves dependent on the different mechanisms of dust polarization. We plot the PI for a wavelength of $1.3\ \mathrm{mm}$ considering the cases of RATs, the combined intensity using RATs and self-scattering, and perfect alignment. The resulting PI relations for the these regions (marked, e.g., in \Fig{SyntheticDustEM}) are shown in \Fig{PIRelation}.  We determined the slope $\alpha$ with a least-squares fit in log-space. When the PI is plotted for the entire map, the value of $\alpha$ for the combined intensity of the polarization $P_{\rm frac}$ is slightly higher than that of RAT alignment, demonstrating a minor contribution of scattered radiation to the total polarization at $1.3\ \mathrm{mm}$. In perfect alignment, the PI relation is completely lost. 
When we compare the values of the degree to those derived from real polarization measurements, we have to take the limited dynamic range (concerning the intensity) and the minimum reliable polarization degree (typically a few $0.1 \%$) into account. 
Consequently, the observed degree is dominated by the densest, that is, (sub-)millimeter brightest regions.

In the bridge alone, all three polarization cases barely depend on intensity. However, for the case of perfect alignment, the degree of polarization is roughly ten times higher than in the other two cases. 

For dense cores, the slope $\alpha$ is typically $0.5 - 1.5$ \citep[][]{Henning2001,Matthews2002}. We see this behavior also for the regions D1 and D2. Here, the exception is the case of perfect alignment, which even shows a slightly positive slope for D1 and the bridge. We note that the exact value of $\alpha$ strongly depends on the upper grain size $a_{\mathrm{max}}$. Because we introduced grains up to $100\ \mu\mathrm{m}$ into the denser regions, we can always  expect some aligned grains with  $a>a_{\mathrm{alg}}$ (see Sect. \ref{sect:RTPostProcessing}), and subsequently, some polarization, even though the radiation field may not fully penetrate these regions. A steeper slope may be achieved by considering smaller grains.

We note that we did not account for a possible detection limit of polarization in \Fig{PIRelation}. However, the trends we show are rather robust even if we were to limit our analysis to a hypothetical detection limit of $P_{\rm frac}>1\ \%$.

\subsubsection{Multiwavelength dust polarimetry}
In order to evaluate the detectability of the magnetic field geometry, we created a series of synthetic observations for $53\ \mu\mathrm{m}$, $89\ \mu\mathrm{m}$, $154\ \mu\mathrm{m}$, $214\ \mu\mathrm{m}$, $550\ \mu\mathrm{m}$, $880\ \mu\mathrm{m,}$ and $1.3\ \mathrm{mm, which    }$ is typical for instruments such as the High-resolution Airborne Wideband Camera Plus (HAWC+) at the Stratospheric Observatory for Infrared Astronomy (SOFIA) \citep[][]{Dowell2010,Harper2018}, Herschel \citep[][]{Pilbratt2010,Rodenhuis2012}, the SubMillimeter Array (SMA) \citep{Ho2004,Marrone2008}, or the Atacama Large (sub-) Millimeter Array (ALMA) \citep[][]{Brown2004}. In \Fig{SyntheticDustEMWave} we present an exemplary map of dust emission considering RATs at a wavelength of $89\ \mu\mathrm{m}$ (extinction-dominated regime; \Fig{DustModel}). Comparing this map with the corresponding $1.3\ \mathrm{mm}$ in \Fig{SyntheticDustEM} we see that the polarization vectors are rotated. This is because dichroic extinction replaces thermal emission as the dominant polarization mechanism. This is especially true in the bridge, D2, and in a patch in the upper right corner of the map, where all vectors are rotated by $90^\circ$. We note that the onset of this rotation starts already in the maps at $214\ \mu\mathrm{m}$. The map of the ratio of scattered ($I_{\mathrm{scat}}$) to emitted radiation ($I_{\mathrm{em}}$) in \Fig{SyntheticDustEM} has values up to unity, indicating that scattering also becomes a considerable polarization factor at $89\ \mu\mathrm{m}$. 

We present these trends more systematically in \Fig{PolWave}. Here, we show the range of polarization caused by RATs, perfect alignment, and scattering as a function of wavelength. Yet again, we evaluated the polarization for the entire map, the bridge, and regions D1 and D2 separately. In all regions we see comparable trends, where perfect alignment tends to overestimate the degree of polarization. From the near-IR to the millimeter regime of wavelengths, the polarization is governed by RATs but scattering contributes only marginally, except for the densest regions. 
We strongly emphasize that because polarization is caused by RAT-aligned grains we cannot infer whether emission or dichroic extinction is most dominant. This needs to be evaluated separately (see below). In the the different regions, scattering starts to considerably affect the polarization pattern between $100\ \mu\mathrm{m}$ and $200\ \mu\mathrm{m}$. This becomes more obvious in \Fig{RatTauWave} where we show the ratio $I_{\mathrm{scat}}/I_{\mathrm{em}}$ and the optical depth $\tau_\lambda$ over wavelength. In the millimeter regime, only region D2 is partly optically thick and emission is mostly due to aligned dust grains. At $200\ \mu\mathrm{m,}$ most of the regions start to become optically thick. Consequently, scattering or dichroic extinction contributes most to the polarization signal.

\subsubsection{Angle between the intensity gradient and magnetic field}
\begin{figure*}
\begin{center}
     \includegraphics[width=0.325\textwidth]{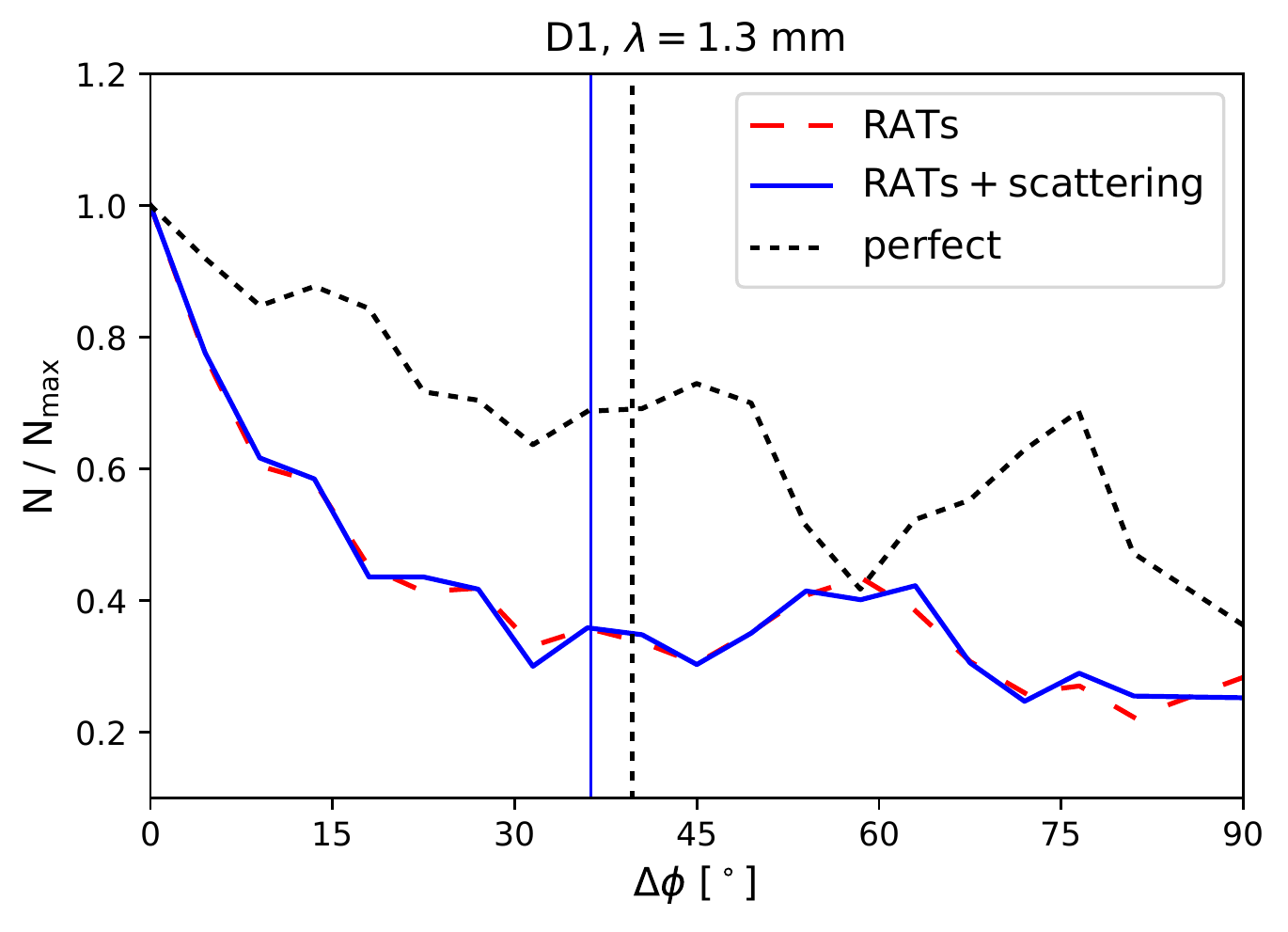}
     \includegraphics[width=0.325\textwidth]{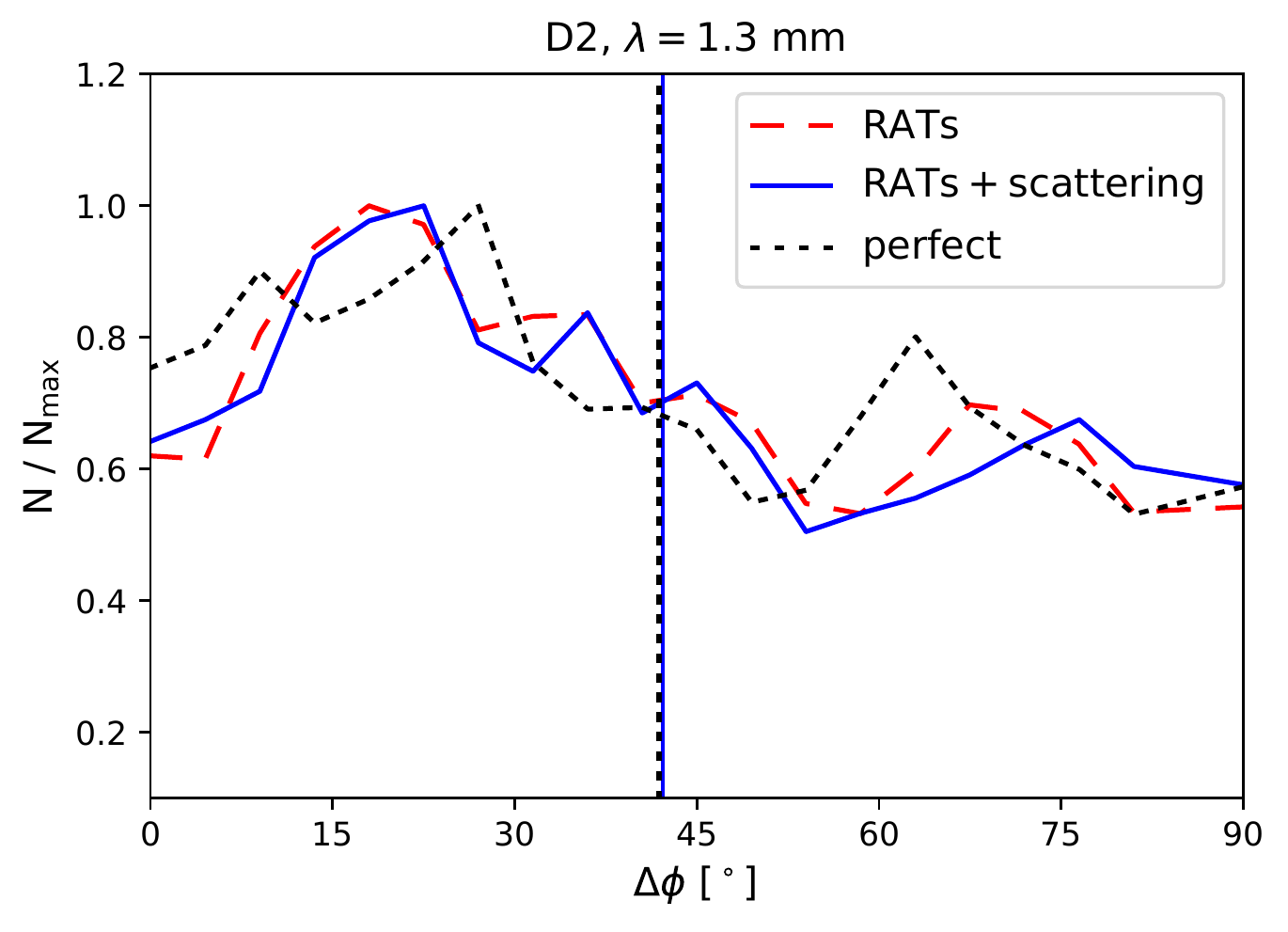}
     \includegraphics[width=0.325\textwidth]{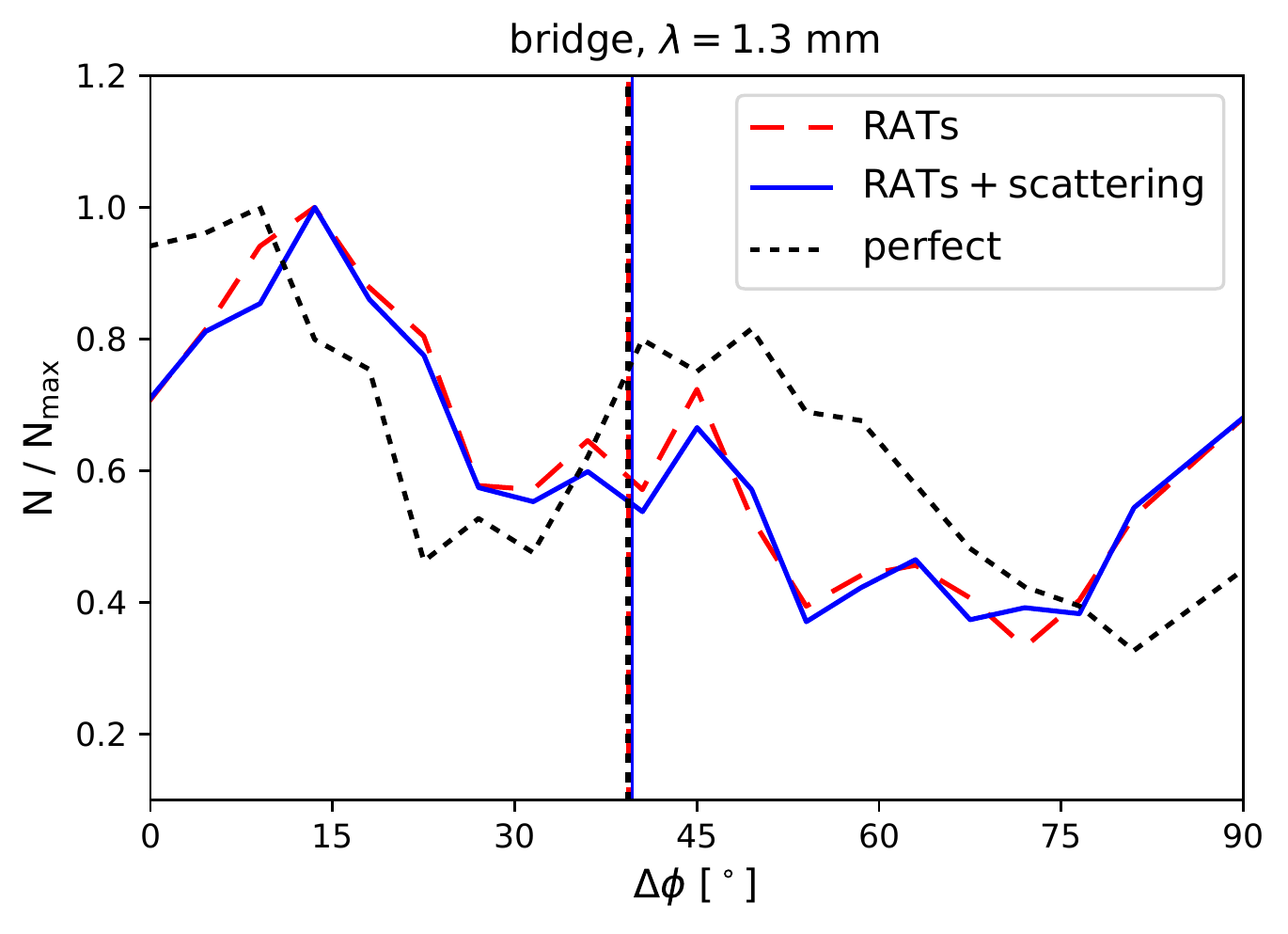}
\end{center}
\begin{center}
     \includegraphics[width=0.325\textwidth]{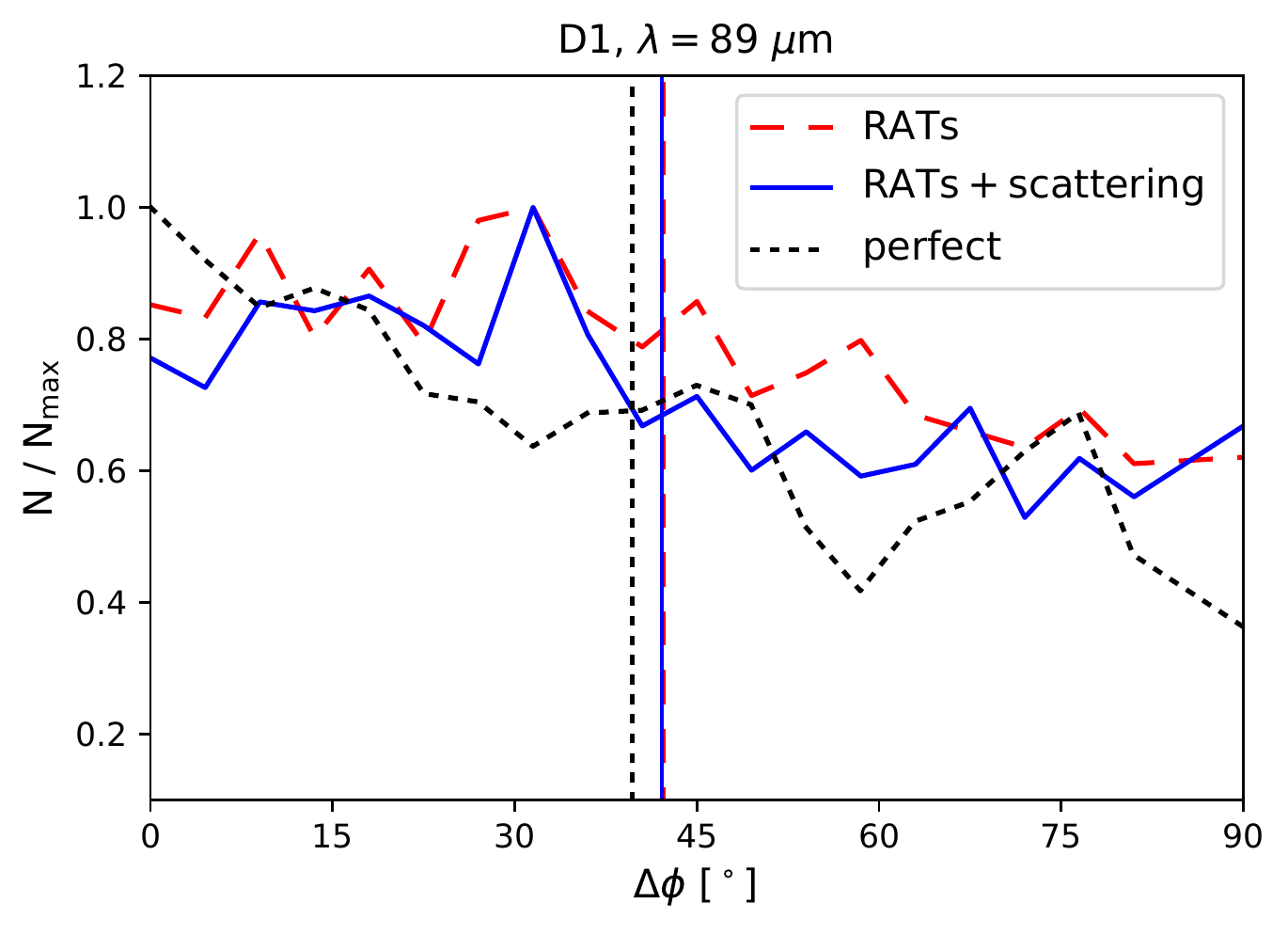}
     \includegraphics[width=0.325\textwidth]{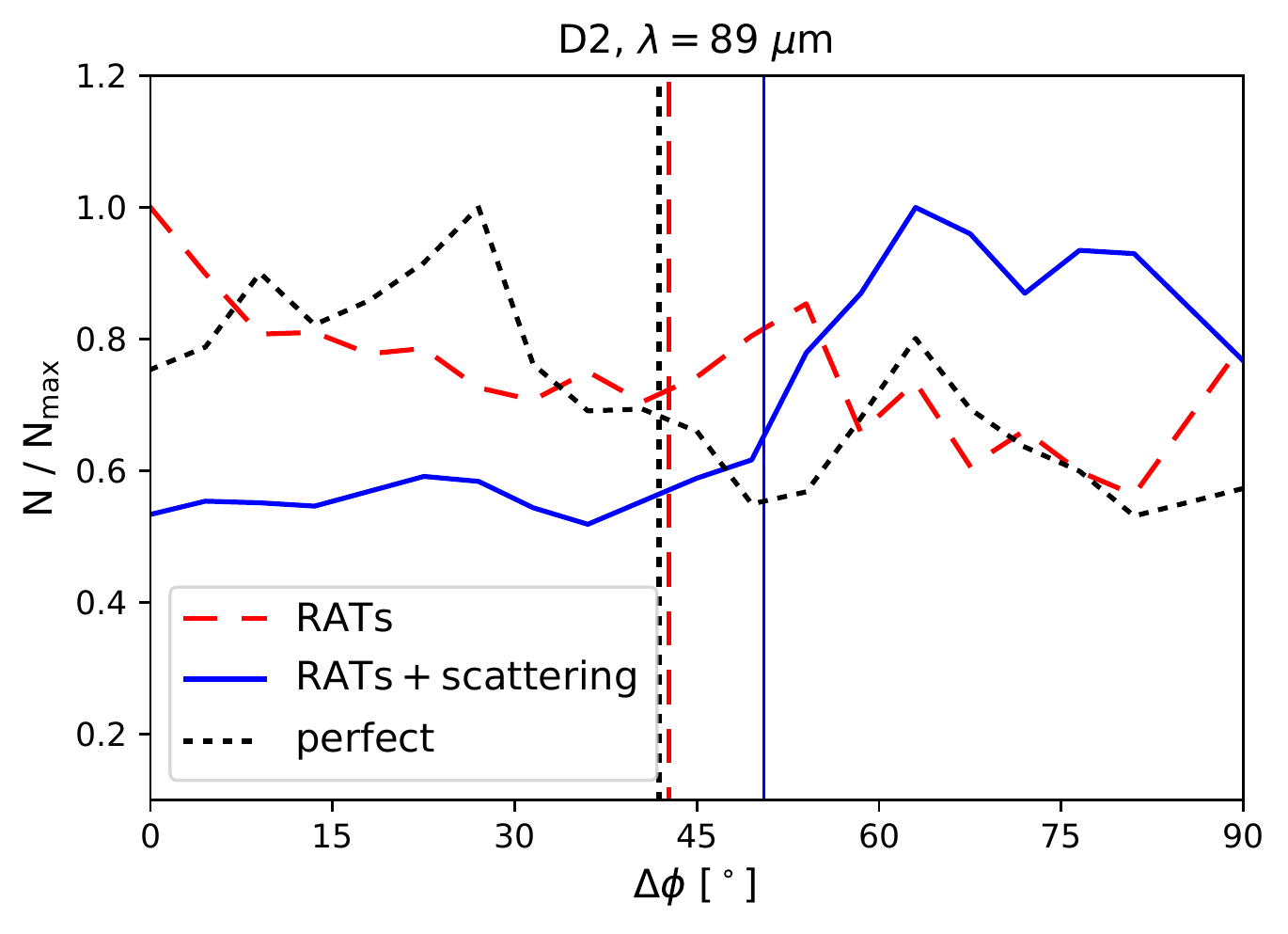}
     \includegraphics[width=0.325\textwidth]{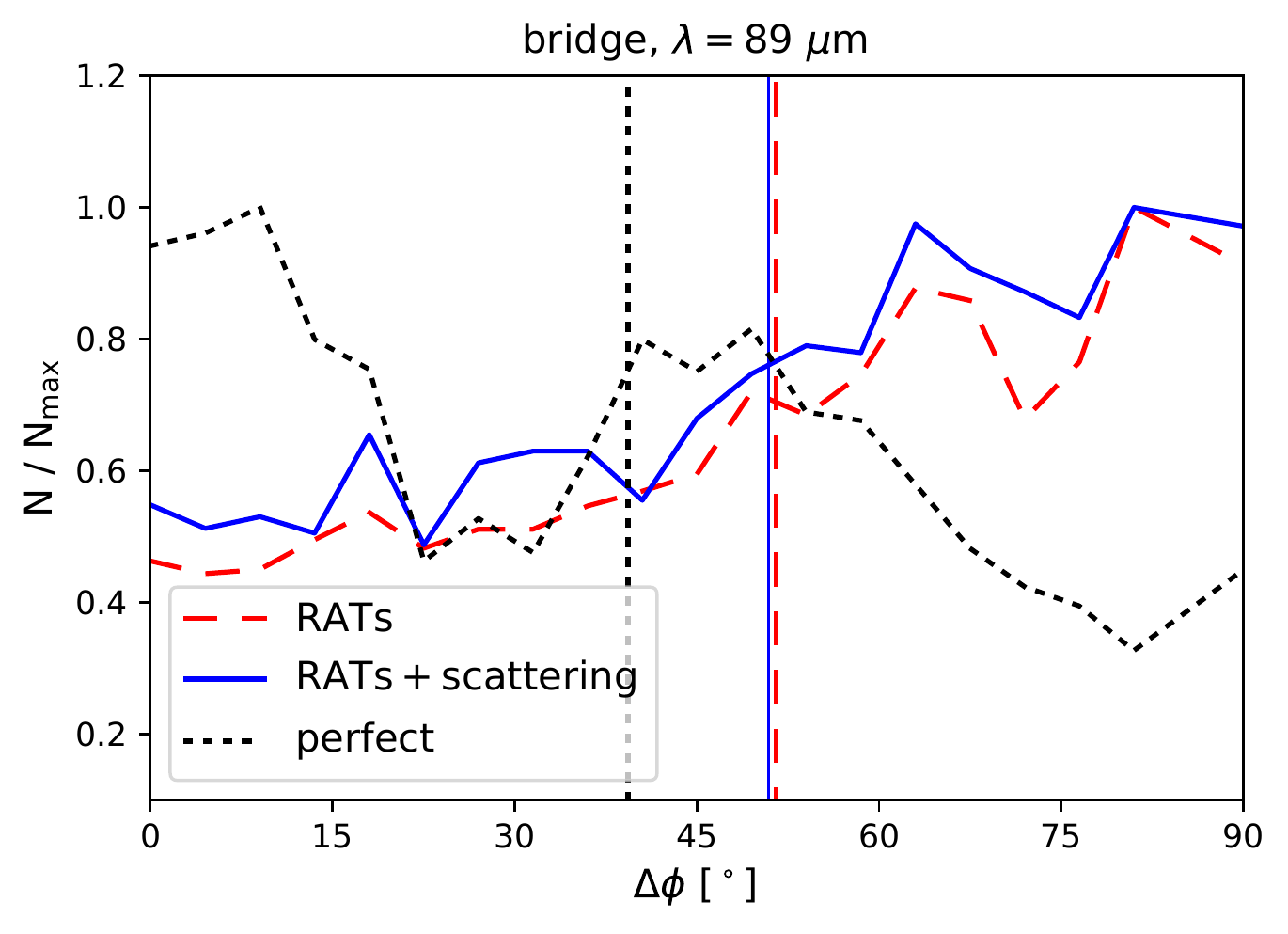}
\end{center}
\caption{Distribution of the angle $\Delta \phi$ between intensity gradient and the magnetic field direction in regions D1 (left column) and D2 (middle column), and the bridge (left). We compare $\Delta \phi$ for the cases of RAT alignment (long-dashed red), RAT alignment and scattering (solid blue), and perfect alignment (short-dashed black) for $89\ \mu\mathrm{m}$ (top row) and $1.3\ \mathrm{mm}$ (bottom row) observations (compare  \Fig{SyntheticDustEM}, \Fig{SyntheticDustSCA}, and \Fig{SyntheticDustEMWave}). Vertical lines indicate the mean of each distribution. We note that the mean values and trends are similar in the distinct regions for $1.3\ \mathrm{mm}$  observations, while the $89\ \mu\mathrm{m}$ observations depend heavily on the considered polarization mechanism.}
\label{fig:HistChi}
\end{figure*}

Analogously to Fig.\ 7 in \citet{Sadavoy2018}, \Fig{HistChi} shows the difference ${ \Delta \phi = |\phi_{\mathrm{pol}} -  \phi_{\mathrm{I}}}|$ between the angle $\phi_{\mathrm{I}}$ of the  intensity gradient and the projected magnetic field direction $\phi_{\mathrm{pol}}$ inferred from the synthetic dust maps at $89\ \mu\mathrm{m}$ and $1.3\ \mathrm{mm}$. 
We note that the relative angles may also be analyzed by means of the histogram of relative orientation (HRO) technique \citep{Soler2013} or the projected Rayleigh statistic (PRS) \citep{Jow2018}. These statistical techniques quantify the relative angles purely dependent on column density. However, in this paper we intend to provide some comparison of distinct regions similar to the observations presented in \citet{Sadavoy2018}. We emphasize that the exact choice of data representation does not affect the conclusions drawn in the following sections.

In \Fig{HistChi} we consider the three cases of RAT alignment, RAT alignment and self-scattering combined, and perfect alignment. The analyzed regions D1, D2, and the bridge are marked in \Fig{SyntheticDustEM}. For $1.3\ \mathrm{mm}$ observations, all three cases show a similar pattern. Altogether, $\Delta \phi$ is rather evenly distributed, with a slight bulge toward $\Delta \phi = 0^\circ$. This is consistent with the winding of the magnetic field lines that is associated with the bipolar outflows. This means that intensity gradient and dust polarization are not clearly correlated. D2 and the bridge in turn show an aggregation of $\Delta \phi $ close to $15^{\circ}$. We speculate that the peaks correspond to the increase in intensity from the surrounding area toward the center, where the magnetic field has a relatively strongly ordered component throughout D2 and the bridge region (see \Fig{bridge_iso_vapor}).

A comparison of the three cases of dust polarization (RAT alignment, self-scattering, and perfect alignment) with each other shows that when we consider RATs with or without the effects of self-scattering, the amplitudes of the angle distribution are changed only marginally, while the general pattern is almost identical. The angle distribution corresponding only to perfect alignment is similar to the other two curves, but with a higher amplitude. However, these differences are only minor and indicate that the assumption of perfect alignment of dust grains in the millimeter regime allows us to draw conclusions about the magnetic field structure in disk and bridge-like structures.

The observations at $89\ \mu\mathrm{m}$ draw a vastly different picture than the observations at $1.3$ mm. The distributions of the angular difference $\Delta \phi$ of each polarization mechanism no longer show a consistent trend within the distinct region. The most physical dust polarization model is represented by the case of RATs plus scattering, but the polarization pattern no longer allows us to infer any information about the magnetic field orientation. This is because the contribution of scattering, dichroic extinction, and thermal emission are on the same order at $89\ \mu\mathrm{m}$. We also note that the trends for the perfect alignment case are almost identical at $89\ \mu\mathrm{m}$ and $1.3\ \mathrm{mm}$. Dust polarization modeling assuming perfect alignment seems to fail for wavelengths $< 200\ \mu\mathrm{m}$.

\section{Discussion}

\subsection{Asymmetric outflows}
As pointed out in the description of \Figure{bridge_vapor}, the magnetic fields cause a bipolar outflow from the primary protostar. 
Following the primary protostar during its evolution, we find that outflows are launched intermittently, while the outflow direction evolves dynamically. 
Evolving outflow directions have been observed for instance for L1157, where the outflow is precessing \citep{Tafalla2015,Podio2016}.
Considering the direction of the bipolar outflow, we find that the outflows are usually asymmetric, which is consistent with the perturbations of the prestellar core in the turbulent birth environment. 
Although asymmetric outflows disagree with predictions from symmetrical core-collapse models with an initial alignment of angular momentum and magnetic field vector \citep[e.g.,][]{Matsumoto2004,BanerjeePudritz2006}, asymmetric outflows are consistent with observations of Class 0 objects such as Serpens SMM1-a and b \citep{Hull2016,LeGouellec2019}, or OMC-3 MMS 6 \citep{Takahashi2019}, as well as several sources in Perseus \citep{Stephens2018,Stephens2019}. 
Asymmetric outflows have also been observed for multiple Class II objects, such as jets associated with DG Tauri B \citep{Mundt1987,Podio2011}, RW Aur \citep{Hamann1994,Hirth1994}, AS 353 A \citep{Hamann1994}, L1551-IRS 5 \citep{Mundt1991}, DO Tau \citep{Hirth1997}, or Haro 6-5 B \citep{Mundt1991}. 

\citet{Machida2020} recently studied the properties of outflows in an initial setup of different angles between magnetic field lines and rotational axis. They found similar outflow asymmetries for initially misaligned cases.
Our zoom-in models demonstrate that turbulence in GMCs can in fact cause such deviations from symmetry of prestellar cores and thereby affect the accretion process of forming protostars. 
Moreover, we know from our previous models \citep{Kuffmeier2017} that outflows launched around stars that form from the collapse of more isolated prestellar cores tend to be more symmetric than outflows associated with more embedded protostars such as the protostellar multiple system studied in this paper. 
Asymmetric bipolar outflows are therefore a direct consequence of star formation in locations of GMCs with complex inhomogeneous velocity patterns. 

\subsection{Ideal MHD limit}
Theory \citep{Hennebelle2016} and nonideal MHD simulations \citep{Tsukamoto2015a,Masson2016,Vaytet2018} suggest a characteristic plateau value of $\sim 0.1$ G for the initial collapse phase of a single protostar as a consequence of ambipolar diffusion, considering ionization rates of $\sim 10^{-17} \unit{s}^{-1}$. 
Because only a few cells exceed the plateau value of $0.1$ G (see \Fig{rho_B_r}), we are confidet that accounting for ambipolar diffusion would change the properties of the bridge and the corresponding formation of the protostellar multiple marginally at most.
However, resolving the disk and its inner physical properties requires higher refinement. 
Modeling the disk around the primary with high enough resolution requires studying the structure of the disk and therefore requires incorporating ambipolar diffusion, considering a moderate level of the ionization fraction due to cosmic rays with ionization rates of $10^{-17}$ s$^{-1}$ to $10^{-16}$ s$^{-1}$ \citep{Caselli1998,Padovani2018}. 
For a detailed model with nonideal MHD, we would also need to account for effects of the grain size distribution on the resistivities that dominate on radial distances of $r\lesssim 10$ AU \citep{Zhao2018}. 
Moreover, we consider a different physical state in this study compared to the early-collapse phase of a single star with a multiple system in which the primary protostar is 70 kyr old, the secondary is 27 kyr old, and the tertiary protostar forms about 4 kyr after the snapshot that is analyzed. For a study investigating synthetic maps of polarized dust continuum emission based on spherical collapse simulations that account for nonideal MHD, we refer to \cite{Valdivia2019}.

\subsection{Bridge formation}
The differences in magnetization shown in \Fig{bridge_iso_vapor} cannot solely be explained by flux-freezing in a $|\mathbf{B}| \propto \rho_{\mathrm{gas}}^{1/\xi}$ manner because the differences in $|\mathbf{B}|$ shown in the right panel correspond to the same constant density.    
As shown in \Fig{bridge_vapor}, the strongest field lines do not correlate with the bridge. 
\Figure{bridge_iso_vapor} shows that the bridge is weakly magnetized compared to the gas of equal density located at smaller radial distances of $r \lesssim 100$ AU from the forming protostars.

Previous analyses of colliding flows in 3D setups \citep{ChenOstriker2015,ChenKingLi2016,Chen2019} that used the \athena\, code \citep{Stone2008} showed the formation of dense filamentary structures as a consequence of gas collisions. 
Similarly, the bridge-structures in our model emerge on a smaller scale as a result of colliding flows, leading to the compression of an initially larger filament, as first described in \citet{KuffmeierBridge}. 

To quantify the comparison of the flow to the magnetic field, we computed the average Alfv\'enic Mach number $M_{\rm A} = |\mathbf{v}|/|\mathbf{v}_{\rm A}|$ in a region within 4000 AU from the center of mass of the primary and secondary protostar. 
The velocity $\mathbf{v}$ was computed relative to the velocity of the mass-weighted velocity of the binary star, and the Alfv\'en velocity was defined as $\mathbf{v}_{\rm A} = \frac{\mathbf{B}}{\sqrt{4\pi\rho}}$.
As the speed of the gas in the region is predominantly super-Alf\'enic (the mass-weighted average is $M_{\rm A}\approx 5.3$ and the volume-weighted average is $M_{\rm A}\approx 2.2$), 
we conclude that
the formation of such bridges is driven by the gas dynamics. 
The field lines are dragged along with the gas motion, 
but do not provide substantial magnetic support. 

\subsection{Density and magnetic field strength in the bridge compared to observations}
Our model shows that the typical density in the bridge is ${ \rho_{\mathrm{gas}}\sim 10^{-16} \unit{g}\unit{cm}^{-3} }$. 
When we assume a mean molecular weight of 2.8 \citep{Kauffmann2008}, $\rho_{\mathrm{gas}}$ corresponds to a number density of ${ n_{\rm br} \approx 2.1 \times 10^{7}\unit{cm}^{-3} }$.  
\cite{vanderWiel2019} carried out radiative transfer models accounting for dust continuum and gas molecular line tracers in the bridge of IRAS 16293--2422.
Based on their study, they expect the number density in the bridge of IRAS 16293--2422 to be typically in the range of $4\times 10^4 \unit{cm}^{-3}$ and $3\times10^7 \unit{cm}^{-3}$. 
$n_{\rm br}$ is in good agreement with this estimate, and \cite{vanderWiel2019} also reported higher peak number densities in the bridge of $7.5\times 10^{8} \unit{cm}^{-3}$ , which is consistent with the densest parts in the synthetic bridge. 

In our synthetic bridge, we determined the typical magnetic field strength to be in the range of 1 to 2 mG. 
Using the Davis-Chandrasekhar-Fermi method \citep{DavisGreenstein1951,ChandrasekharFermi1953}, \cite{Sadavoy2018} determined the magnetic field strength in IRAS 16293--2422 to be in the range of $23$ to $78 \unit{mG}$, which is significantly higher than the field strength in our model.
The magnetic field strength is based on the assumption of a typical number density of $5.6 \times 10^8 \unit{cm}^{-3}$ for the bridge, which is more than an order of magnitude 10 higher than $n_{\rm br}$. 
However, \cite{Sadavoy2018} acknowledged that the density may well be overestimated using only a modified blackbody function (Eq.\ 7 in their paper) given the uncertainties on the grain size distribution, the dust temperature, and the dust-to-gas ratio. \citet{vanderWiel2019} argued that the density is in fact lower than $10^8 \unit{cm}^{-3}$ because high-density tracers such as o-H$_2$CO $5_{1,5}$ -– $4_{1,4}$ (critical density of $7.5\times 10^7 \unit{cm}^{-3}$) or H$_{13}$CN 4–3 (critical density of $1.2\times 10^8 \unit{cm}^{-3}$) are not detected.
\cite{Sadavoy2018} pointed out that when a lower number density of $3\times 10^7 \unit{cm}^{-3}$ were assumed, the field strength would be expected to be about 6 to 20 mG, which is only a factor of a few more than the field strength in our model. 
The analysis therefore suggests that the environment, in particular the bridge structure, of IRAS 16293--2422 is more magnetized by a factor of a few than the protostellar triple considered in the model.

A higher magnetization of IRAS 16293--2422 than in the model is also consistent with a smoother pattern of the polarization vectors in IRAS 16293--2422. 
For the lower magnetic field strength in the model, the turbulent motions that cause the formation of the bridge can more easily drag the magnetic field lines with them, and hence lead to the more perturbed polarization pattern in the bridge of the model compared to the more magnetized region in IRAS 16293--2422.   
Furthermore, we speculate that IRAS 16293--2422 is in general more magnetized than other sources of similar age that show a weaker and more chaotic pattern of dust polarization, as suggested by \cite{Sadavoy2019}. 
This interpretation is also in agreement with the polarization pattern obtained in models by \citet{HullMocz2017}, where the polarization pattern becomes more regular for increasing levels of magnetization (see Fig.\ 2 in their paper). 

Recently, other studies have emphasized the difference in polarization fraction depending on the orientation of the mean magnetic field with respect to the viewing angle of the observer \citep{King2018}. 
In this study, we predominantly analyzed the bridge along one line of sight, motivated by the resemblance of the structure with IRAS 16293--2422 when seen from this viewing angle. 
In our synthetic observation, 
the polarization pattern and fraction also depends on the viewing angle. 
However, the magnetic field structure in the bridge is generally more perturbed and toroidal (as illustrated in \Fig{bridge_iso_vapor}) than in more idealized parameter studies. 
This shows that the viewing angle only mildly affects the polarization fraction in the bridge, although the bridge structure  becomes less visible when seen from an angle along the elongation. 

\subsection{Origin of dust polarization}

The degree of polarization for IRAS 16293--2422 reported in \cite{Sadavoy2018} may reach peak values of more than $20\ \%$ near source B at $1.3$ mm wavelength. This is somewhat higher than the peak polarization of $8\ \%$ we find around the protostars and the bridge in our synthetic observations when RATs and self-scattering are included. One factor may be a difference in the radiation field. The models of \cite{Jacobsen2018-IRAS16293} suggest sources with luminosities up to $18\ L_\odot$  within IRAS 16293--2422, whereas sources S1 and S2 have only $2.54-4.27\ L_\odot$. However, \polaris\ runs with higher luminosities reveal that the radiation field is still dominated by the ISRF alone because S1 and S2 are embedded in dense core-like structures. By evaluating the 3D dust temperature distribution and the intensity maps presented in this paper, we estimate that the range of influence of S1 and S2 is between $50 - 200\ \mathrm{AU}$ .

%Furthermore, a size distribution with grains up to $a_{\mathrm{max}}=3\ \mathrm{mm}$ is already a favorable assumption for maximizing the dust polarization in emission. 
Perfect grain alignment would in turn overestimate the degree of polarization. 
This finding concerning grain alignment is consistent with the parsec-scale synthetic radiative transfer observations of the ISM presented in \cite{Seifried2019} and  \cite{Reissl2020}. 
Because it is known that under realistic conditions, grains are not perfectly aligned, previous models have typically accounted for the imperfection by multiplying the degree of polarization with an efficiency factor of about $0.1$ to $0.2,$  which is introduced ad hoc as a proxy to match observations \citep[e.g.,][]{FiegePudritz2000}. 
However, by solving the underlying equation of grain alignment, \polaris\, consistently accounts for the efficiency of grain alignment. 
To ensure an appropriate comparison of perfect alignment to the effects of RATs, we assumed $100 \%$ efficiency in the perfect alignment scenario.
Because of the higher estimated degree of polarization, perfect alignment would violate the PI relation, that is, depolarization towards high-density regions. \cite{King2018} reported that the perfect alignment case (or `homogeneous grain alignment' in their nomenclature)  cannot reproduce the correlation between column density and polarization. \cite{King2018} highlighted the importance of properly treating grain alignment physics in modeling synthetic observations. Corrections for mimicking RAT alignment were later investigated in \cite{King2019} without invoking the full complexity of RAT physics. 

We note a somewhat lower polarization fraction in our synthetic observations than in the observations of the IRAS 16293--2422 system (see section above). We attribute the difference in polarization to a higher magnetization in IRAS 16293--2422. Because our bridge is less magnetized than IRAS 16293--2422, it shows a lower resistance to perturbations. Consequently, we have more twisted magnetic field lines in the bridge, and they are even more twisted in the vicinity of the stars on scales $\lesssim 100 \unit{AU}$ from the individual sources. The huge effect of twisted magnetic field lines on the polarization was demonstrated in \cite{Reissl2020} by comparing synthetic dust observations of regular and irregular fields. Hence, the emission may become depolarized along its way to the observer, leading to the overall lower degree of polarization. Moreover, polarization fractions of a few up to $\sim 10 \%$ are in agreement with observations by \cite{Galametz2018}, who used the SMA for envelopes of 12 Class 0 objects.

The synthetic observations at $1.3\ \mathrm{mm}$ wavelength show that self-scattering causes additional radiation toward the observer mostly from the dense regions around the protostars. For most of the simulation domain, we find dust temperatures $T_{\mathrm{dust}}$ of about $6-18\ \mathrm{K}$. The emission in sources S1 and S2 with their effective temperatures $T_*$ of ${ 4776\ \mathrm{K} }$ and ${ 4336\ \mathrm{K} }$, respectively,   peaks at about  $600\ \mathrm{nm}$, and this emission would be $\sim450$ times higher than the $1.3$ mm dust emission for $T_{\rm dust} = 15$ K.
However, this stellar radiation is well shielded by the dense envelope in which S1 and S2 are embedded. This means that most of the radiation shown in \Fig{SyntheticDustSCA} is due to dust self-scattering from the interstellar radiation field. Scattering of stellar radiation on dust is only dominant within radial distances of a few AU from the individual sources. However, compared to polarized dust emission, self-scattering is not the dominant factor of polarization in our synthetic observations.

This contradicts the finding reported in \cite{Sadavoy2018} that the polarization in the disk associated with source B in IRAS 16293--2422 is mostly due to self-scattering. However, this may not be universally true. The multiwavelength study of HL Tau by \cite{Stephens2017} suggests wavelength-dependent transitions between regimes where distinct polarization mechanisms may become dominant. We note that such transitions would also be highly dependent on the grain properties. However, determining the exact boundaries of parameters where aligned dust grains may trace the magnetic field lines is beyond the scope of this paper. For the time being, studies like that of \cite{Stephens2017} raise serious doubts about the reliability of dust for inferring the magnetic field geometry on scales $\lesssim 100 \unit{AU}$ from individual stars. 
In disks, it can be particularly misleading to us dust polarization at these low wavelengths as a tracer for the magnetic field structure, as was shown by \cite{Kataoka2015}.

\subsection{Ideal observational conditions}
\label{sect:ObsConditions}
The importance of considering dichroic extinction and thermal emission simultaneously in multiwavelength observations has been emphasized in \cite{Reissl2014,Reissl2016,Reissl2017}. In particular, \cite{Reissl2014} presented a simple criterion for estimating the transition regime between dichroic extinction and emission. When the quantities ${ \tau_\lambda=\rho_{\mathrm{dust}}\ell\times\kappa_\lambda }$ and ${ \Delta\tau_\lambda=\rho_{\mathrm{dust}}\ell\times\Delta\kappa_\lambda }$ are introduced, the inequality ${ 1 \lessgtr \tau_\lambda^{-1}\tanh^{-1}(\Delta\tau_\lambda/\tau_\lambda) }$ indicates where emission ($>$) and extinction ($<$) become dominant. Here, $\Delta\kappa_\lambda$ is the difference of extinction along the major and minor principal axis of a nonspherical dust grain. A similar relation for fluxes instead of optical depth was discussed in \cite{Brauer2016}. However, the inequality assumes that the dust density $\rho_{\mathrm{dust}}$ is constant along the path $\ell$. The corresponding regimes are shown in \Fig{DustModel} for the two considered dust components assuming a constant product of $\rho_{\mathrm{dust}}\times \ell$. We note that this inequality just estimated the most dominant mechanism. It does not imply that other mechanisms may not also contribute to polarization. Furthermore, for any complex model, the inequality may change its sign multiple times along a single line of sight. The exact polarization vector may therefore only be determined by a radiative transfer simulation.

Altogether, polarization vectors already start to rotate for a wavelength of $200\ \mu\mathrm{m}$ for our particular MHD simulation. This is consistent with the synthetic observations of a molecular outflow presented in \cite{Reissl2017}. However, this case study did not include scattering. In this paper we quantified that for wavelengths $<200\ \mu\mathrm{m,}$ scattering increasingly contributes to polarization, while for $>500\ \mu\mathrm{m,}$ the polarization is completely due to emission. Consequently, the polarization pattern remains stable. Hence, the magnetic field structure can best be probed in the far-IR to millimeter regime. Neither the SOFIA/HAWC+ bands nor the lower HERSCHEL bands seem to be suitable for this task.

First and foremost, this result emphasizes the importance of a realistic dust polarization modeling for an accurate prediction of polarimetric observations. 
Especially at low wavelengths, modeling the structure based on a simplified alignment assumption may lead to fundamentally incorrect conclusions about the magnetic field structure. 

\section{Conclusion}
We have investigated the magnetic properties of a bridge-like structure that is associated with the formation of a protostellar triple system in an MHD zoom-in simulation.
To compare our results with observations, we postprocessed the simulation data with the radiative transfer code \polaris\ to produce synthetic dust polarization maps. 
We determined the density of the bridge to a typical value of $\rho_{\mathrm{gas}} \sim10^{-16} \unit{g}\unit{cm}^{-3}$ , corresponding to a number density $n_{\rm br}$ in the range of 2 to $3 \times 10^7 \unit{cm}^{-3}$. 
This density agrees well with estimates for the density in observed bridge-like structures, such as the protostellar multiple system IRAS 16293--2422 \citep{vanderWiel2019}.
We find that the bridge in our model is only weakly magnetized with 1 to 2 mG, while in the vicinity of the forming protostars, field strengths are rising to $\sim$100 mG, which is consistent with flux-freezing.
The magnetic field in the bridge has an elongated toroidal morphology that
is nonuniform because the sources are embedded in a turbulent environment of the GMC.
The results indicate that the bridge in IRAS 16293--2422 is more magnetized than the bridge in our model. 
Therefore the polarization pattern is smoother in IRAS 16293--2422 because the magnetic field lines provide more resistance to perturbations than the more weakly magnetized case of our model.

When we consider that the bridge structure in our model has a similar density to the structure in IRAS 16293--2422, but most likely a different magnetization, this emphasizes that transient bridge structures are primarily a result of larger-scale colliding flows that are caused by turbulence in GMCs. 
Apparently, the magnetic field is dynamically unimportant for the formation of these structures. 
However, magnetic fields play an important role on smaller scales because they are responsible for the magnetic braking of disks and for the launching of bipolar outflows. 
Although we only marginally resolved the outflows in the model, our results show that asymmetric bipolar outflows such as are observed for Serpens SMM1-a and b, or OMC-3 MMS 6, are launched during embedded star formation in the more complex environments of GMCs.    

For observations at larger wavelength ($\gtrsim$200 $\mu$m, i.e., SMA and ALMA bands), we find that assuming perfect alignment is a good approximation for estimating the orientation of the magnetic field on scales of $\sim$100 to $\sim$1000 AU from the protostars. 
However, the synthetic observations demonstrate that compared to the scenarios that account for RATs, assuming only perfect alignment leads to an overestimate of the polarization fraction of a factor of 2 to 3.
Especially accounting for RATs in the denser parts, that is, inside the bridge, is crucial for estimating the polarization fraction appropriately. 
Moreover, considering perfect alignment violates the relation in the PI diagram, whereas the slope of the PI relation is consistent with observations for the scenario with RATs. 
We also tested the effect of self-scattering and found that its contribution to the polarization of the dust grains in the bridge is $\approx 10 \%$, which is minor but non-negligible.

At smaller wavelength ($\lesssim$100 $\mu$m, i.e., short-wavelength bands of SOFIA/HAWC+ as well as the lower HERSCHEL bands), scattering and dichroic extinction have to be considered when constraints of the underlying magnetic field are derived. 
The synthetic observations show that the smaller wavelengths predominantly trace the alignment of dust grains that is induced by scattering and dichroic extinction in the bridge. 
Scattering and dichroic extinction contribute more substantially to the alignment at shorter wavelength because the optical depth is about unity or higher for wavelengths $\lesssim 200 \mu$m, while it is optically thin for (sub-)millimeter wavelengths in the denser regions. 

In general, these results show the difficulties and possible confusion in interpreting the results from dust polarization measurements. However, the results of this study also demonstrate the prospects of multiwavelength dust polarization because the different wavelengths trace different physical processes. For the magnetic field structure, our results show that observations of dust polarization at $\sim1$ mm wavelength are good tracers in star-forming regions on scales beyond $\sim100$ AU from the protostars.

\begin{acknowledgements}
We thank the anonymous referee for constructive comments and suggestions to an earlier version of the manuscript.
We thank Lars E. Kristensen for his comments and suggestions that helped to improve the manuscript.
Also, we thank Daniel Seifried for spotting and informing us about transposed digits at one occasion in an earlier version of the manuscript. 
MK thanks Troels Haugb{\o}lle and {\AA}ke Nordlund for their development of the zoom-in technique for the modified version of \ramses. 
MK thanks the developers of the python-based analyzing tool YT http://yt-project.org/ \citep{yt-reference}, which simplified the analysis significantly. 
The research of MK is supported by a research grant of the Independent Research Foundation Denmark (IRFD) (international postdoctoral fellow, project number: 8028-00025B). We acknowledge PRACE for awarding access to the computing resource CURIE based in France at CEA for carrying out part of the simulations. Thanks to a research grant from Villum Fonden (VKR023406), MK could use archival storage and computing nodes at the University of Copenhagen HPC centre to carry out essential parts of the simulations and the post-processing.
MK acknowledges the support of the DFG Research Unit `Transition Disks' (FOR 2634/1, DU 414/23-1).  S.R. acknowledges funding from the Deutsche Forschungsgemeinschaft (DFG, German Research Foundation) -- Project-ID 138713538 -- SFB 881 ``The Milky Way System'' (sub-projects B1, B2, and B8)), from the Priority Program SPP 1573 ``Physics of the Interstellar  Medium''  (grant  numbers  KL  1358/18.1,  KL  1358/19.2), and acknowledges also support from the DFG via the Heidelberg Cluster of Excellence {\em STRUCTURES} in the framework of Germany’s Excellence Strategy (grant EXC-2181/1 - 390900948). SW acknowledges funding by the DFG (project number: W0857/18-1).  
\end{acknowledgements}

%-------------------------------------------------------------------

\bibliography{general}
\bibliographystyle{aa}

\appendix
\section{Wavelength dependence of polarization measurements}
\label{app:A} 
\begin{figure*}
\begin{center}
     \includegraphics[width=0.195\textwidth]{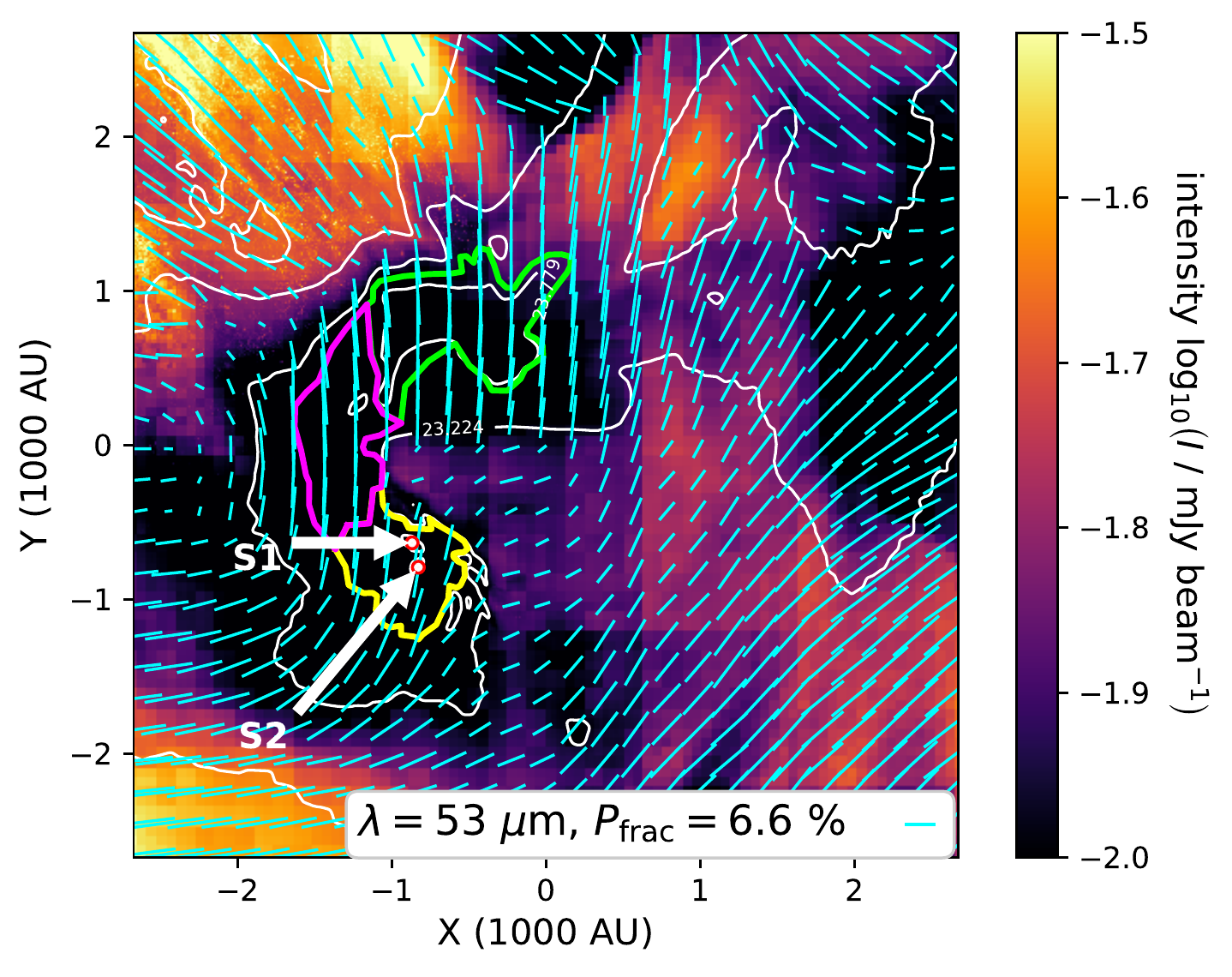}
     \includegraphics[width=0.195\textwidth]{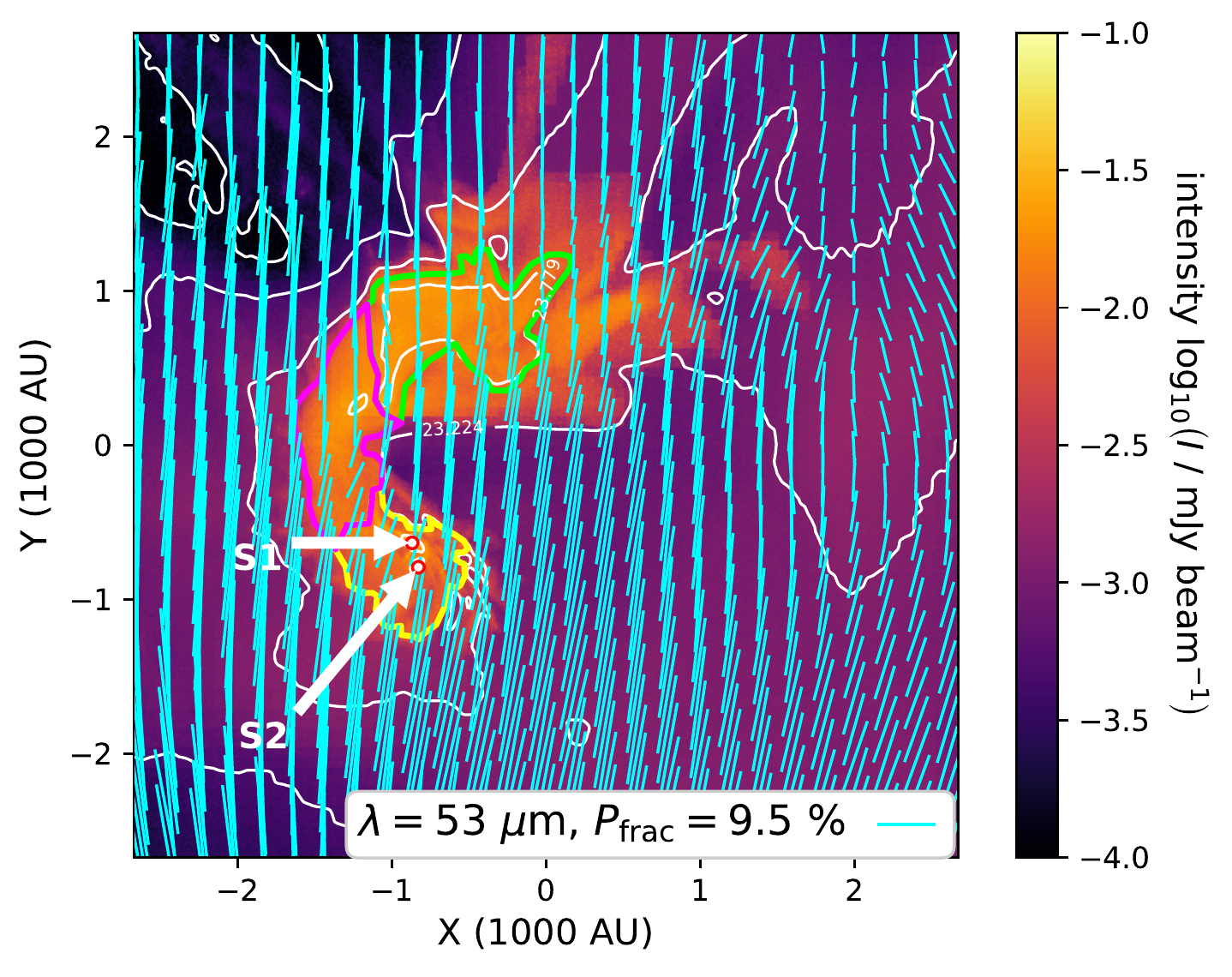}
     \includegraphics[width=0.195\textwidth]{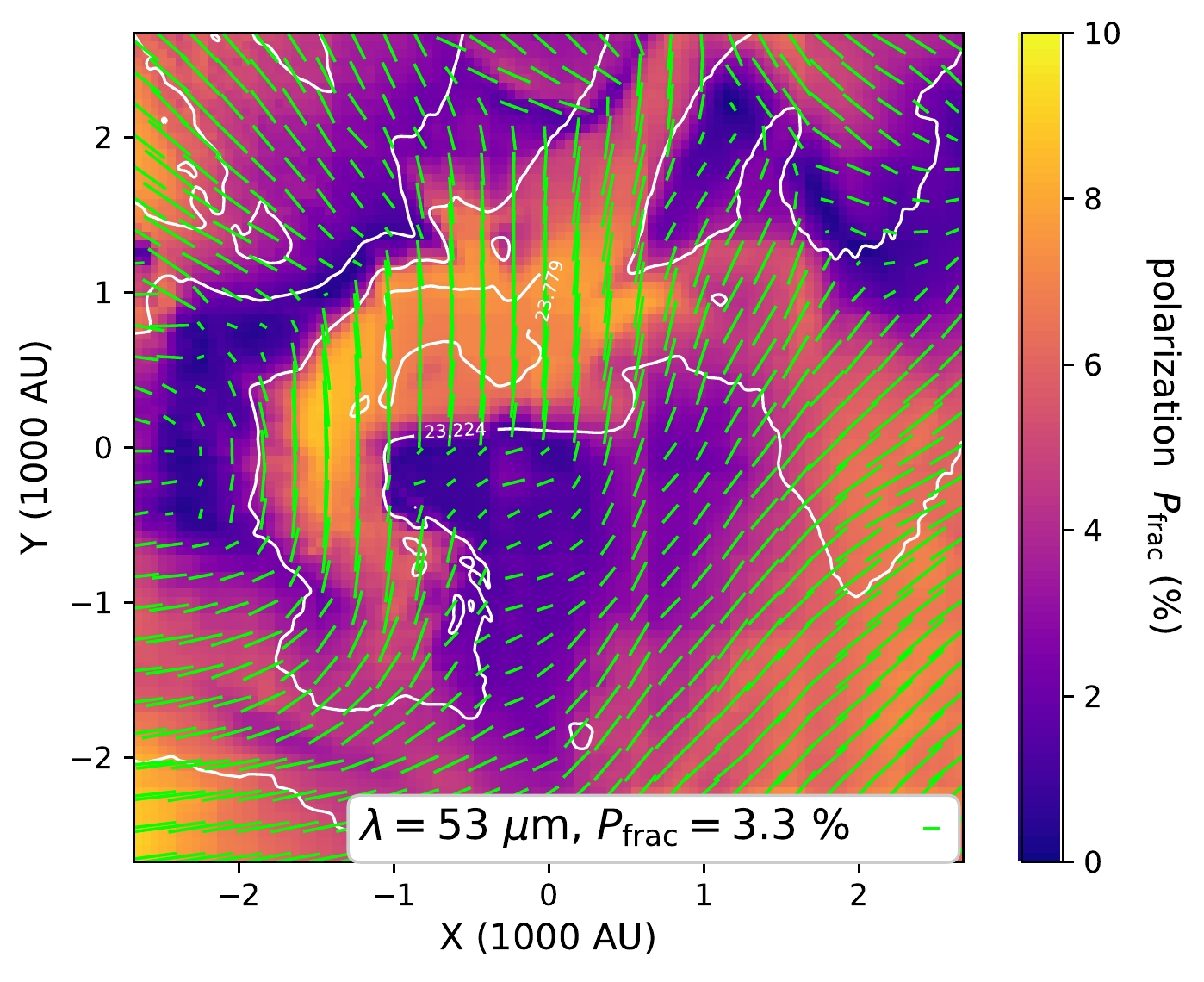}
     \includegraphics[width=0.195\textwidth]{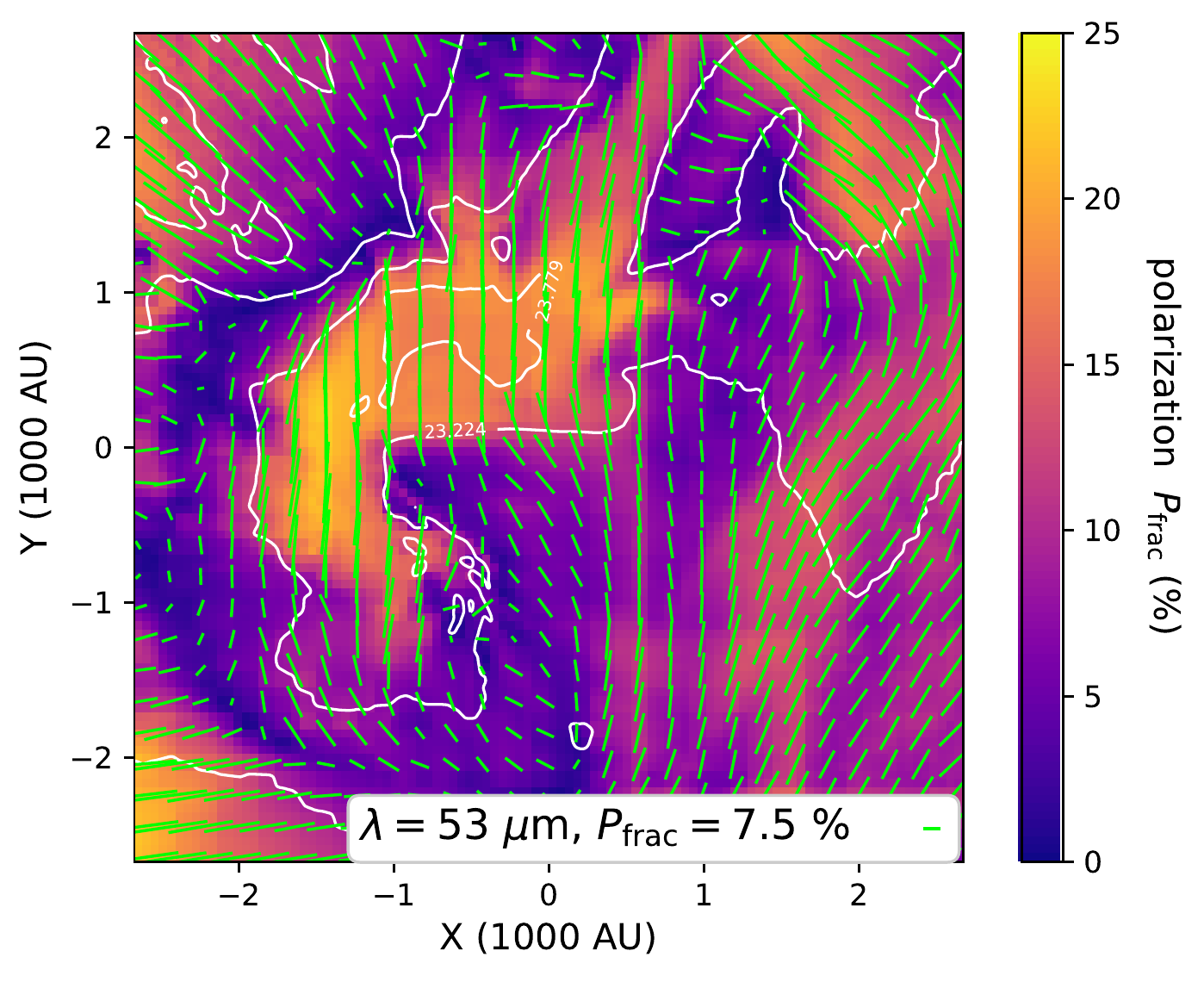}
     \includegraphics[width=0.195\textwidth]{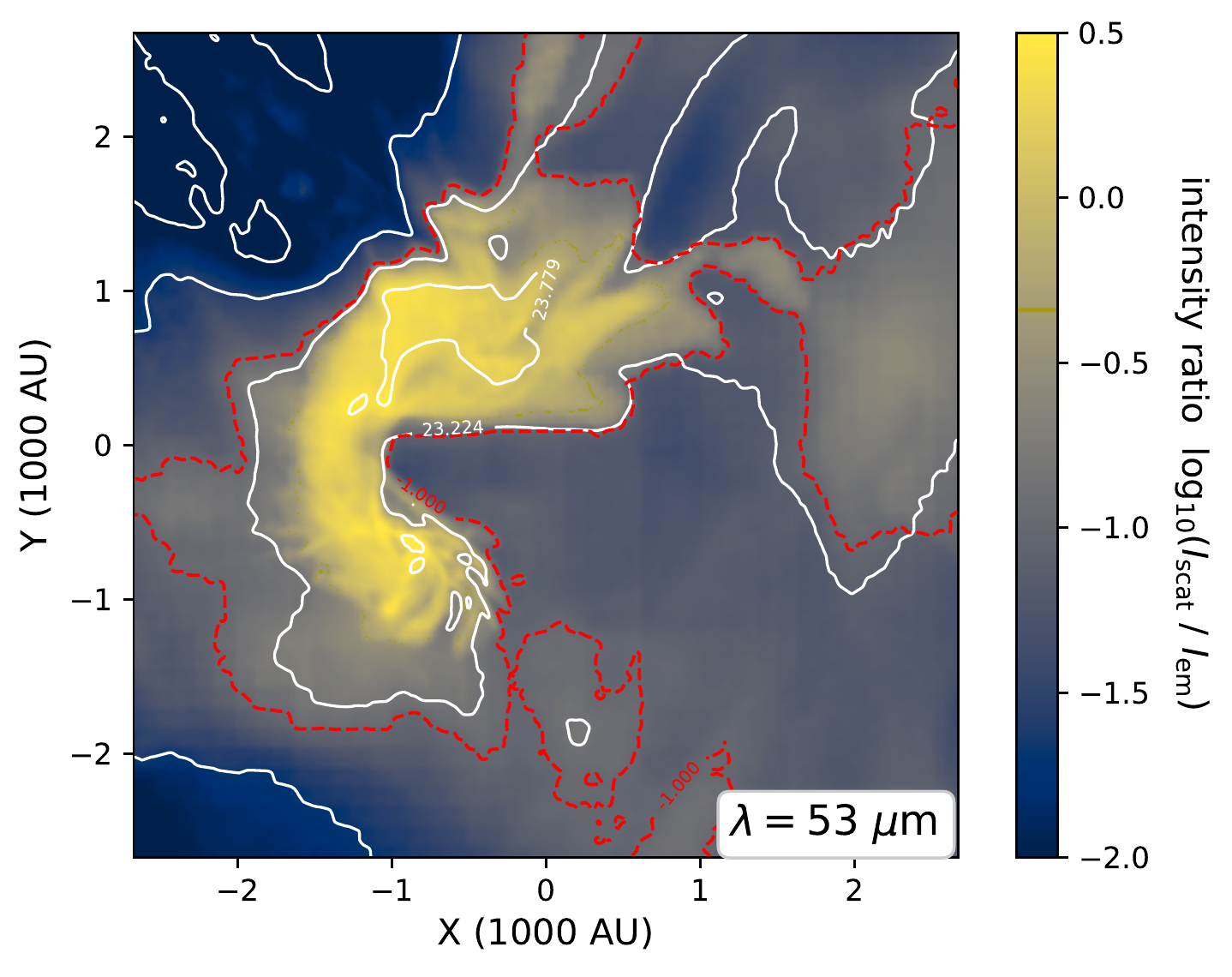}
\end{center}

\begin{center}
     \includegraphics[width=0.195\textwidth]{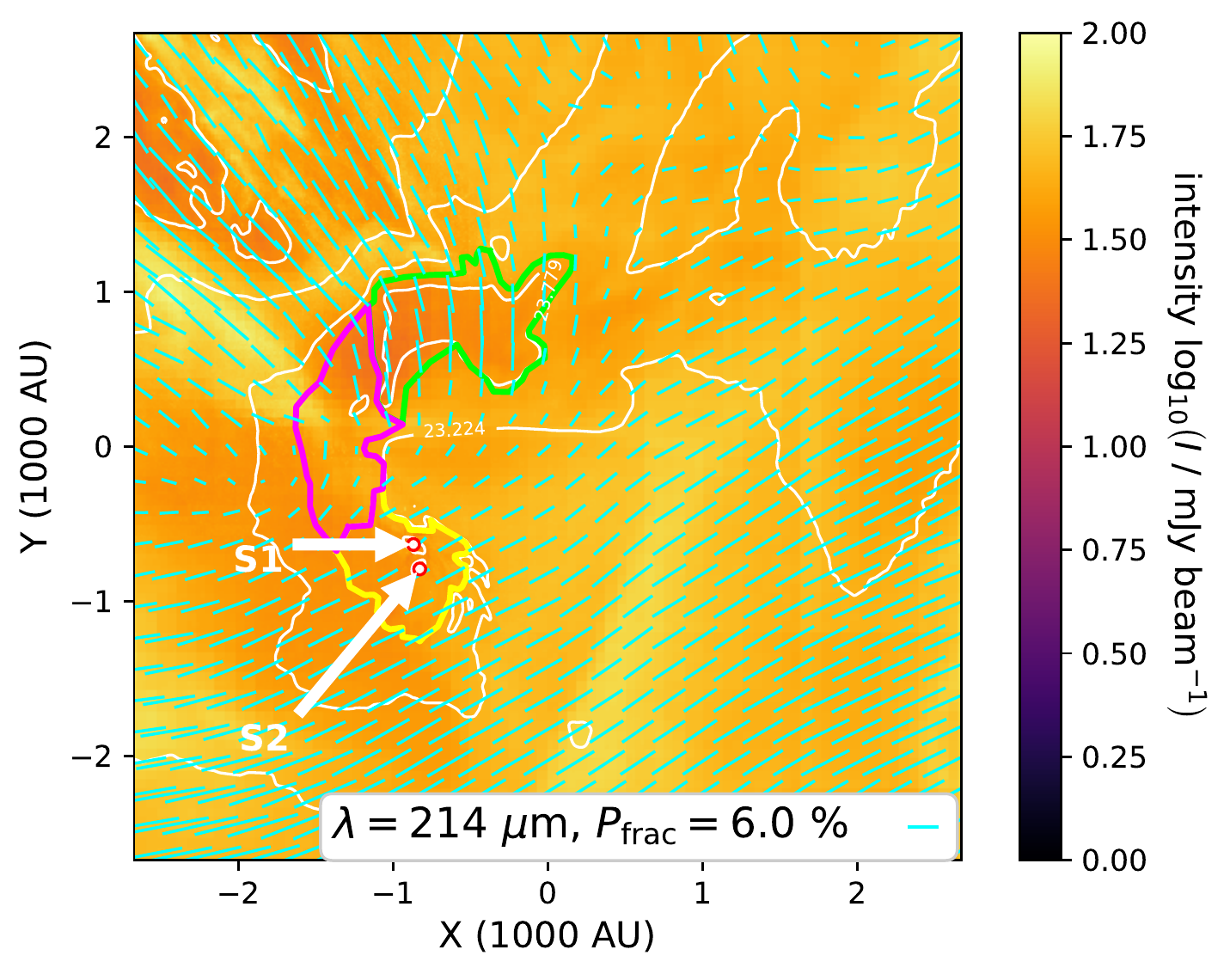}
     \includegraphics[width=0.195\textwidth]{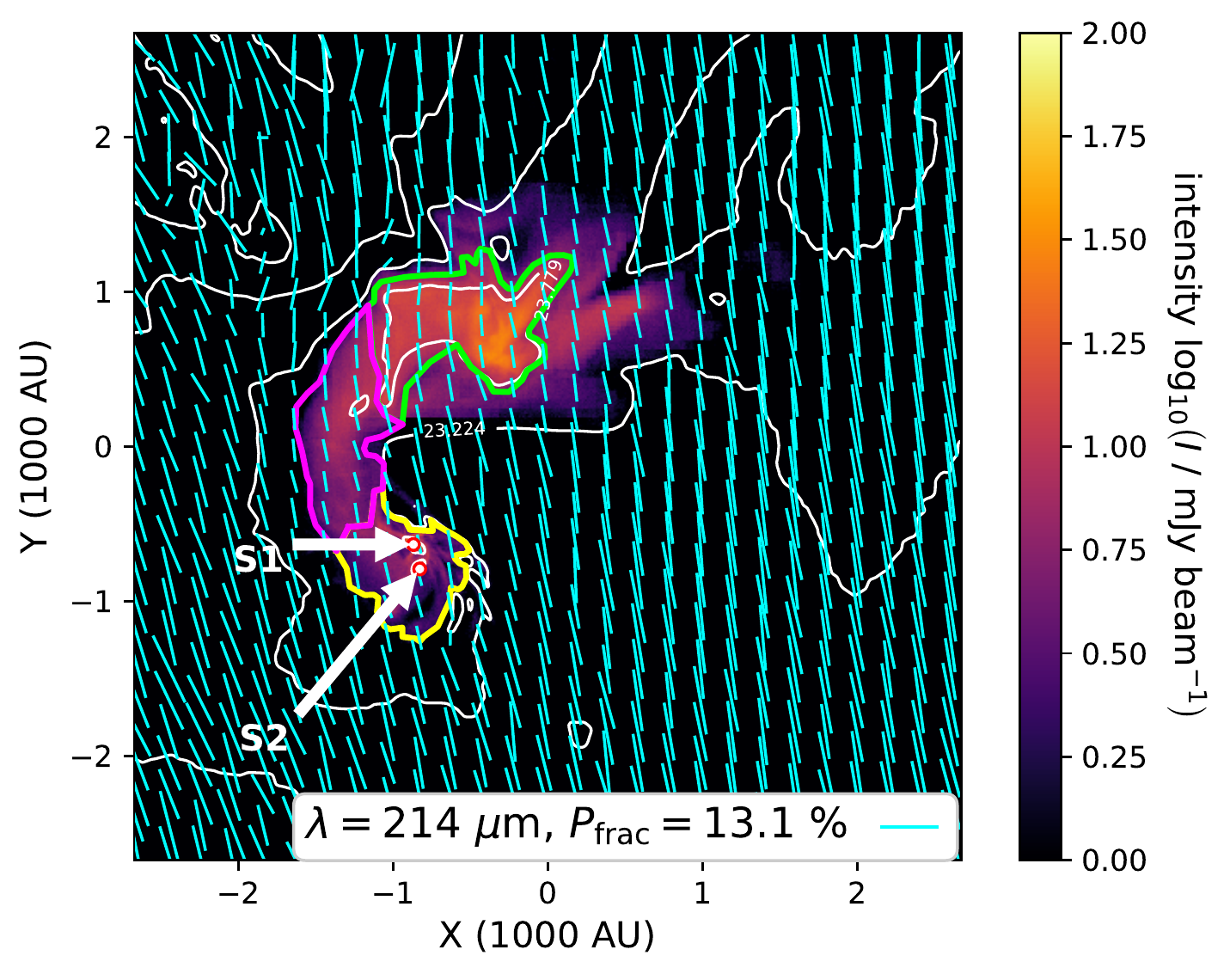}
     \includegraphics[width=0.195\textwidth]{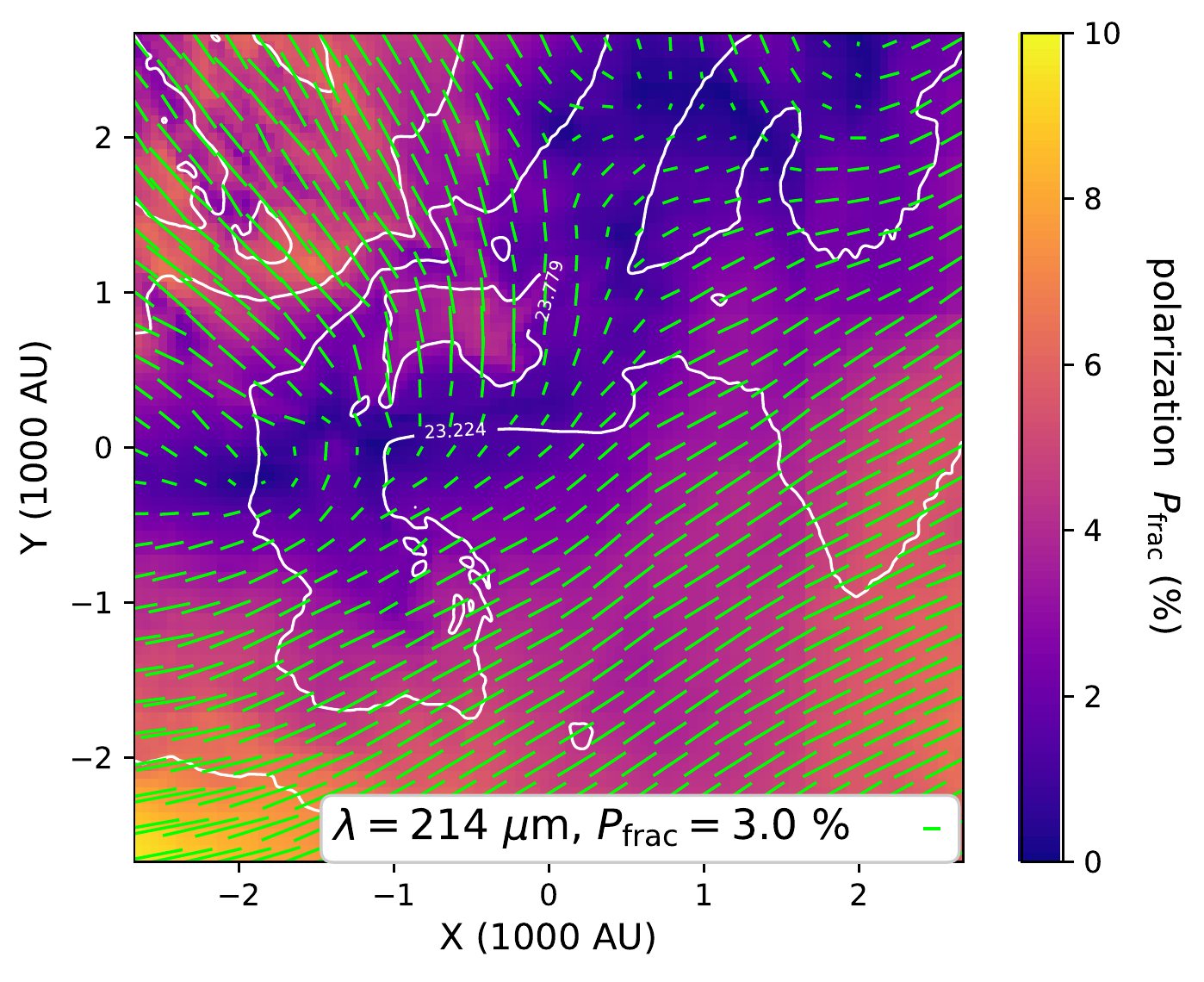}
     \includegraphics[width=0.195\textwidth]{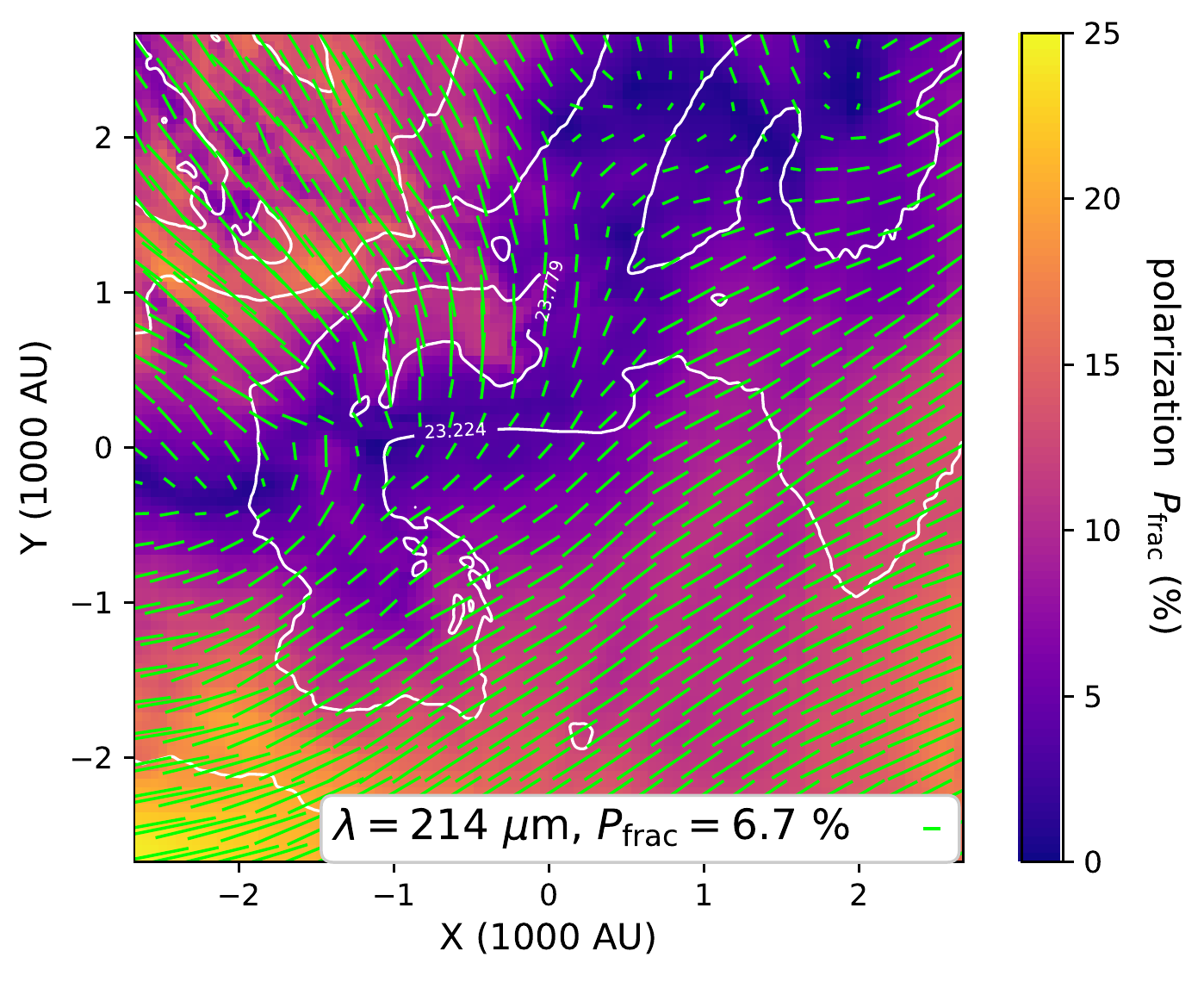}
     \includegraphics[width=0.195\textwidth]{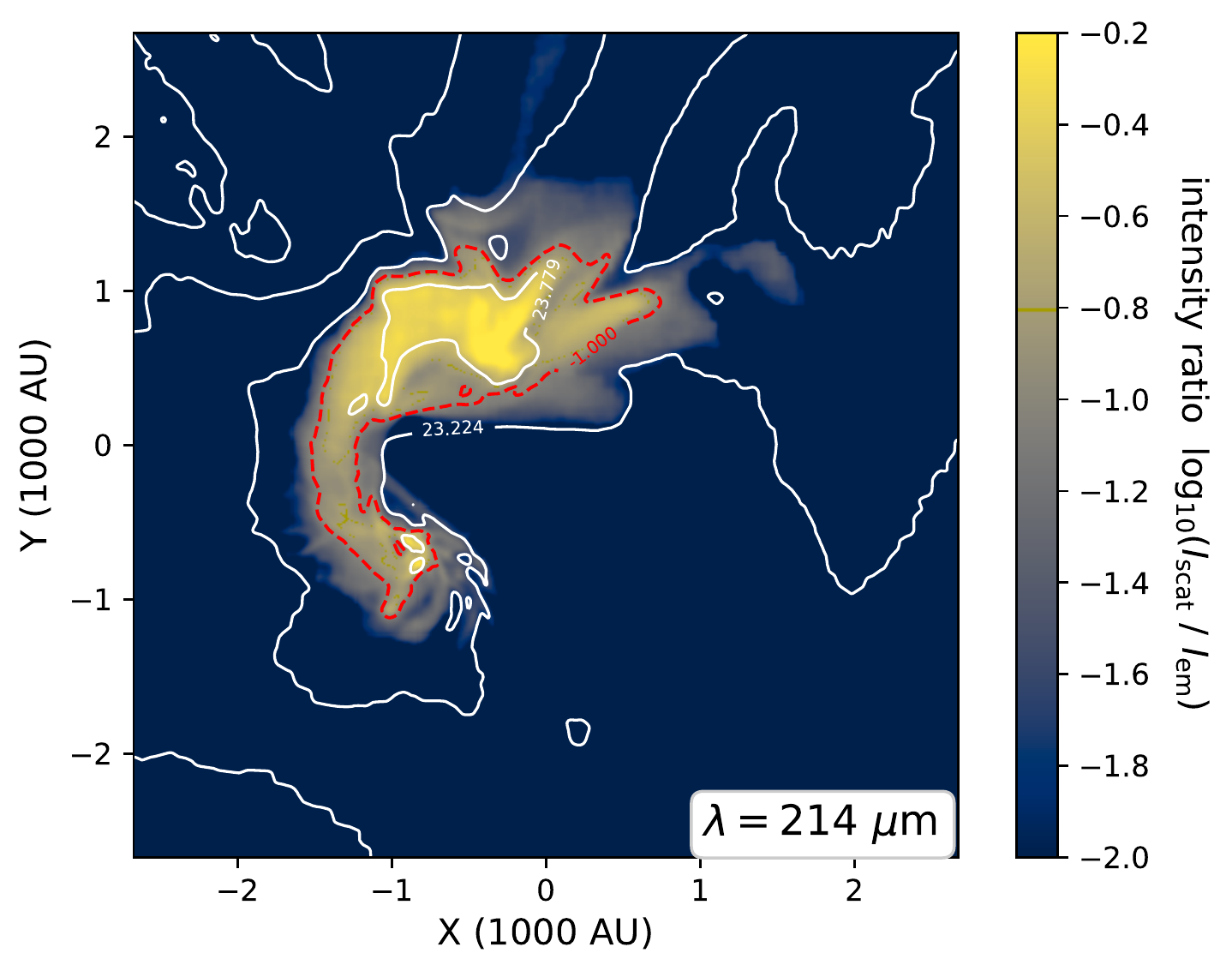}
\end{center}

\begin{center}
    \includegraphics[width=0.195\textwidth]{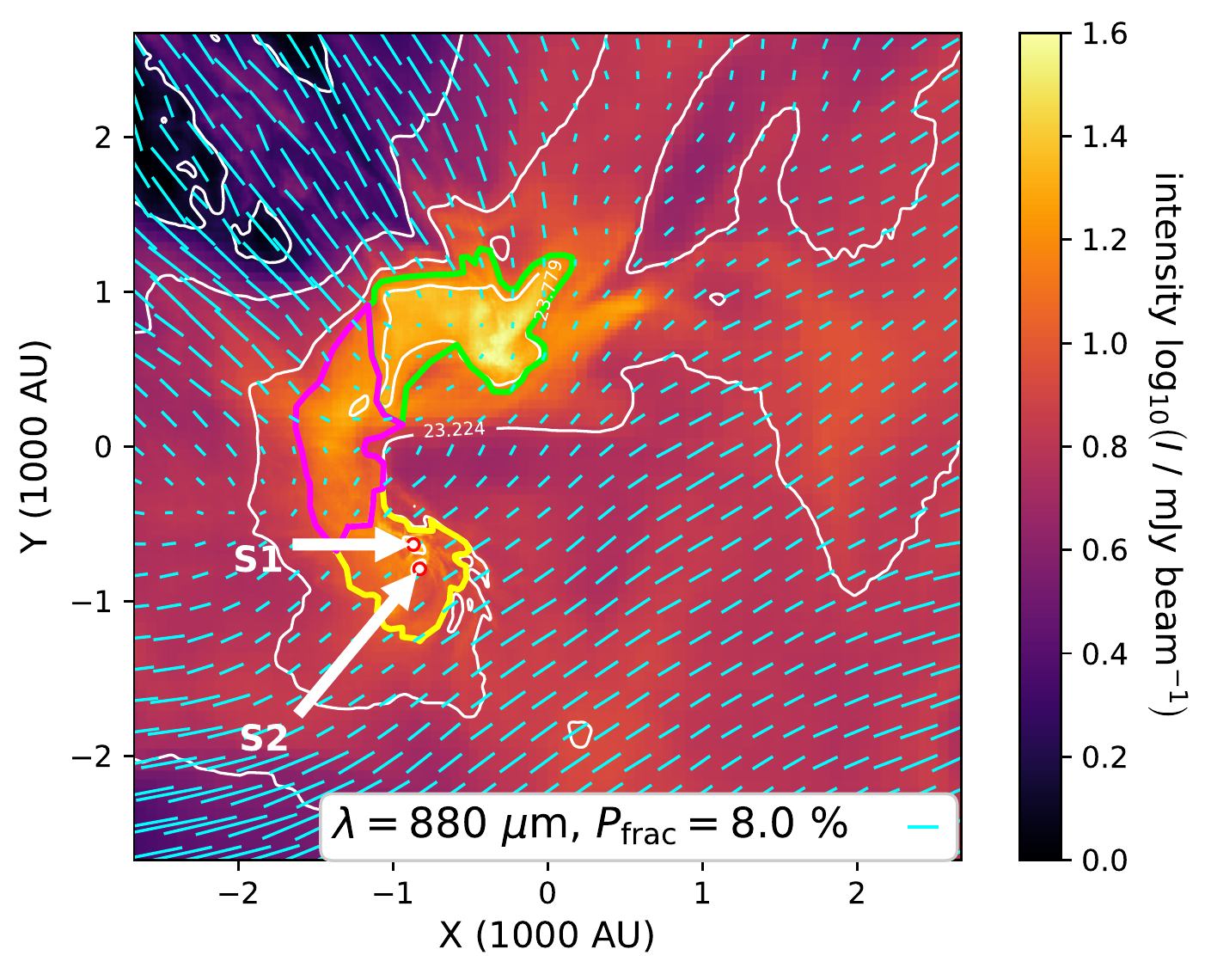}
    \includegraphics[width=0.195\textwidth]{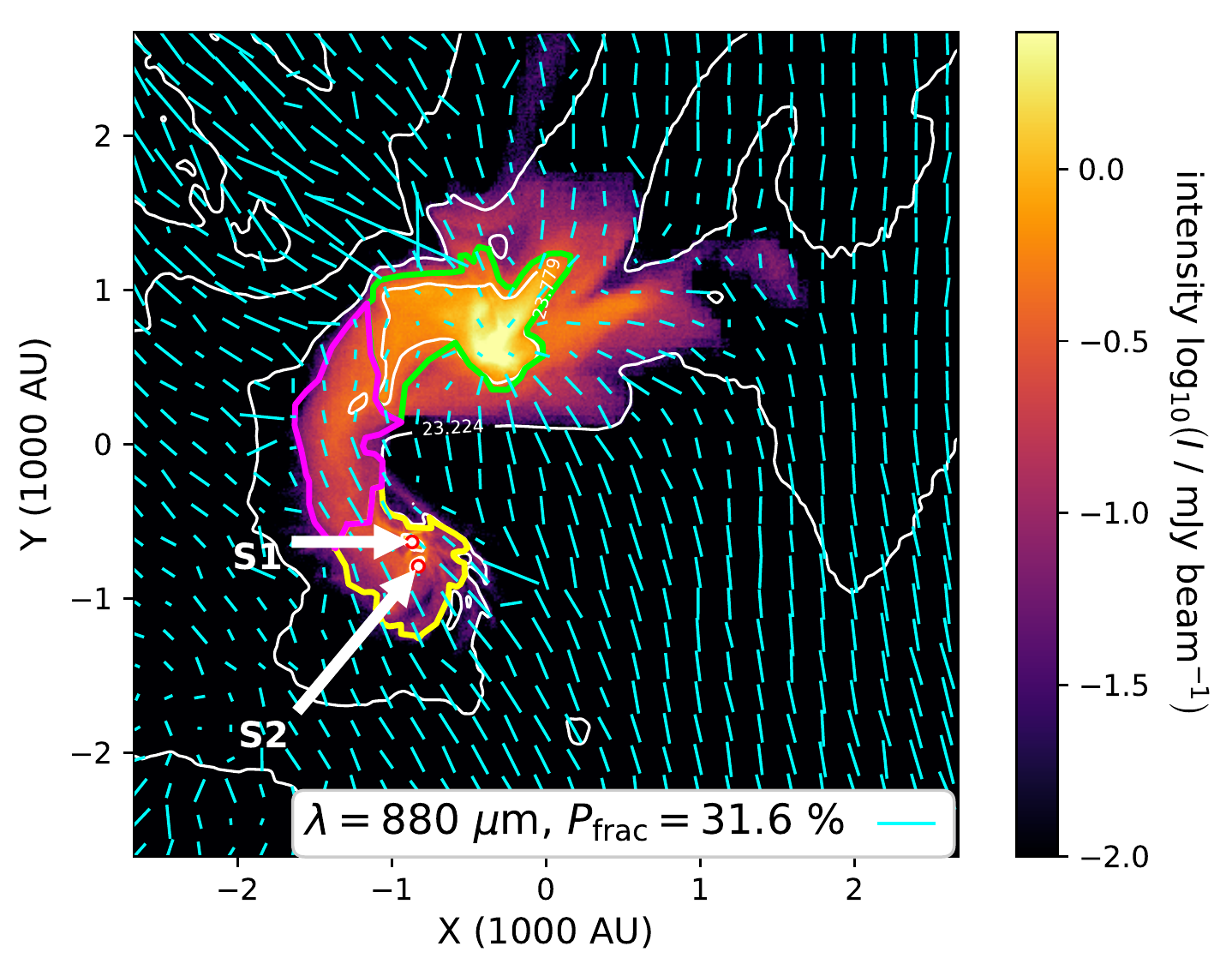}
    \includegraphics[width=0.195\textwidth]{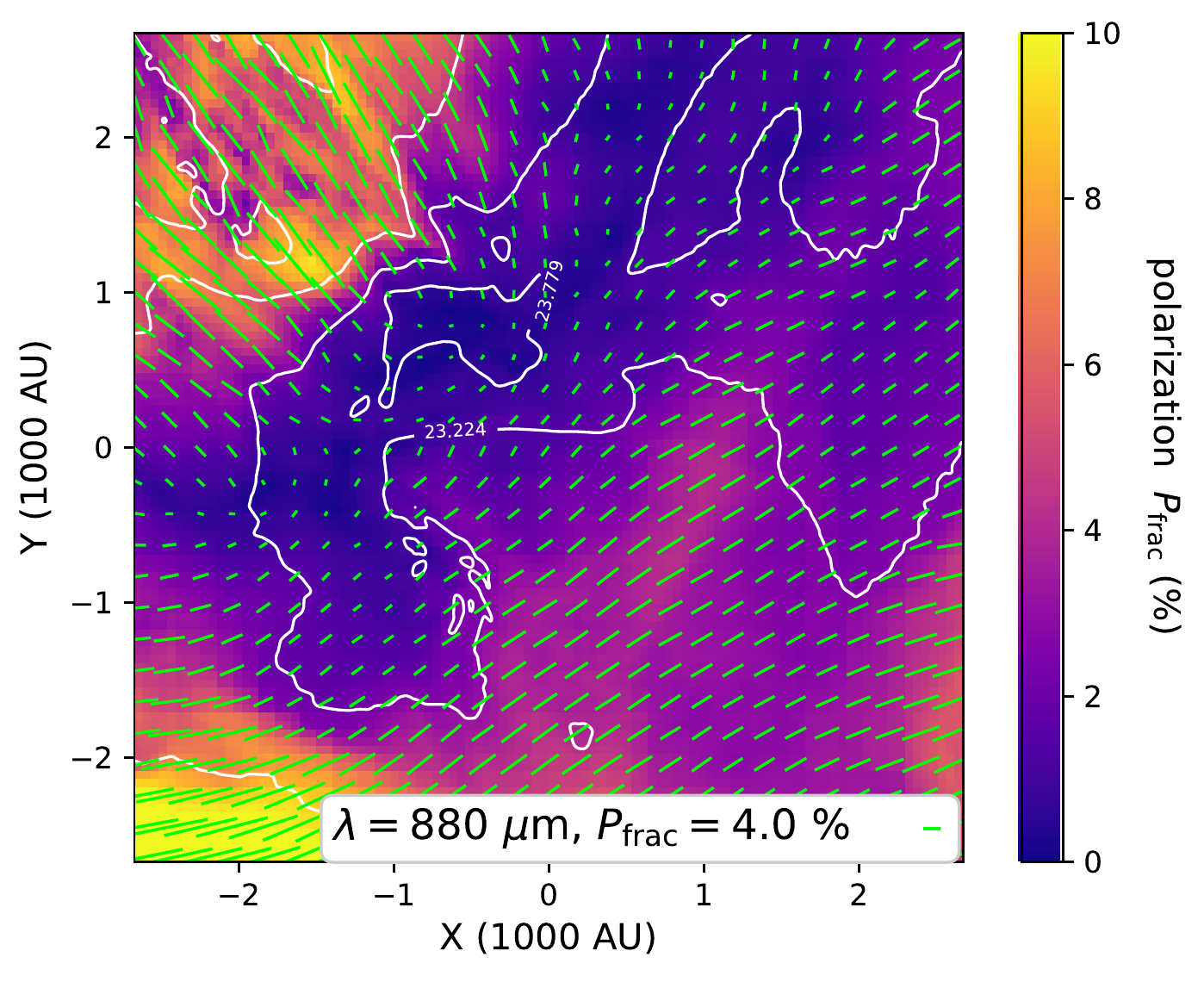}
    \includegraphics[width=0.195\textwidth]{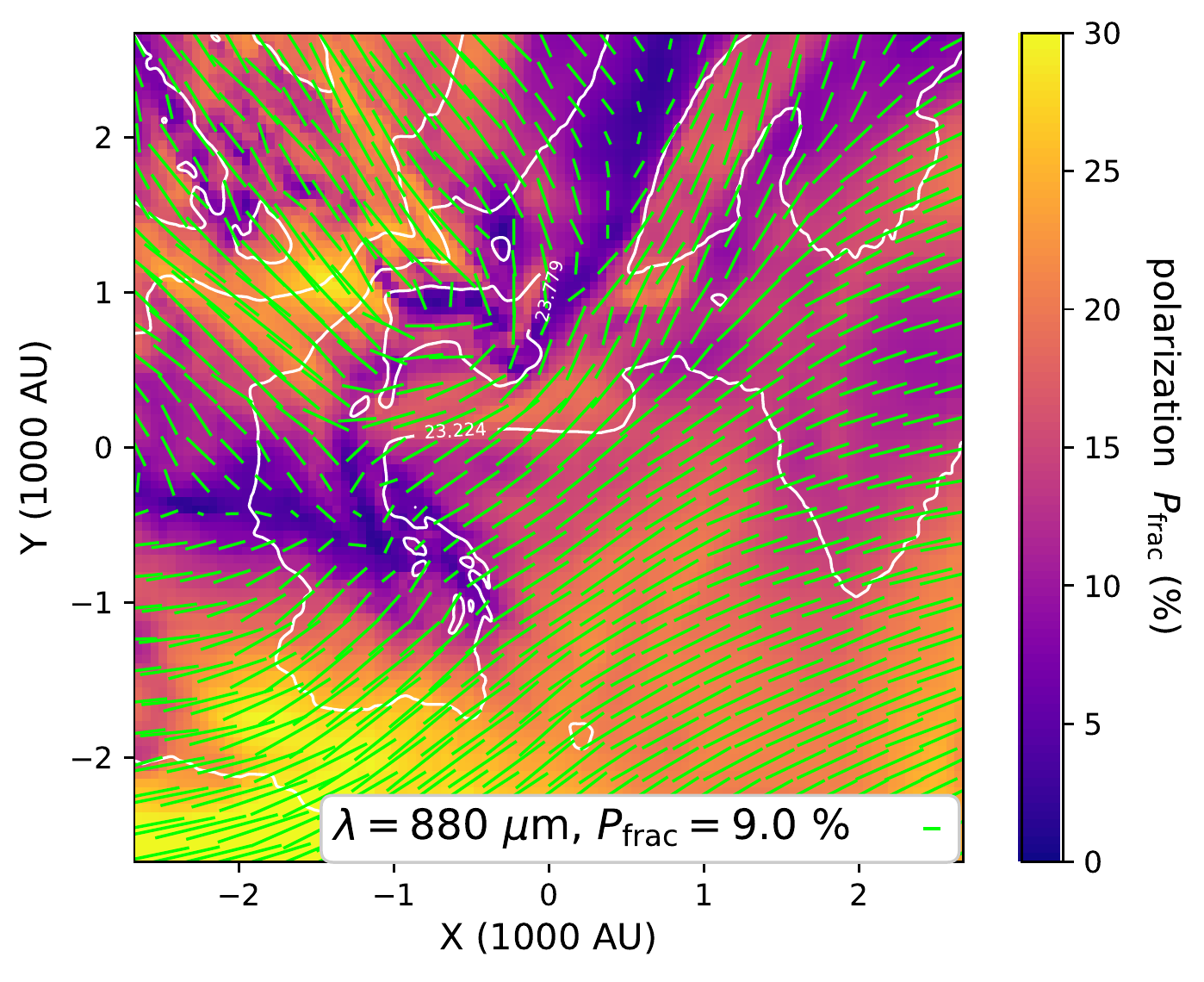}
    \includegraphics[width=0.195\textwidth]{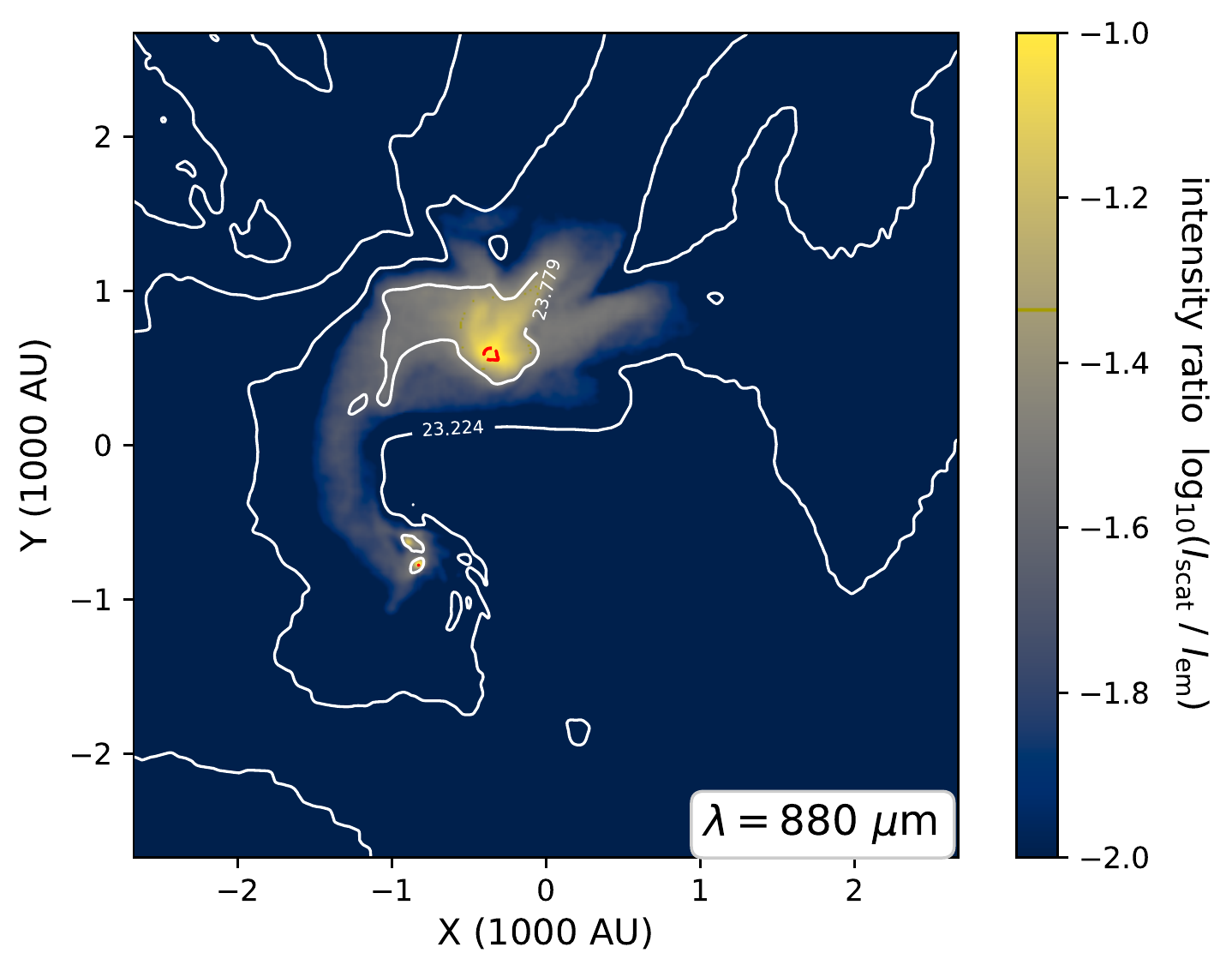}
\end{center}

\caption{Emitted radiation (outer left), scattered radiation (center left), dust polarization assuming RATs (center), polarization assuming perfect grain alignment  (center right), and the ratio of scattered to emitted radiation (outer right) observed at a wavelength of $53\ \mu\mathrm{m}$ (top row), $214\ \mu\mathrm{m}$ (bottom row), and $880\ \mu\mathrm{m}$ (bottom row), respectively. Here, the maximum radius is $a_{\mathrm{max}}=100\ \mu\mathrm{m}$ for the dense component.}
\label{fig:AppSyntheticDust100}
\end{figure*}

\begin{figure*}
\begin{center}
     \includegraphics[width=0.195\textwidth]{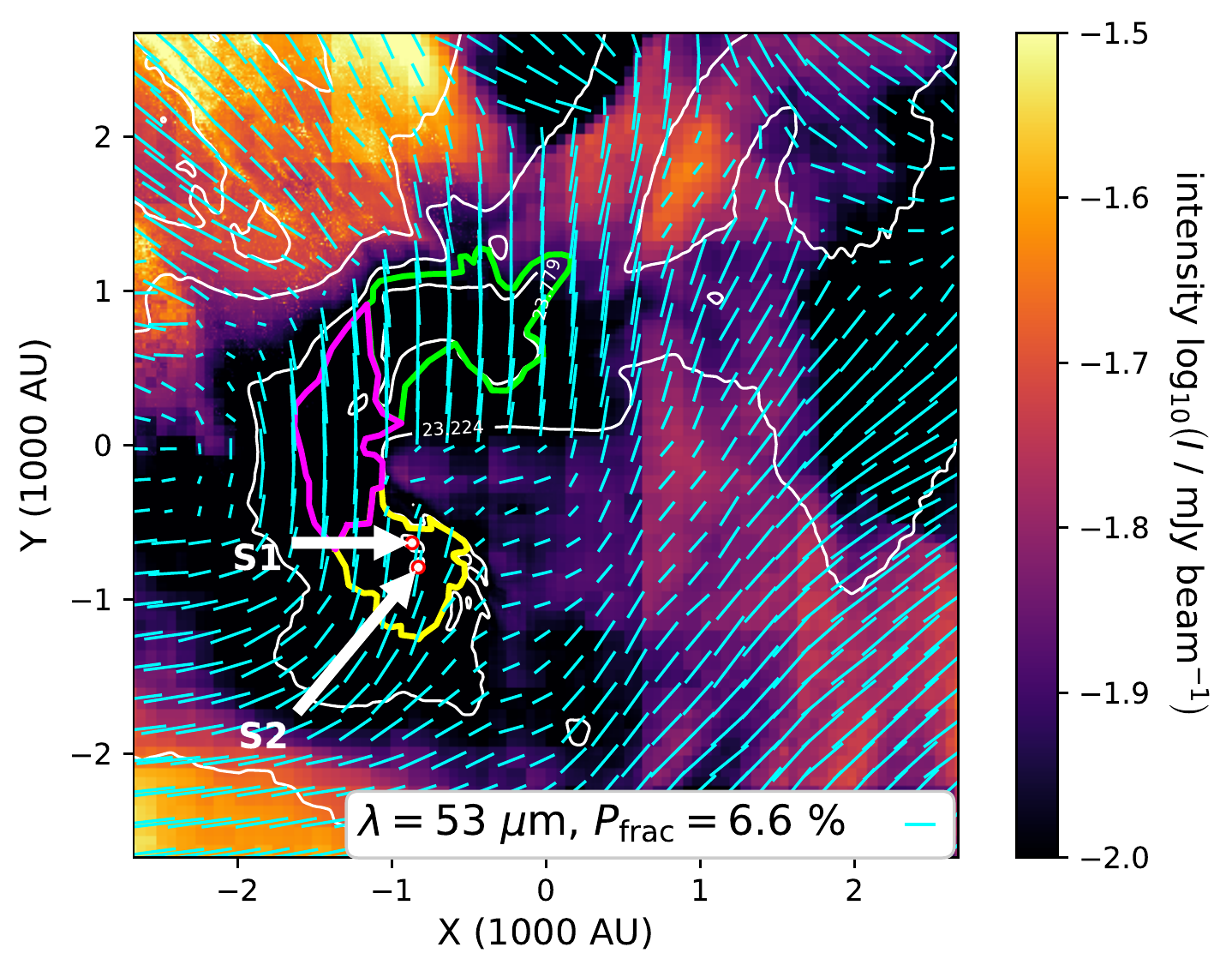}
     \includegraphics[width=0.195\textwidth]{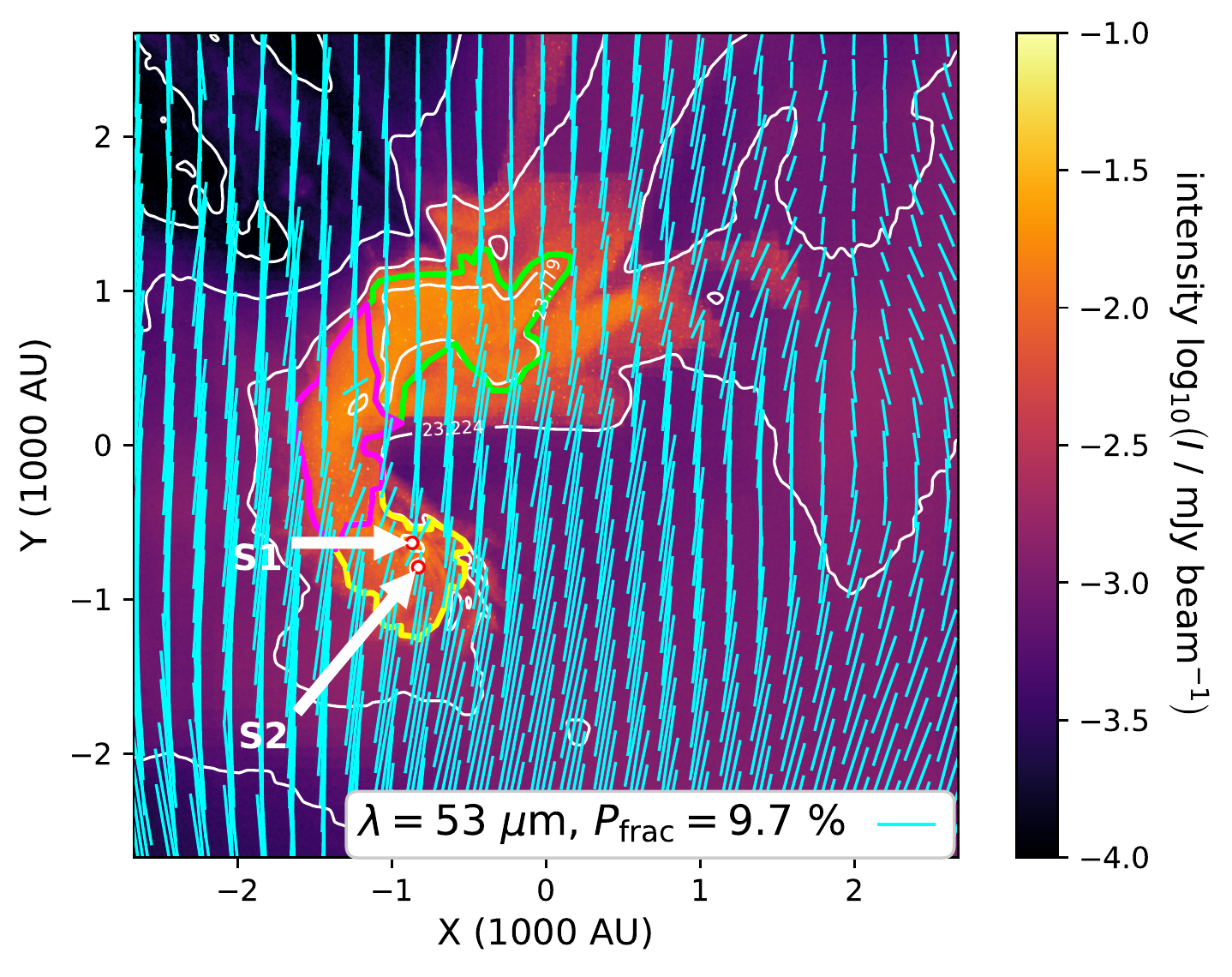}
     \includegraphics[width=0.195\textwidth]{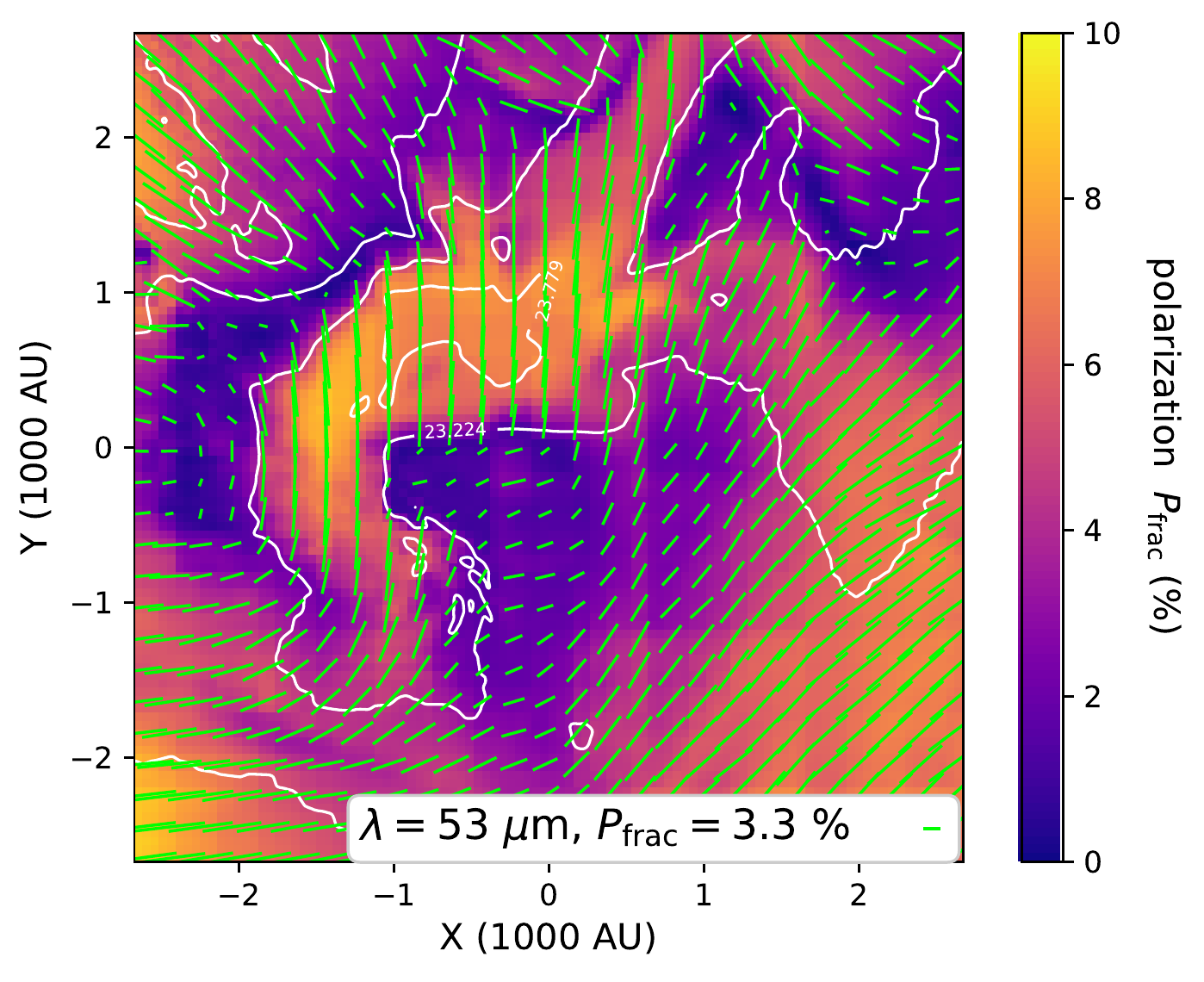}
     \includegraphics[width=0.195\textwidth]{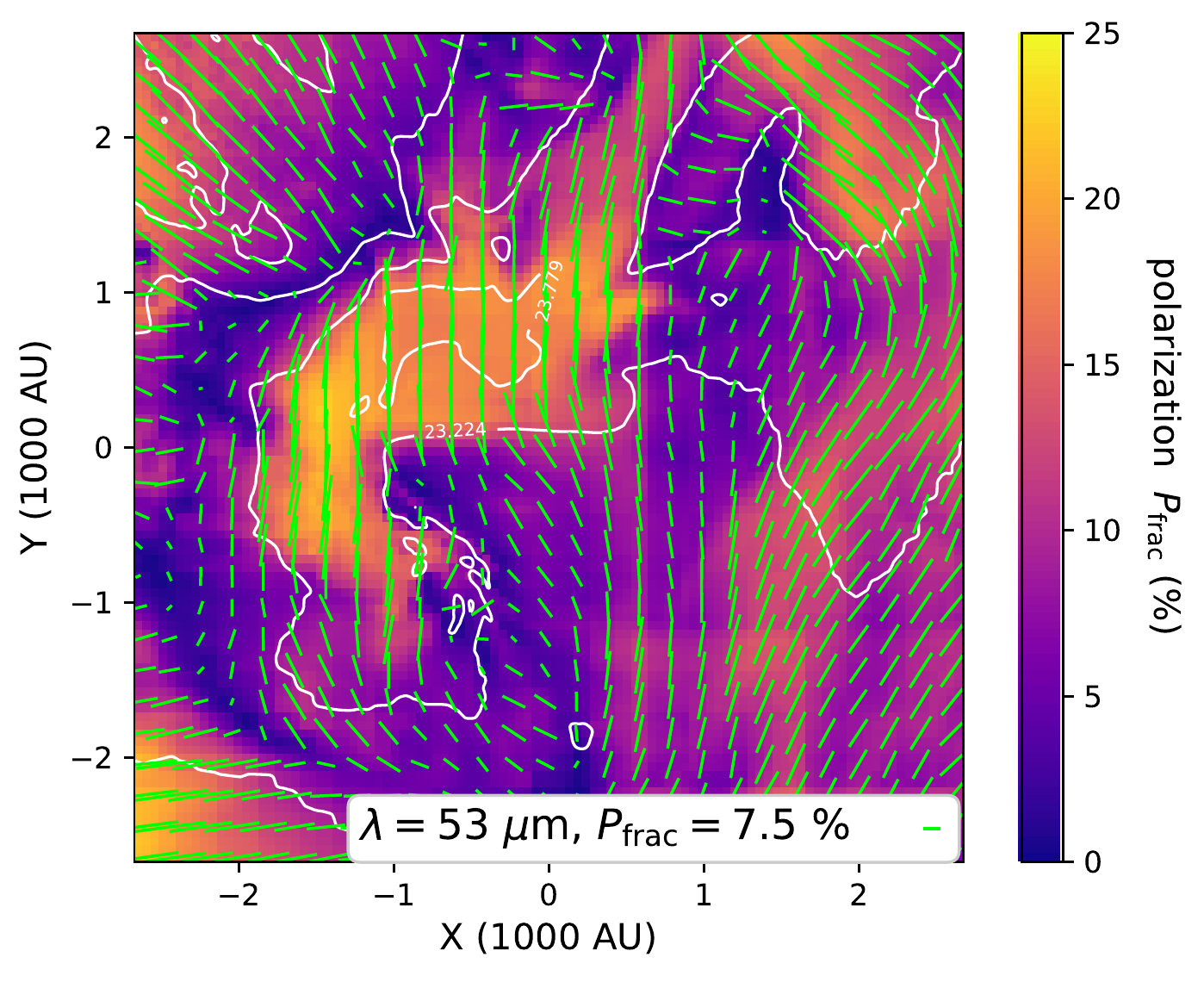}
     \includegraphics[width=0.195\textwidth]{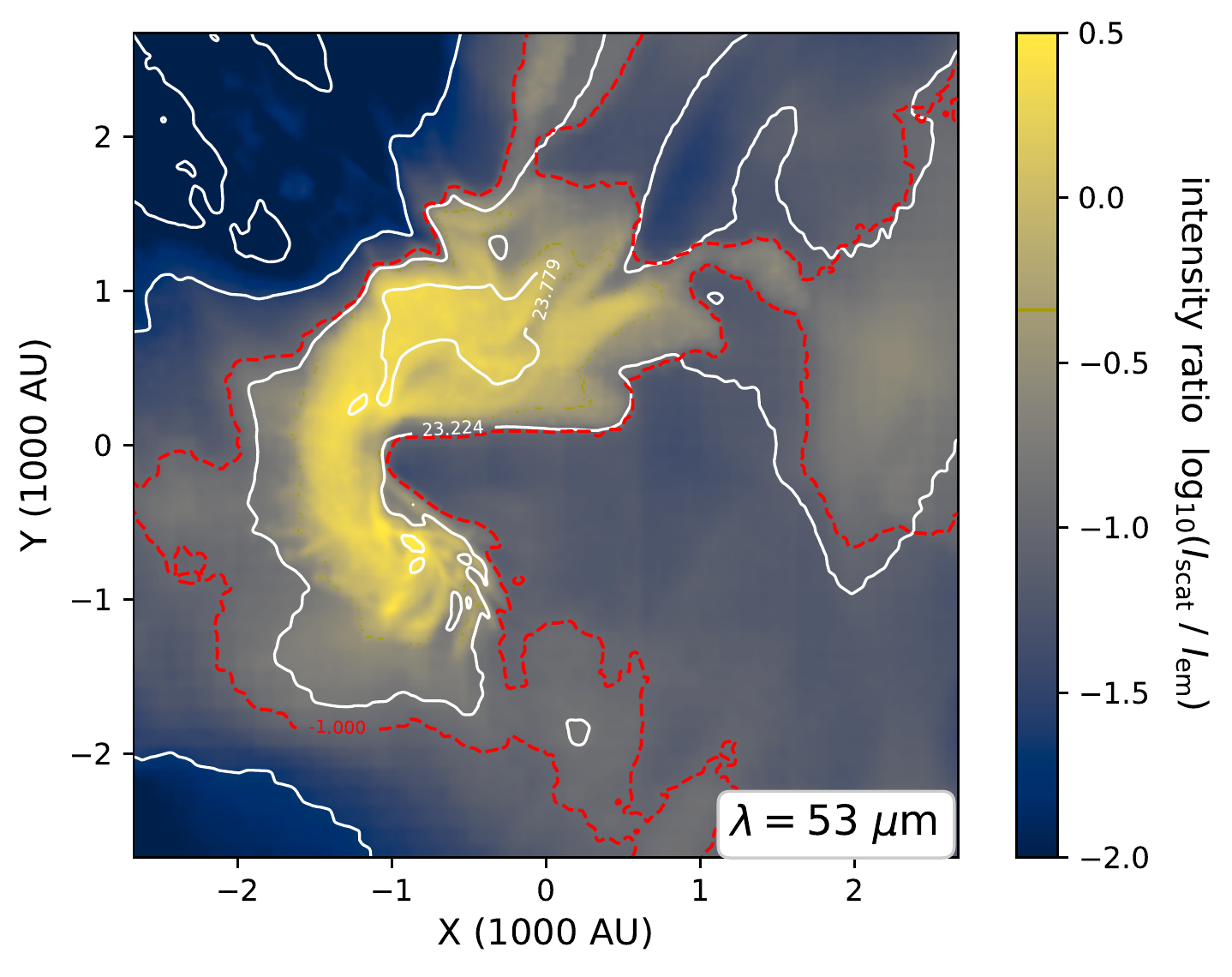}
\end{center}

\begin{center}
     \includegraphics[width=0.195\textwidth]{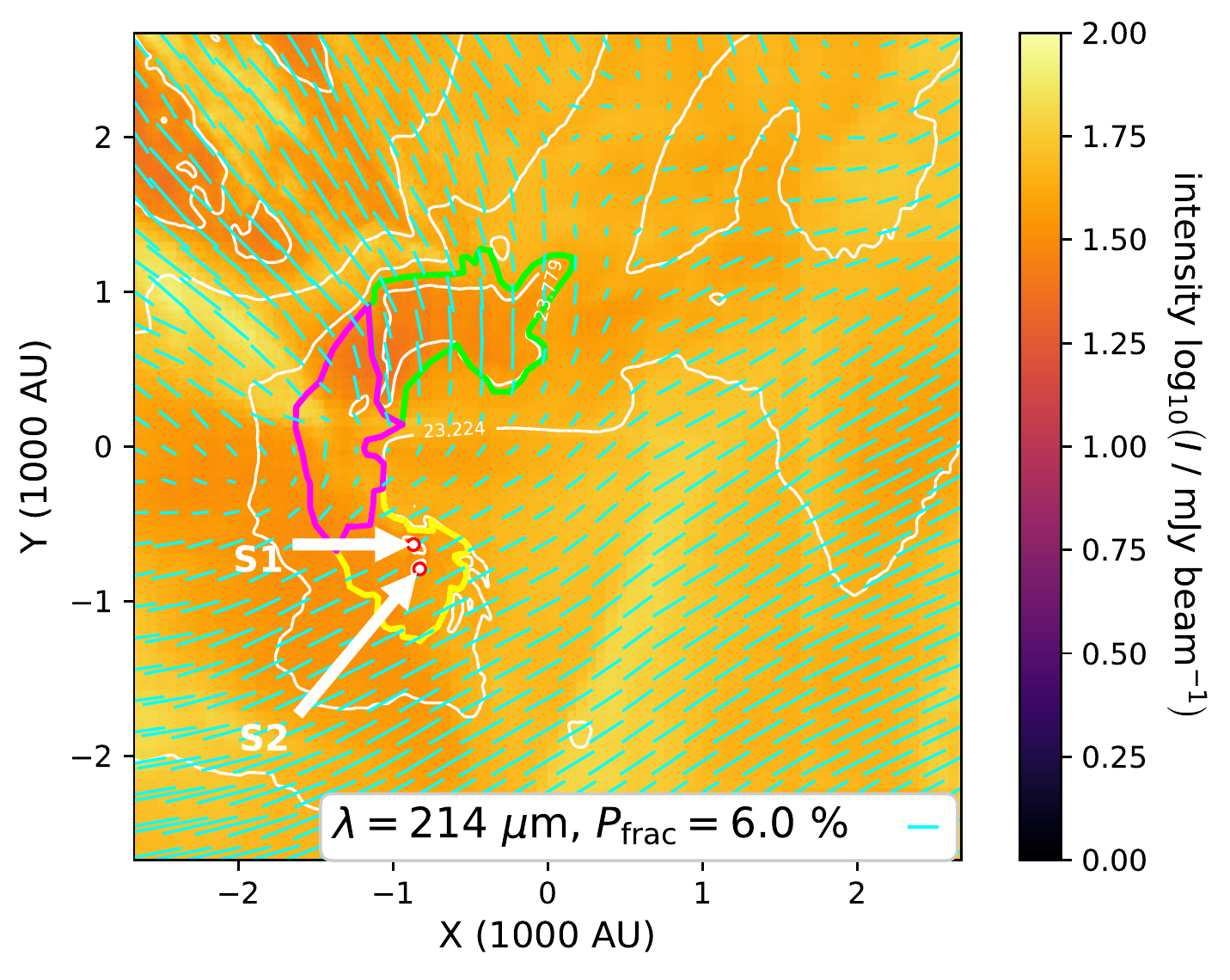}
     \includegraphics[width=0.195\textwidth]{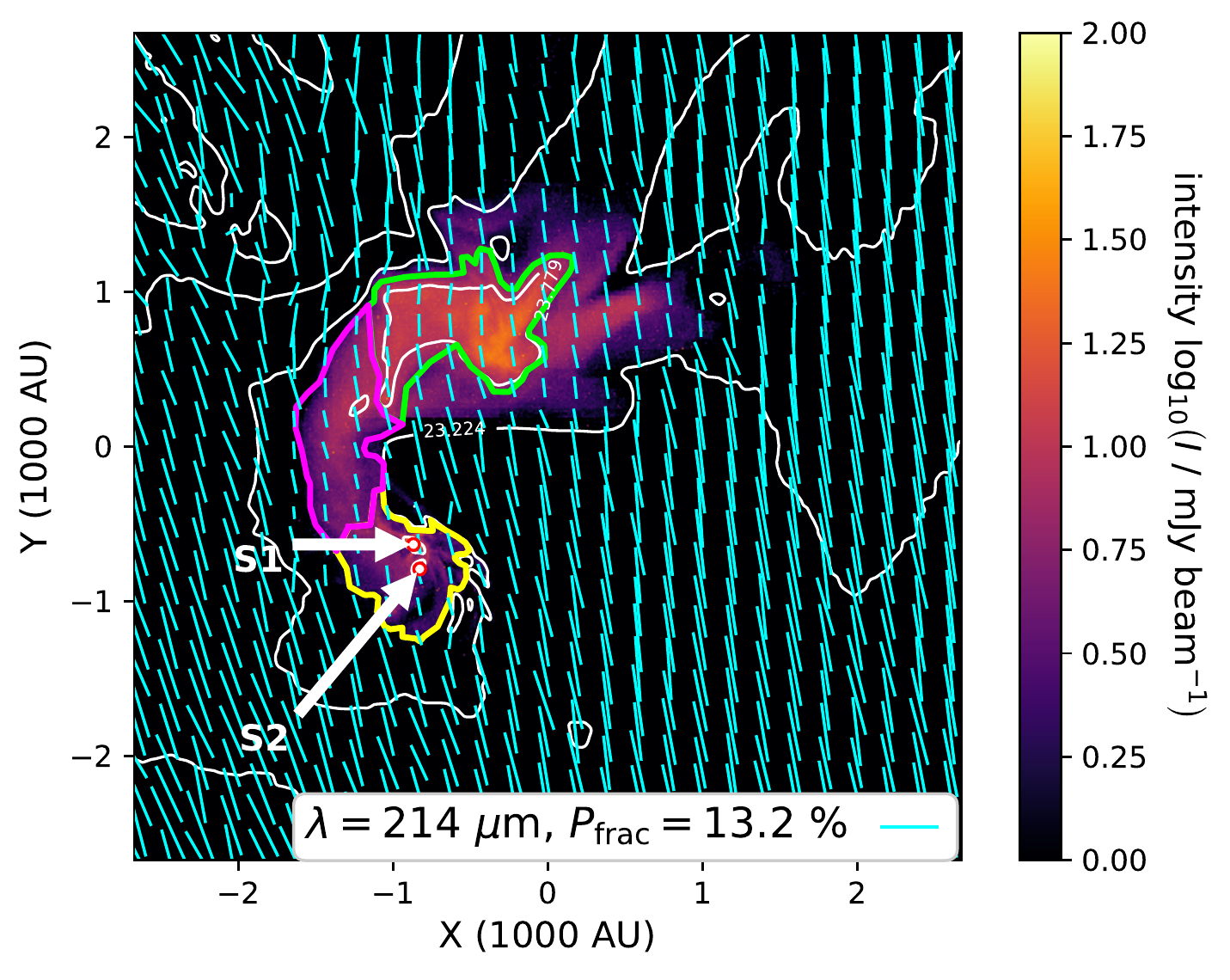}
     \includegraphics[width=0.195\textwidth]{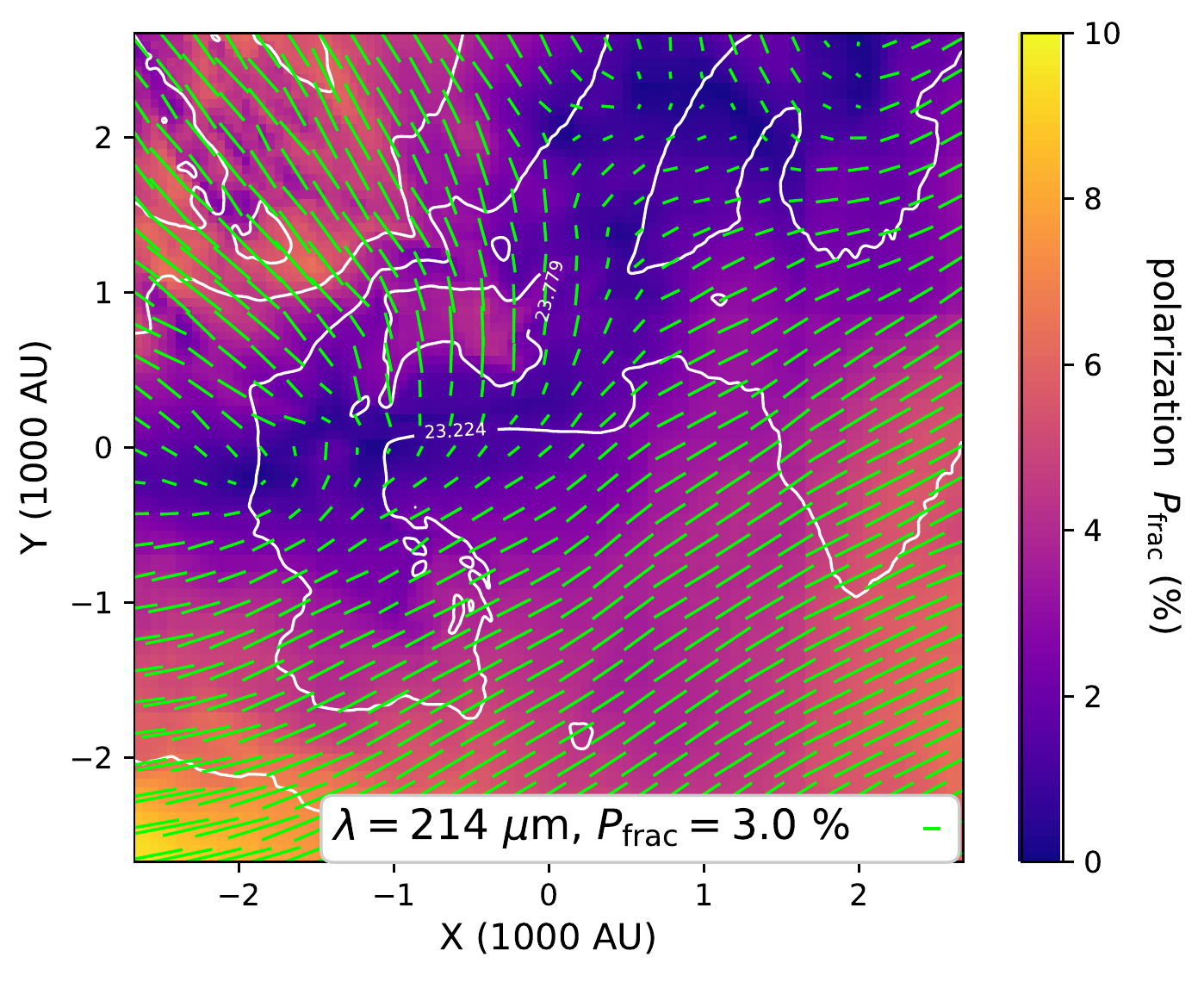}
     \includegraphics[width=0.195\textwidth]{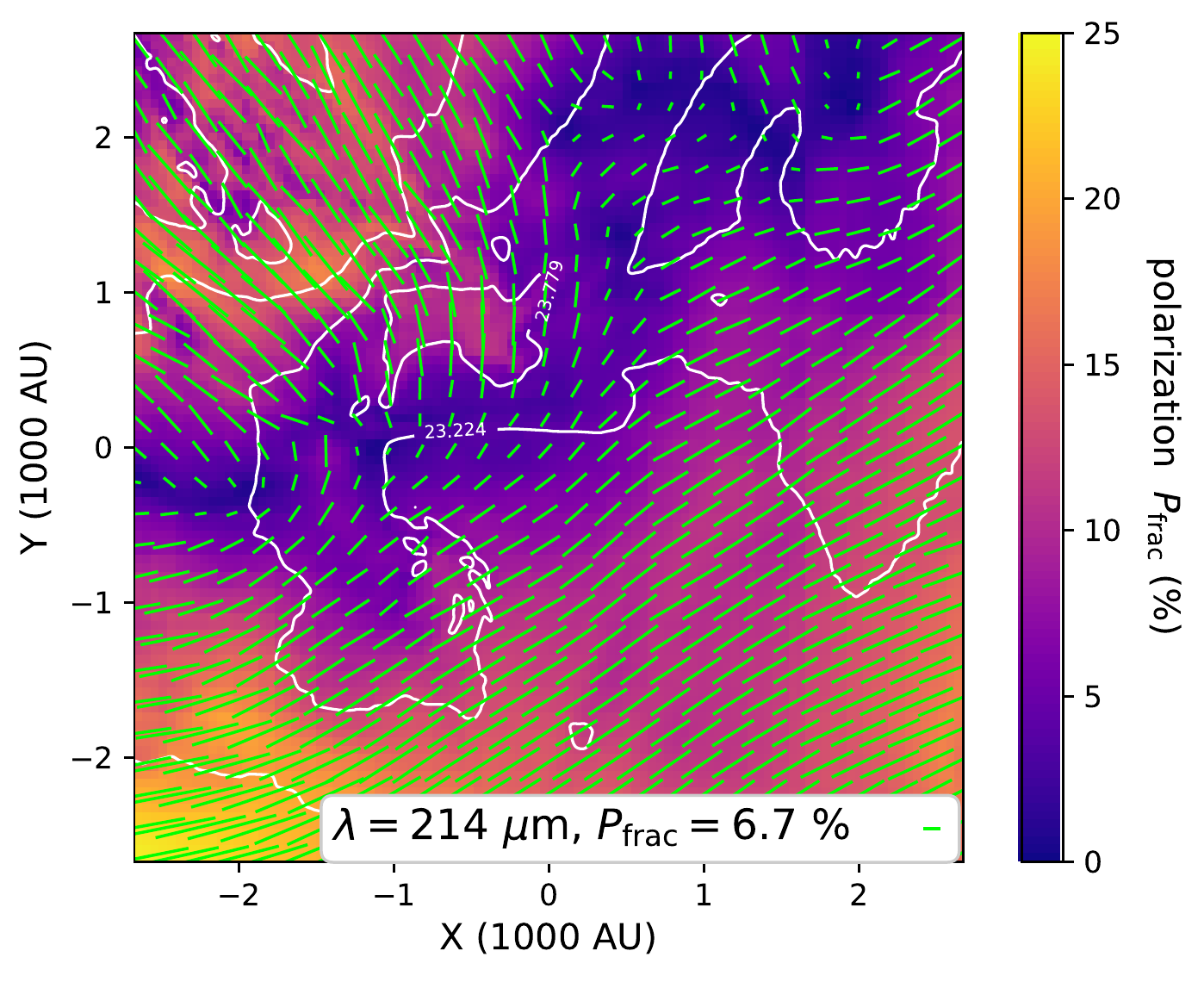}
     \includegraphics[width=0.195\textwidth]{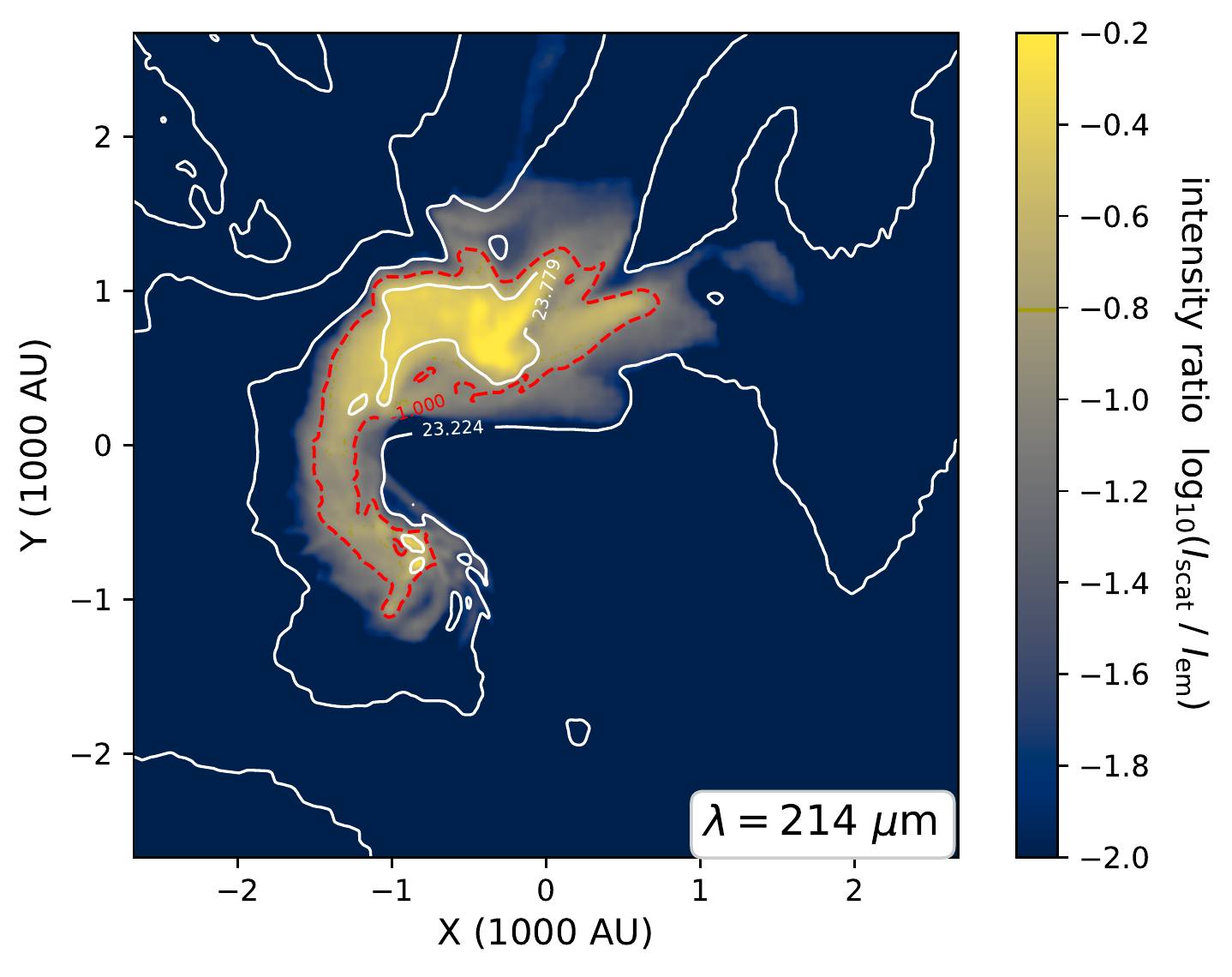}
\end{center}

\begin{center}
    \includegraphics[width=0.195\textwidth]{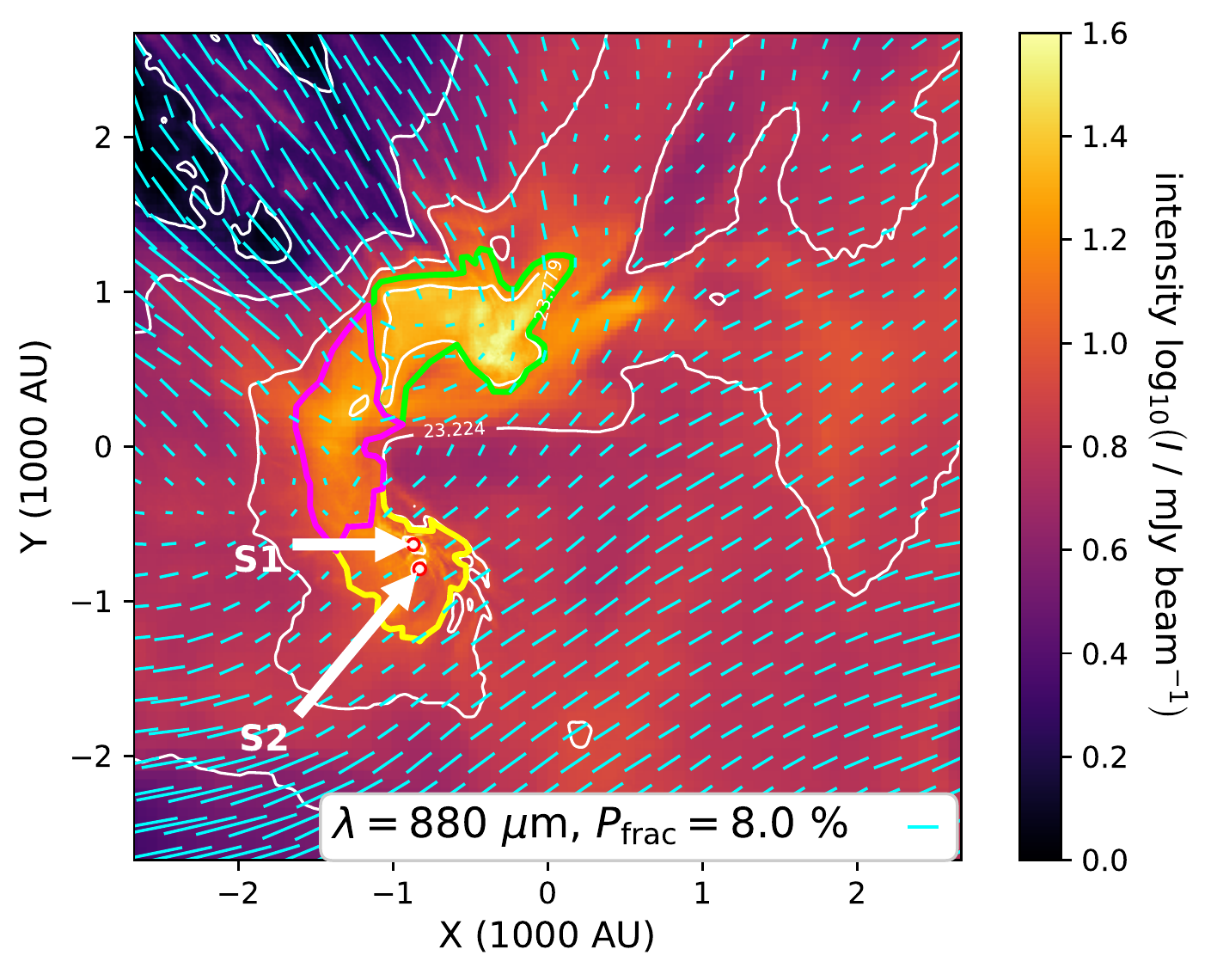}
    \includegraphics[width=0.195\textwidth]{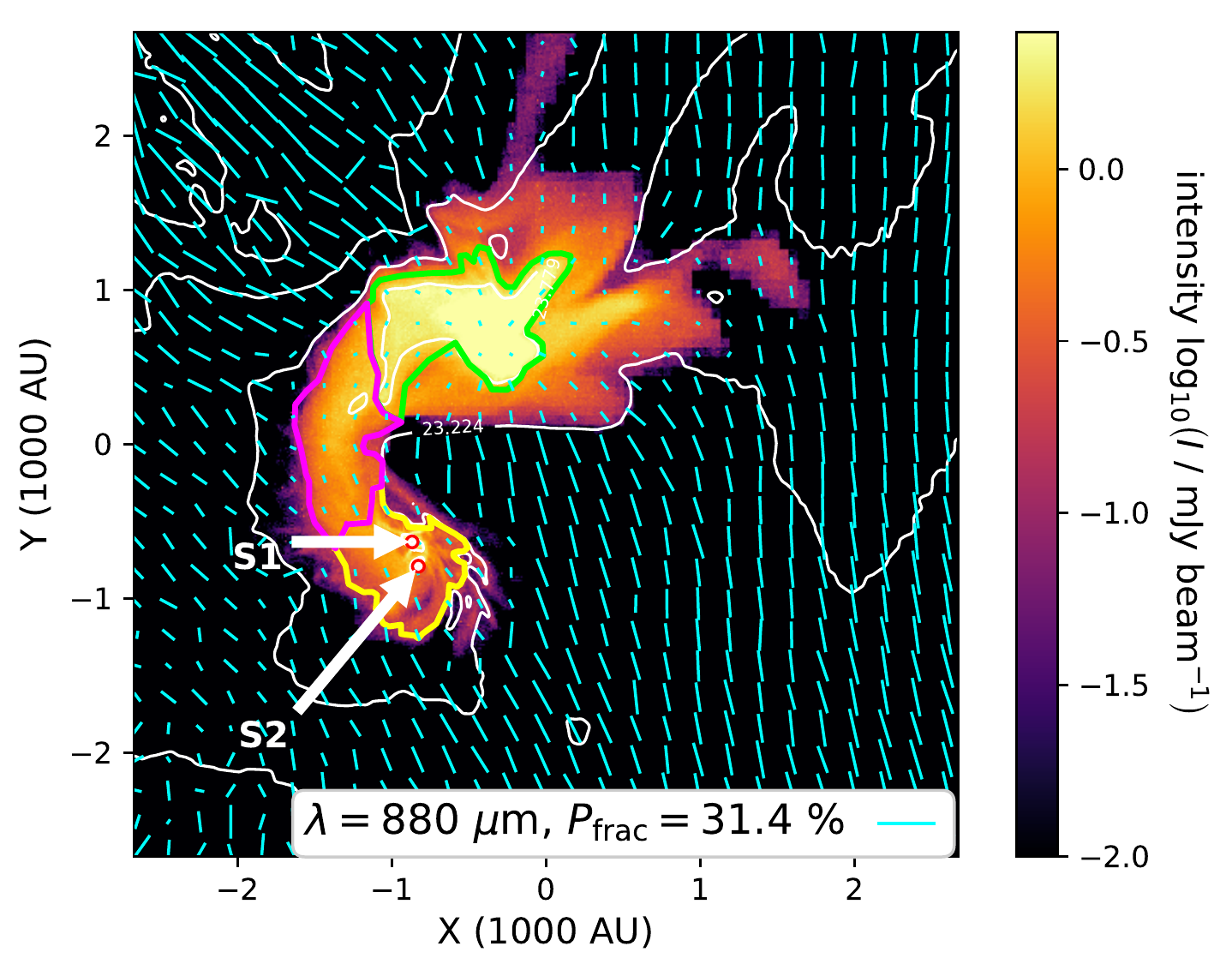}
    \includegraphics[width=0.195\textwidth]{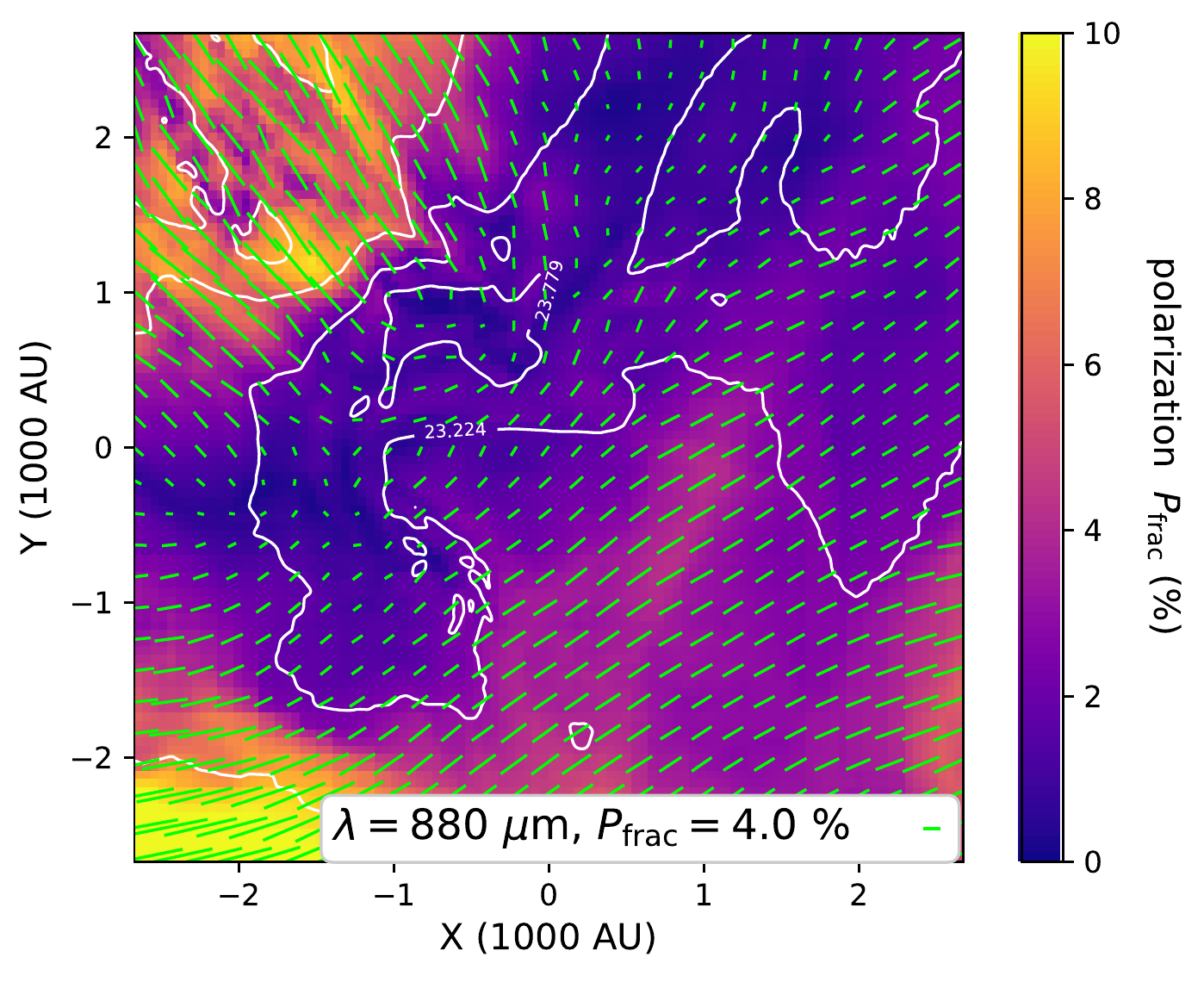}
    \includegraphics[width=0.195\textwidth]{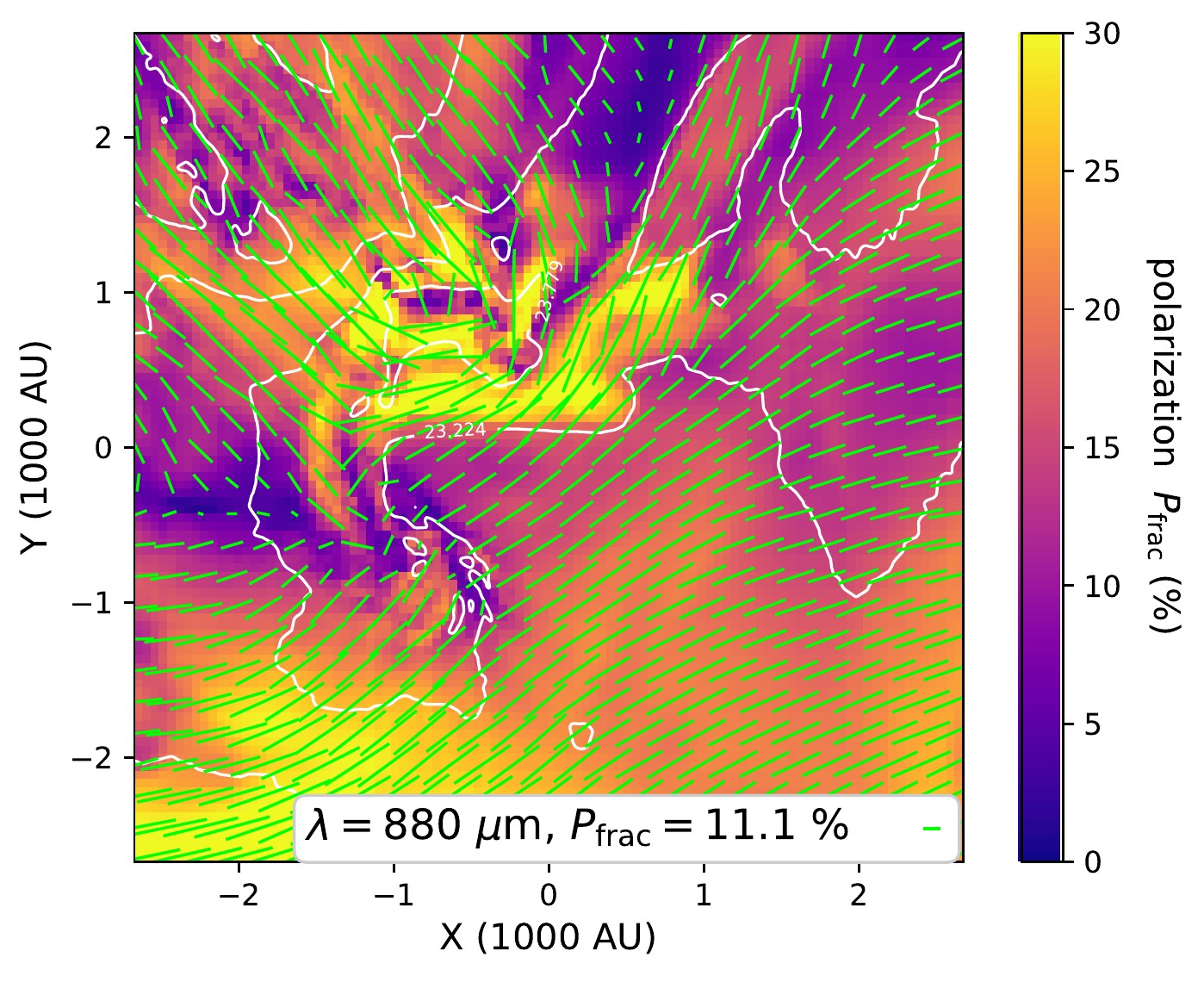}
    \includegraphics[width=0.195\textwidth]{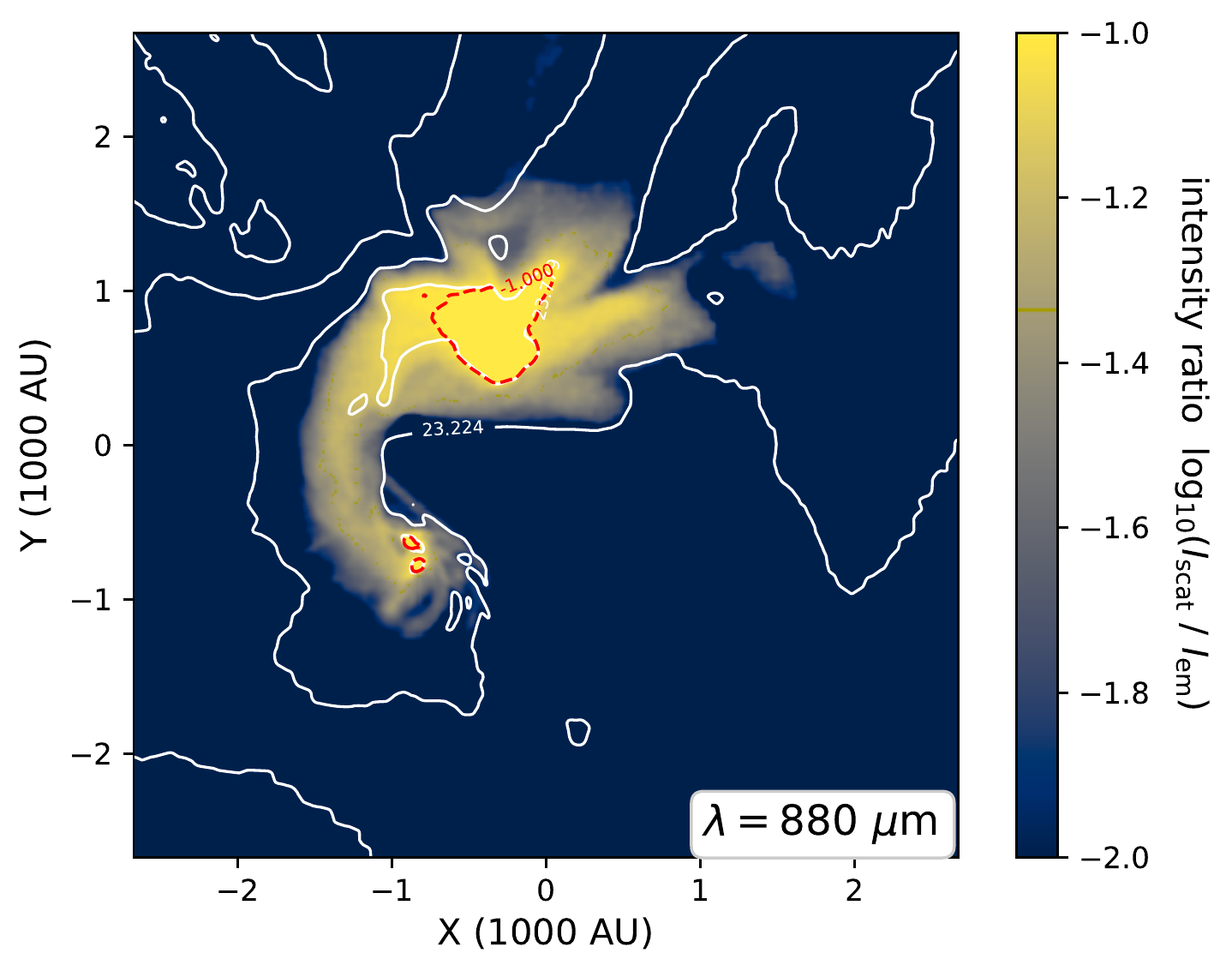}
\end{center}

\caption{Same as \Fig{AppSyntheticDust100} but with $a_{\mathrm{max}}=3\ \mathrm{mm}$ for the dense dust component.}
\label{fig:AppSyntheticDust3mm}
\end{figure*}

To study the effect of the assumed grain size distribution, we also carried out radiative transfer models assuming an upper dust grain size of 3 mm instead of $100\ \mu$m for the dense region. 
\Fig{AppSyntheticDust100} shows results for an upper dust grain size of $100\ \mu$m as assumed in the paper, and \Fig{AppSyntheticDust3mm} shows the results for an upper grain size of 3 mm. 
Both figures show synthetic maps of emitted radiation, scattered radiation, dust polarization accounting for RATs, polarization assuming perfect alignment, and the ratio of scattered to emitted radiationas observed at $53\ \mu$m wavelength, $214\ \mu$m wavelength,  and $880\ \mu$m wavelength. 
The figures show that for a larger maximum grain size $a_{\rm max}$, the radiation at larger wavelength (here $880 \mu$m) is higher, especially the scattering part. We emphasize that the polarization patterns of scattering and RATs are rather similar for the two grain sizes. In contrast to this, the perfect alignment case predicts much more polarization for larger grains in the dense regions.

\section{$\rho-B$ relation in the region}
\label{app:B} 
\begin{figure}
    \includegraphics[width=\columnwidth]{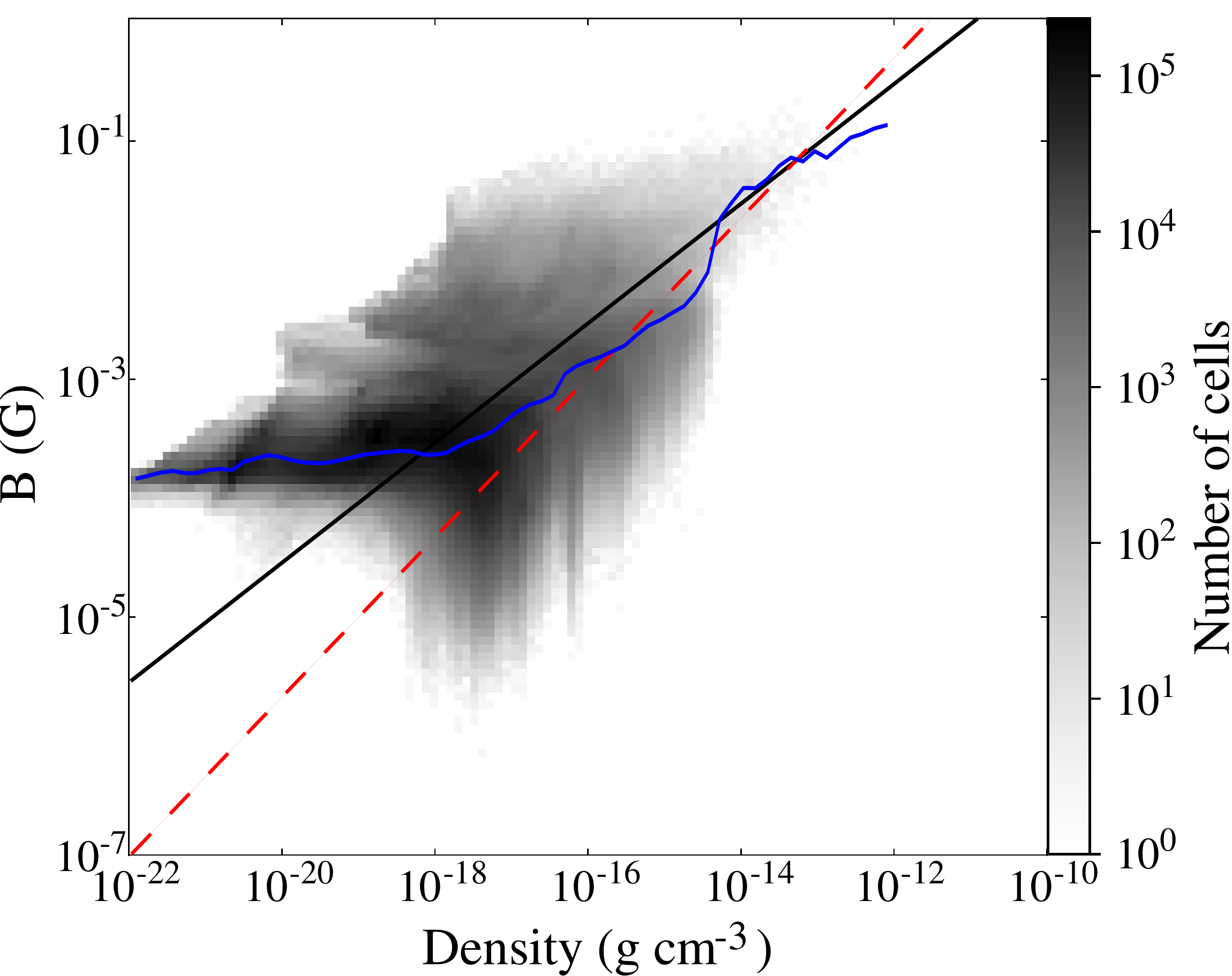} 
        \caption{Same as \Fig{rho_B_r}, but instead of the radius, the number of cells for each combination of density $\rho$ and magnetic field strength $B$ is shown in gray.}
        \label{fig:rho_B_count}
\end{figure}

In \Fig{rho_B_r} we show the magnetic field strength $B$ over density $\rho$ within a region of $10^4$ au. 
The colors in the plot illustrate the radial distance from the primary protostar. 
Because multiple cells can have the same combination of $\Delta \rho$ and $\Delta B$, not all cells are displayed in this figure.
To show the number of cells per combination of $\rho$ and $B$, \Fig{rho_B_count} displays the number of cells per bin. 
Note that the blue solid line shows the volume-averaged average as the cells in our model can have varying size depending on their level of refinement. 
In the selected region within $10^4$ au from the primary protostar, the smallest cells are $\Delta x \approx 4$ au, and the largest cells are $\Delta x \approx 504$ au in length.  

\end{document}